\documentclass[12pt,aps,showpacs]{revtex4}
\usepackage{amsmath,amssymb,amsfonts,epsfig}
\newcommand{\beeq}{\begin{equation}}
\newcommand{\eneq}{\end{equation}}
\newcommand{\be}{\begin{eqnarray}}
\newcommand{\ee}{\end{eqnarray}}
\newcommand{\bpic}{\begin{picture}}
\newcommand{\epic}{\end{picture}}
\newcommand{\bs}{\begin{scriptsize}}
\newcommand{\es}{\end{scriptsize}}

\def\la{\raise.16ex\hbox{$\langle$} \, }
\def\ra{\, \raise.16ex\hbox{$\rangle$} }

\def\a{\alpha}

\def\l{\lambda}

\def\Box{\kern1pt\vbox{\hrule height 1.2pt\hbox{\vrule width 1.2pt\hskip 3pt
   \vbox{\vskip 6pt}\hskip 3pt\vrule width 0.6pt}\hrule height 0.6pt}\kern1pt}
\def\gtwid{\mathrel{\raise.3ex\hbox{$>$\kern-.75em\lower1ex\hbox{$\sim$}}}}
\def\ltwid{\mathrel{\raise.3ex\hbox{$<$\kern-.75em\lower1ex\hbox{$\sim$}}}}

\begin{document}


\title{Vacuum Fluctuations of a Scalar Field during Inflation:\\Quantum versus Stochastic Analysis}
\author{V. K. Onemli}\email{onemli@itu.edu.tr}

\affiliation{$^{\ast}$Department of Physics, Istanbul Technical
University, Maslak, Istanbul 34469, Turkey}
\begin{abstract}
We consider an infrared truncated massless minimally coupled scalar field with a quartic self-interaction in the locally de Sitter background of an inflating universe. We compute the two-point correlation function of the scalar at one and two-loop order applying quantum field theory. The tree-order correlator at a fixed comoving separation (that is at increasing physical distance) freezes in to a nonzero value. At a fixed physical distance, it grows linearly with comoving time. The one-loop correlator, which is the dominant quantum correction, implies a negative temporal growth in the correlation function, at this order, at a fixed comoving separation and at a fixed physical distance. We also obtain quantitative results for variance in space and time of one and two-loop correlators and infer that the contrast between the vacuum expectation value and the variance becomes less pronounced when the loop corrections are included. Finally, we repeat the analysis of the model applying a stochastic field theory and reach the same conclusions.
\end{abstract}


\maketitle \vskip 0.2in \vspace{.4cm}

\section{Introduction}
Quantum field theory (QFT) of the massless, minimally coupled (MMC) scalar with a quartic self-interaction on a locally de Sitter background predicts \cite{OW1,OW2,KOW1,BOW,KO,O} several intriguing enhanced quantum effects of cosmological interest. A locally de Sitter background provides the simplest framework for an inflating spacetime. The cosmological constant $\Lambda$ drives inflation in this framework and the scalar is just a spectator field. The operation of inflation on the fluctuating quantum vacuum is to rip virtual particle pairs out of the vacuum by the Hubble flow before they find time to annihilate each other. After they emerge from the vacuum, they become real. As in flat spacetime, particles with smaller masses persist longer. Any massless particle which happens to emerge from the vacuum with sufficiently small wave number $k\!\lesssim\!Ha(t)$, where $H$ is the expansion rate, $a(t)$ is the cosmic scale factor, and $t$ is the comoving time, can persist forever \cite{RPBW1, rev1, rev2}. However, the particle production rate is suppressed for conformally invariant particles. Hence, this quantum effect  during inflation is enhanced for the particles that are classically conformally non-invariant and effectively massless. The MMC scalars that we consider in this paper are an example of such particles.

We computed \cite{OW1,OW2,KOW1} the renormalized
vacuum expectation value (VEV) of the stress-energy tensor for the MMC scalars endowed with a quartic self-interaction during inflation, at one
and two-loop order. We showed that the renormalized energy density $\rho_{{\rm ren}}$
and pressure $p_{{\rm ren}}$ of the scalar violate the classical
weak energy condition $\rho + p\geq0$ on
cosmological scales at two-loop order. The equation of state parameter $w\equiv
p_{{\rm ren}}/\rho_{{\rm ren}}\!<\!-1$. As a result, a
phase of super-acceleration is induced. This effect, however, is temporary since the field
develops~\cite{BOW} a growing self-mass squared due to quantum processes and
the particle production is cut off. The induced mass remains perturbatively small and does not go tachyonic which ensures the stability of the model \cite{KO}. We studied~\cite{KOW2} how such a spectator scalar field might affect the measured curvature power spectrum $\Delta^2_{\mathcal{R}}(k,t)$ defined in terms of the Fourier transform
of the equal time {\it two-point correlation function} of~the derivative stripped three-curvature scalar field ${\mathcal{R}}(t,\vec{x})$
\beeq
\Delta^2_{\mathcal{R}}(k,t)\!\equiv\!\frac{k^3}{2 \pi^2}\!\int \!\!
d^3x \, e^{-i \vec{k} \cdot \vec{x}} \langle \Omega
\vert \mathcal{R}(t, \vec{x}) \mathcal{R}(t, \vec{0})
\vert \Omega\rangle \; .
\eneq
More recently, we computed \cite{O} the one and two-loop corrected power spectrum
\beeq
\Delta^2_{\varphi}(k,t)\!=\!\frac{k^3}{2 \pi^2}\!\int\!\! d^3x\,
e^{-i\vec{k}\cdot\vec{x}}\langle\Omega|\varphi(t, \vec{x})
\varphi(t, \vec{0})|\Omega\rangle\; ,\eneq
of the spectator scalar in our model. Correlations in various scalar potential models during inflation have been studied \cite{vascfl}. In this paper, we compute the one and two-loop corrected two-point correlation function of the spectator scalar\beeq
\langle\Omega|\bar\varphi(t, \vec{x})
\bar\varphi(t'\!, \vec{x}\,')|\Omega\rangle\; ,\label{twopoint}
\eneq
for $t'\!\!\leq\!t$ and  $\vec{x}\,'\!\neq\!\vec{x}$, in our model. Bar on top of the scalar indicates that the ultraviolet modes with comoving wave number $k\!>\!Ha$ are cut off the field. Hence, the IR modes with $H\!<\!k\!<\!Ha(t)$ are retained. The IR truncated field is guaranteed to reproduce the leading IR logarithms of scalar potential models in QFT. The advantage of using $\bar\varphi(t, \vec{x})$ is that it is a simple way of recovering the most important contributions; the disadvantage is that it leaves out sub-leading contributions.

Physical wavelengths $\lambda_{\rm phys}$ of the IR modes range from $2\pi/H$ to $2\pi a(t)/H$. Had the lower limit of the $k$ for the retained modes taken to be zero rather than $H$, the model would have suffered from IR divergences. To regulate these divergences we take the spacetime topology as $T^{D-1}\!\times\!R$ where the coordinate toroidal radii $r^i\!=\!2\pi/H$. Here, the spatial index $i=1, 2, \dots, (\!D\!-\!1)$. The VEV of the field $\bar{\varphi}(t, \vec{x})$ is zero at any event, hence the VEV of variation $\Delta \bar{\varphi}(t, t' ; \vec{x}, \vec{x}\,') \!\equiv\!
\bar{\varphi}(t, \vec{x})-\bar{\varphi}(t'\!, \vec{x}\,')$ vanishes. We calculate the non-vanishing~variance
\be
&&\hspace{-1.cm}\sigma^2_{\Delta \bar{\varphi}}(t, t' ; \vec{x}, \vec{x}\,')\!\equiv\!\langle\Omega|\!
\left[\Delta\bar{\varphi} \!-\!\langle\Omega|\Delta\bar{\varphi}|\Omega\rangle\right]^2\!|\Omega\rangle\!=\!\langle\Omega|
\left(\Delta\bar{\varphi}\right)^2\!|\Omega\rangle\; ,
\ee
at one and two-loop order, using the two-point correlation function we compute in this paper. We also examine the stochastic properties of the IR truncated scalar in the same model repeating the computations in the context of stochastic field theory (SFT) which is known to be successful in recovering the most important secular effects \cite{recover} in inflationary QFT. Stochastic formulations of inflation have also proven useful in studying initial conditions \cite{initial}, global structure \cite{global}, non-Gaussianity \cite{nG}, power spectrum of a spectator scalar \cite{Kuhnel} and conditions for eternal inflation \cite{sari}. Starobinsky developed \cite{Staro} a stochastic formalism which reproduces the leading IR logs at each order in perturbation theory in scalar potential models. Starobinsky's formalisms has recently been extended \cite{W1,W4,WRGflow} to include scalars which interact with fermions \cite{W3} and with gauge particles \cite{Wstocsqed} but the inclusion of derivative interactions such as those of quantum gravity \cite{MiaoWood} and the nonlinear sigma model \cite{Kit1,Kit2} is a major unsolved problem. In this paper, we use the SFT approach of Ref.~\cite{W0} where the annihilation and creation operators in a quantum field are regarded as two complex conjugate random variables whose real and imaginary parts follow the Gaussian probability distribution with mean zero and standard deviation one. Results we achieved by applying the SFT are in perfect agreement with the ones we obtained applying the QFT. See also Refs~\cite{Rig1,Rig2} comparing different aspects of quantum and stochastic formalisms.

A technique \cite{SY, Wstocqgrav, FMSVV1, FMSVV2} to compute expectation values of functionals of a stochastic field is to integrate the functional weighted with a probability density function which obeys a Fokker-Planck equation. Employing this technique, we finally note a stochastic check of our two-point correlation function in the equal spacetime limit.

This introduction is the first of five sections. The outline of the remainder is as follows. In Sec.~\ref{sec:model} we specify the background spacetime geometry and present the Lagrangian of the model. In Sec.~\ref{sec:Quant} we analyze the model applying the QFT: We compute the two-point correlation function and the variance of the IR truncated scalar field with a quartic self-interaction at one and two-loop order during inflation. In Sec.~\ref{sec:Stoch} we analyze the model applying the SFT.
We summarize our conclusions in Sec.~\ref{sec:conclusions}. The Appendices comprise the details of some computations in the paper.

\section{The Model}
\label{sec:model}

We consider a MMC scalar with a self-interaction potential $V(\varphi)$. The Lagrangian density of the model is
\begin{equation}
\mathcal{L} = -\frac{1}{2} (1 \!+\! \delta Z) \partial_{\mu} \varphi \partial_{\nu}
\varphi g^{\mu\nu} \sqrt{-g} \!-\! V(\varphi) \sqrt{-g} \; ,\label{lagden}
\end{equation}
where $\delta Z$ is the field strength counterterm and $g_{\mu\nu}$ is the background metric of locally de~Sitter spacetime. The invariant
line element can be expressed in comoving
coordinates~as\beeq ds^2\!=\! g_{\mu\nu} dx^{\mu}
dx^{\nu}\!=\!-dt^2 \!+\! a^2(t) d\vec{x} \cdot d\vec{x}
\; , \eneq where the scale factor \beeq a(t)=e^{H t} \; ,\eneq is normalized to unity at the initial time $t_i\!=\!0$. We work in a $D$-dimensional spacetime,
hence the indices $\mu,\nu\!=\!0,1,2, \dots,(\!D\!-\!1)$. Thus, $x^\mu\!=\!(x^0\!,\vec x)$, $x^0\!\equiv\!t$, and
$\partial_\mu\!=\!(\partial_0,\vec\nabla)$ in our
notation. We use the open conformal coordinate patch of the full de Sitter manifold where each of the $(\!D\!-\!1)$ spatial comoving coordinates $x^i$ lies in the range $0\!\leq\!x^i\!\leq\!H^{-1}$.

The equation of motion for the scalar field
\beeq
\ddot{\varphi}(t,\vec{x})\!+\!(\!D\!-\!1) H \dot{\varphi}(t,\vec{x})\!-\!
\frac{\nabla^2}{a^2} \varphi(t,\vec{x})\!+\!
\frac{V'(\varphi)(t,\vec{x})}{1\!+\!\delta Z}= 0 \; , \label{feq}
\eneq
where a dot denotes derivative with respect to comoving time $t$, has the solution
\beeq
\varphi(t,\vec{x})\!=\!\varphi_0(t,\vec{x})
\!-\!\!\int_0^t\!\!dt' a^{D-1}(t')\!\!\int\!d^{D-1}x'G(t,
\vec{x};t'\!, \vec{x}\,')\frac{V'(\varphi)(t'\!,\vec{x}\,')}{1\!+\!\delta Z} \; .
\label{fullfield}
\eneq
The free field $\varphi_0(t, \vec{x})$ in Eq.~(\ref{fullfield}) obeys
the linearized equations of motion and agrees with the full field
at $t\!=\!t_i\!=\!0$. The Green's function ${G}(t,
\vec{x};t'\!, \vec{x}\,')$, on the other hand, is any solution of the equation
\beeq
\ddot{G}(t,
\vec{x};t'\!, \vec{x}\,')\!+\!(\!D\!-\!1)H \dot{G}(t,
\vec{x};t'\!, \vec{x}\,')\!-\!\frac{\nabla^2}{a^2} G(t,
\vec{x};t'\!, \vec{x}\,')\!=\!\delta(t\!-\!t')\delta^{D-1}(\vec{x}\!-\!\vec{x}\,')\; ,\label{greeneqn}\eneq
which obeys the retarded boundary conditions. We continue studying the model applying the QFT in the next section. We study it applying the SFT in Sec.~\ref{sec:Stoch}.

\section{Quantum Field Theoretical Analysis}

\label{sec:Quant}
In this section, we analyze the model described in Sec.~\ref{sec:model} in the QFT point of view. Section~\ref{subsect:free} is reserved for the non-interacting limit of the model. Section~\ref{subsect:int} considers the model with a quartic self-interaction potential.

\subsection{Free Theory}
\label{subsect:free}

Recall that the background is homogeneous and isotropic with a topology $T^{D-1}\!\times\!R$ and we work on a comoving coordinate patch where the $D\!-\!1$ spatial coordinates $x^i$ lie in the finite range $0\!\leq\!x^i\!\leq\! H^{-1}$. The modes of fields are, therefore,
{\it discrete} as they are for any finite spatial manifold. Hence, we expand the free scalar field in a spatial Fourier {\it series} \cite{W0},
\begin{equation}
\varphi_0(t,\vec{x}) \!=\! H^{\frac{D-1}{2}} \Bigg\{\!Q\!-\!\frac{P}{(\!D\!-\!1) H a^{D-1}(t)}
+\!\!\sum_{\vec{n} \neq 0} \!\left[ u(t,k) e^{i \vec{k} \cdot \vec{x}}
\hat{A}_{\vec{n}} +\! u^*(t,k) e^{-i \vec{k} \cdot \vec{x}} \hat{A}^{\dagger}_{\vec{n}}
\right]\!\!\Bigg\} \; , \label{freefield}
\end{equation}
where the comoving wave vector $\vec{k}\!=\!2 \pi H \vec{n}$ and the comoving wave number $k\equiv\|\vec{k}\|$.
The mode with physical wave number $k_{\rm
phys}(t)\!=\!k/a(t)$ crosses the horizon for the first time at $t\!=\!t_k$. Hence, its physical wavelength
$\lambda_{\rm phys}(t_k)\!=\!2\pi k^{-1}_{\rm phys}(t_k)\!=\!H^{-1}(t_k)$. The expansion rate $H$, however, is
constant during de Sitter inflation, $H(t_k)\!=\!H$. Therefore, this mode has
$k\!=\!2\pi H a(t_k)\!=\!2\pi H e^{Ht_k}$. The Bunch-Davies mode function $u(t, k)$ in Eq.~(\ref{freefield}) is a solution of the spatial Fourier transform of the linearized field equation,
\be
\widetilde{\ddot\varphi}_0(t, \vec{k})\!+\!(\!D\!-\!1)H\widetilde{\dot\varphi}_0(t, \vec{k})\!+\!\frac{k^2}{a^2(t)}\widetilde{\varphi}_0(t, \vec{k})=0\; ,
\label{LinFiEq}
\ee
for the modes with $k\!\neq\!0$. It can be given in terms of the Hankel function of the first kind as
\beeq
u(t, k) \!=\! i\sqrt{\frac{\pi}{4Ha^{D-1}(t)}}\,
\mathcal{H}^{(1)}_{\frac{D-1}{2}}\! \Bigl(\frac{k}{H a(t)}\Bigr)
\; .
\label{udef}\eneq
For the $k\!=\!0$ mode, the two solutions of Eq.~(\ref{LinFiEq}) are a constant and $\frac{a^{-(\!D-1)}}{(D-1)H}$. Hence, the first two terms in Eq.~(\ref{freefield}) represent the zero mode of the quantum field $\varphi_0(t,\vec{x})$. The operators $Q$ and $P$ satisfy the commutation relation\beeq
\left[Q, P\right] = i \; .
\eneq They can be expressed in terms of the Fourier transforms of the field and its first time~derivative for $k\!=\!0$, evaluated at $t=0$,
\be
Q\!\!&=&\!\!\frac{\widetilde{\dot\varphi}_0(0, 0)
\!+\!(\!D\!-\!1)H{{\widetilde{\varphi}}}_0(0,0)}{(\!D\!-\!1)H^{\frac{D+1}{2}}}\; ,\\
P\!\!&=&\!\!\frac{\widetilde{\dot\varphi}_0(0, 0)}{H^{\frac{D-1}{2}}}\; .
\ee The annihilation and creation
operators, $\hat{A}_{\vec{n}}$ and $\hat{A}^{\dagger}_{\vec{n}}$ in Eq.~(\ref{freefield}), may change as different Green's functions are used in
Eq.~(\ref{fullfield}), but their nonzero commutation relation
\begin{equation}
\left[\hat{A}_{\vec{n}} ,
\hat{A}^{\dagger}_{\vec{m}} \right] \!=\! \delta_{\vec{n} , \vec{m}} \; ,\label{aadagger}
\end{equation}
remains fixed. They can also be expressed in terms of $\widetilde{\varphi}_0(0, \vec{k})$ and $\widetilde{\dot\varphi}_0(0, \vec{k})$. Recall that Eq.~(\ref{aadagger}) and the canonical commutator \beeq
\left[\widetilde{\varphi}_0(t, \vec{k}), \widetilde{\dot\varphi}_0(t, \vec{k}\,')\right]\!=\!\frac{i}{a^{D-1}(t)}H^{D-1}\delta_{\vec{n},\vec{n}\,'}\; ,
\eneq fixes the Wronskian
\beeq
u(t, k)\dot{u}^*(t, k)\!-\!u^*(t, k)\dot{u}(t, k)\!=\!\frac{i}{a^{D-1}(t)}\; .\label{Wronskian}
\eneq
Using Eqs.~(\ref{freefield}) and (\ref{Wronskian}) we obtain
\be
\hat{A}_{\vec{n}}\!\!&=&\!\!\frac{-i}{H^{\frac{D-1}{2}}}\!\left[\dot{u}^*(0, k)\widetilde{\varphi}_0(0, \vec{k})\!-\!u^*(0, k)\widetilde{\dot\varphi}_0(0, \vec{k})\right]\; ,\nonumber\\
\hat{A}^\dagger_{\vec{n}}\!\!&=&\!\!\frac{i}{H^{\frac{D-1}{2}}}\!\left[\dot{u}(0, k)\widetilde{\varphi}_0(0, -\vec{k})\!-\!u(0, k)\widetilde{\dot\varphi}_0(0, -\vec{k})\right]\; .
\ee
The state $\vert \Omega \rangle$ annihilated by all $\hat{A}_{\vec{n}}$
is known as Bunch-Davies~vacuum. What we assume about the dependence of vacuum state $\vert \Omega \rangle$ on the zero mode operators $Q$ and $P$ does not matter much because there is only one zero-mode as opposed to an ever-increasing number of nonzero modes. Thus, we neglect them in free field expansion~(\ref{freefield}).

Free field expansion~(\ref{freefield}) also includes arbitrarily large wave numbers which cause ultraviolet divergences. Therefore, we may consider the expansion obtained by cutting out the sub-horizon (ultraviolet) modes and retaining only the super-horizon (infrared) ones
\beeq
\varphi_0(t,\vec{x}) \!=\! H^{\frac{D-1}{2}}\! \sum_{\vec{n} \neq 0} \Theta(\!Ha(t)\!-\!k)\!\left[ u(t,k) e^{i \vec{k} \cdot \vec{x}}
\hat{A}_{\vec{n}} +\! u^*(t,k) e^{-i \vec{k} \cdot \vec{x}} \hat{A}^{\dagger}_{\vec{n}}
\right]\; .\label{expantruncwithu}
\eneq
In the above expansion, we may further take the IR limit of the mode function~(\ref{udef}) given in terms of the Hankel function\beeq
\mathcal{H}^{(1)}_{\frac{D-1}{2}}\!(z)\!=\!{J}_{\frac{D-1}{2}}\!(z)\!+\!i{Y}_{\frac{D-1}{2}}\!(z)\; .\label{HankelinBessel}
\eneq
Here $z\!\equiv\!\frac{k}{H a(t)}$, ${J}_{\frac{D-1}{2}}$ is the Bessel function of order $\frac{D-1}{2}$ and
\beeq
{Y}_{\frac{D-1}{2}}\!(z)\!=\!
{J}_{\frac{D-1}{2}}\!(z)\cot\!\left(\!{\frac{(\!D\!-\!1)\pi}{2}}\!\right)
\!-\!{J}_{-\left(\!\frac{D-1}{2}\!\right)}\!(z)\csc\!\left(\!{\frac{(\!D\!-\!1)\pi}{2}}\!\right)\; ,\label{Neumann}
\eneq
is the Neumann function of order $\frac{D-1}{2}$. Using the series expansion of the Bessel function\beeq
{J}_{\frac{D-1}{2}}\!(z)\!=\!\left(\frac{z}{2}\right)^\frac{D-1}{2}\!\!
\left[\frac{1}{\Gamma\!\left(\frac{D-1}{2}\!+\!1\right)}\!+\!\mathcal{O}(z^2)\right]\; ,\label{BesselSeries}
\eneq
the reflection formula for the Gamma function
\beeq
p\pi\csc(p\pi)\!=\!\Gamma(1\!-\!p)
\Gamma(1\!+\!p)\; ,\label{gammareflection}
\eneq
with $p\!=\!\frac{D-1}{2}$ and the identity
\beeq
\Gamma\!\left(\!\frac{D\!-\!1}{2}\!\right)
\!=\!\frac{\sqrt{\pi}}{2^{D-2}}\frac{\Gamma(\!D\!-\!1)}{\Gamma\left(\!\frac{D}{2}\right)}\; ,\label{gammaidentity}
\eneq
one gets \cite{Wstocqgrav, Wstocsqed, W3} the leading IR limit of the mode function
\beeq
u(t, k)\longrightarrow
\frac{\Gamma(\!D\!-\!1)}{\Gamma(\!\frac{D}{2})}\frac{H^{\frac{D}{2}-1}}{(2k)^{{(D-1)}/{2}}} \Biggl\{\!1\!+\!
\mathcal{O}\Bigl(\frac{k^2}{H^2 a^2(t)}\Bigr)\!\Biggr\}\; .\label{modeleadingorder}
\eneq
Then, substituting $u(t, k)$ in Eq.~(\ref{modeleadingorder}) into Eq.~(\ref{expantruncwithu}) we obtain the IR truncated free field in $D$-dimensions,
\be
{\bar\varphi}_0(t, \vec{x})\!=\!\sum_{\vec{n} \neq 0} \!\frac{\Gamma(\!D\!-\!1)}{\Gamma(\!\frac{D}{2})}\frac{H^{D-3/2}}{(2 k)^{{(D-1)}/{2}}}
\Theta\Bigl(\!H a(t) \!-\! k\!\Bigr)\!\left[e^{i \vec{k} \cdot \vec{x}}
\hat{A}_{\vec{n}} \!+\! e^{-i \vec{k} \cdot \vec{x}} \hat{A}^{\dagger}_{\vec{n}}
\right] \; .\label{whitef}
\ee
Let us define the first time derivative of the field ${\bar\varphi}_0(t,\vec{x})$ as
\be
f_{\!\bar\varphi_0}(t,\vec{x})\!\equiv\!\dot{{\bar\varphi}}_0(t,\vec{x})\!=\!\sum_{\vec{n} \neq 0} \!\frac{\Gamma(\!D\!-\!1)}{\Gamma(\!\frac{D}{2})}
\frac{H^{D-\frac{1}{2}}}{2^{{(D-1)}/{2}}k^{{(D-3)}/{2}}}
\,\delta_{Ha(t_n)\, ,\,k}\Bigl[ e^{i \vec{k} \cdot \vec{x}} \hat{A}_{\vec{n}} \!+\!
 e^{-i \vec{k} \cdot \vec{x}} \hat{A}^{\dagger}_{\vec{n}}
\Bigr] \; , \label{stoksour} \ee
where $t_n=\frac{\ln(2\pi n)}{H}$ and $n\equiv\|\vec{n}\|$. Its stochastic analog represents stochastic noise in Sec.~\ref{subsec:check}.

We now want to calculate the two-point correlation function of the IR truncated free field
\beeq\langle\Omega|{\bar\varphi}_0(t,\vec{x}) {\bar\varphi}_0(t'\!,\vec{x}\,')|\Omega\rangle\; ,\eneq
for $t'\!\!\leq\!t$ and  $\vec{x}\,'\!\neq\!\vec{x}$. It will be needed in Sec.~\ref{subsect:int} where we compute
the correlation function of the IR truncated full field. Using Eqs.~(\ref{aadagger}) and (\ref{whitef}), we obtain
\be
&&\hspace{-1.3cm}\langle\Omega|{\bar\varphi}_0(t,\vec{x}) {\bar\varphi}_0(t'\!,\vec{x}\,')|\Omega\rangle\nonumber\\
&&\hspace{1.5cm}=\!\frac{\Gamma^2(\!D\!-\!1)}{\Gamma^2(\!\frac{D}{2})}
\frac{H^{2D-3}}{2^{D-1}}\sum_{\vec{n}\neq 0}\!\frac{\Theta(\!H\!a(t)\!-\! k)}{k^{{(D-1)}/{2}}}e^{i\vec{k}\cdot\vec{x}}\sum_{\vec{m}\neq 0}\!\frac{\Theta(\!H\!a(t')\!-\! k')}{k'^{{(D-1)}/{2}}}e^{-i\vec{k}\,'\cdot\vec{x}\,'}\!\delta_{\vec{n},\vec{m}}\; .
\ee Here $\vec{k}'\!=\!2\pi H\vec{m}$. When $\vec{m}\!=\!\vec{n}$ we have $\vec{k}'\!=\!\vec{k}$, therefore,
\be
&&\hspace{-1cm}\langle\Omega|{\bar\varphi}_0(t,\vec{x}) {\bar\varphi}_0(t'\!,\vec{x}\,')|\Omega\rangle\nonumber\\
&&\hspace{0.5cm}=\frac{\Gamma^2(\!D\!-\!1)}{\Gamma^2(\!\frac{D}{2})}
\frac{H^{D-2}}{2^{2D-2}\pi^{D-1}}\!\sum_{\vec{n}\neq 0}\!\frac{\Theta(\!H\!a(t)\!-\! H2\pi n)\,\Theta(\!H\!a(t')\!-\!H2\pi n)}{n^{D-1}}e^{i2\pi H\vec{n}\cdot(\vec{x}-\vec{x}\,')}
\; .\label{sumtheta}
\ee
The discrete sum over $\vec{n}$ in Eq,~(\ref{sumtheta}) can be approximated as a $(\!D\!-\!1)$-dimensional integral
\be
&&\hspace{-1cm}\sum_{\vec{n}\neq 0}\!\frac{\Theta(\!H\!a(t)\!-\! H2\pi n)\,\Theta(\!H\!a(t')\!-\!H2\pi n)}{n^{D-1}}e^{i2\pi H\vec{n}\cdot(\vec{x}-\vec{x}\,')}\nonumber\\
&&\hspace{4cm}\simeq\!\!\!\int\!\!d\Omega_{D-1}\!\!\int_0^\infty\!\frac{d k}{k}\,\Theta(\!Ha(t)\!-\!k)\,\Theta(\!Ha(t')\!-\!k)e^{ik \Delta x \cos{(\theta)}}\; ,\label{angkeyeqn}
\ee
where $\Delta x\equiv\parallel\!\Delta\vec{x}\!\parallel=\parallel\!\vec{x}\!-\!\vec{x}\,'\!\!\parallel$ and $\theta$ is the angle between the vectors $\vec{n}$ and $\Delta\vec{x}$. Evaluating the angular integrations on the right side of Eq.~(\ref{angkeyeqn}) yields
\be
&&\hspace{-1cm}2^{D-1}\pi^{\frac{D}{2}-1}\frac{\Gamma\left(\!\frac{D}{2}\right)}{\Gamma(\!D\!-\!1)}\int_0^\infty\!\!\frac{d k}{k}\Theta(\!Ha(t)\!-\!k)\,\Theta(\!Ha(t')\!-\!k)\frac{\sin(k\Delta x)}{k\Delta x}\, .\label{sumthetaintegral}
\ee
Because $0\!\leq\!t'\!\leq\!t$, we have $Ha(t')\!\leq\!Ha(t)$ and hence,
\beeq
\Theta(\!H\!a(t)\!-\!k)\,\Theta(\!H\!a(t')\!-\!k)\!=\!\Theta(\!H\!a(t')\!-\!k)\; .
\label{thetaatatprime}
\eneq
Therefore, the integral over $k$ in Eq.~(\ref{sumthetaintegral}),
\be
\int_0^\infty\!\!\frac{d k}{k}\Theta(\!Ha(t')\!-\!k)\frac{\sin(k\Delta x)}{k\Delta x}\!=\!\!\int_H^{Ha(t')}\!\!\frac{d k}{k}\frac{\sin(k\Delta x)}{k\Delta x}\!=\!\!\int_{H\Delta x}^{Ha(t')\Delta x}\!\!\frac{d y}{y}\frac{\sin(y)}{y}\nonumber\\
={\rm ci}\!\left(a(t')H\!\Delta x\right)\!-\!\frac{\sin\left(a(t')H\!\Delta x\right)}{a(t')H\!\Delta x}\!-\!{\rm ci}(\!H\!\Delta x)\!+\!\frac{\sin(\!H\!\Delta x)}{H\!\Delta x}\, ,\label{logar}
\ee
where the cosine integral
\be
{\rm ci}(\a) \!\equiv\! -\!\int_\a^{\infty} \!\!dt {\cos(t)
\over t} \!=\! \gamma \!+\! \ln(\a) \!+\!\!\int_0^\a \!\!dt {\cos(t)
\!-\!1 \over t}\!=\!\gamma\!+\ln(\a)\!+\!\!\sum_{n=1}^\infty
                 \frac{(-1)^{n}\a^{2n}}{2n\,(2n)!} \; ,\label{cipowerseries}
\ee is an {\it entire} function. The Euler's constant $\gamma \approx .577$. Employing Eqs.~(\ref{thetaatatprime}) and (\ref{logar})  in Eq.~(\ref{sumthetaintegral}) we obtain
\be
&&\hspace{-1.5cm}\sum_{\vec{n}\neq 0}\!\frac{\Theta(\!H\!a(t)\!-\! H2\pi n)\,\Theta(\!H\!a(t')\!-\!H2\pi n)}{n^{D-1}}e^{i2\pi H\vec{n}\cdot(\vec{x}-\vec{x}\,')}\nonumber\\
&&\hspace{-0.5cm}\simeq2^{D-1}\pi^{\frac{D}{2}-1}
\frac{\Gamma\!\left(\!\frac{D}{2}\right)}{\Gamma\!\left(\!D\!-\!1\right)}\!\!\left[{\rm ci}\!\left(a(t')H\!\Delta x\right)\!-\!\frac{\sin\left(a(t')H\!\Delta x\right)}{a(t')H\!\Delta x}\!-\!{\rm ci}(\!H\!\Delta x)\!+\!\frac{\sin(\!H\!\Delta x)}{H\!\Delta x}\right]\; .\label{sumthetafinal}
\ee
We can power expand the time dependent terms in Eq.~(\ref{sumthetafinal}),
\beeq
{\rm ci}(\a)\!-\!\frac{\sin(\a)}{\a}\!=\!\gamma\!-\!1\!+\!\ln(\a)\!+\!\!\sum_{n=1}^\infty
                 \frac{(-1)^{n}(\a)^{2n}}{2n\,(2n\!+\!1)!}\; ,\label{ciminussin}
\eneq
with $\a\equiv a(t')H\!\Delta x$. Substituting Eq.~(\ref{ciminussin}) into Eq.~(\ref{sumthetafinal}) yields
\be
&&\hspace{-1.5cm}\sum_{\vec{n}\neq 0}\!\frac{\Theta(\!H\!a(t)\!-\! H2\pi n)\,\Theta(\!H\!a(t')\!-\!H2\pi n)}{n^{D-1}}e^{i2\pi H\vec{n}\cdot(\vec{x}-\vec{x}\,')}\nonumber\\
&&\hspace{-0.5cm}\simeq2^{D-1}\pi^{\frac{D}{2}-1}
\frac{\Gamma\!\left(\!\frac{D}{2}\right)}{\Gamma\!\left(\!D\!-\!1\right)}\!\!\left[\mathcal{C}(\!\Delta x)\!+\!\ln(a(t'))\!+\!\!\sum_{n=1}^\infty\!
                 \frac{(-1)^{n}(a(t')H\!\Delta x)^{2n}}{2n\,(2n\!+\!1)!}\right]\; ,\label{sumthetafinalseries}
\ee
where we define the comoving spatial separation dependent function\be
\mathcal{C}(\!\Delta x)\!&\equiv&\!\gamma\!-\!1\!+\!\ln(\!H\!\Delta x)\!-\!{\rm ci}(\!H\!\Delta x)\!+\!\frac{\sin(\!H\!\Delta x)}{H\!\Delta x}\label{C}\\
\!&=&\!-\!\sum_{n=1}^\infty\!\frac{(-1)^{n}(\!H\!\Delta x)^{2n}}{2n\,(2n\!+\!1)!}   \; .\label{Cpower}
\ee

Combining Eq.~(\ref{sumtheta}) and the analytic form of the sum given in Eq.~(\ref{sumthetafinal}) gives the two-point correlation function of the IR truncated free field as
\be
&&\hspace{-1.5cm}\langle\Omega|{\bar\varphi}_0(t,\vec{x}) {\bar\varphi}_0(t'\!,\vec{x}\,')|\Omega\rangle\nonumber\\
&&\hspace{-0.5cm}\simeq\!\frac{H^{D-2}}{2^{D-1}\pi^{{D}/{2}}}
\frac{\Gamma\!\left(\!D\!-\!1\right)}{\Gamma\!\left(\!\frac{D}{2}\right)}\!\!\left[{\rm ci}\!\left(a(t')H\!\Delta x\right)\!-\!\frac{\sin\left(a(t')H\!\Delta x\right)}{a(t')H\!\Delta x}\!-\!{\rm ci}(\!H\!\Delta x)\!+\!\frac{\sin(\!H\!\Delta x)}{H\!\Delta x}\right]\; .\label{expphisq}
\ee
We can use this analytic form of the correlator to infer its behavior for a {\it fixed comoving separation} (increasing physical distance) at late $t'$ when $a(t')$ becomes large. Equation~(\ref{expphisq}) and the asymptotic form of the function ${\rm ci}(\alpha)$ for large argument
\be
\hspace{-0.5cm}{\rm ci}(\alpha)\!\!&=&\!\!\frac{\sin(\alpha)}{\alpha}\!\left(1\!-\!\frac{2!}{\alpha^2}
\!+\!\frac{4!}{\alpha^4}\dots\right)
\!-\!\frac{\cos(\alpha)}{\alpha}\!\left(\frac{1}{\alpha}\!-\!\frac{3!}{\alpha^3}
\!+\!\frac{5!}{\alpha^5}\dots\right)\!\longrightarrow\!\frac{\sin(\alpha)}{\alpha}\; ,\label{ciasym}
\ee
imply that the correlator at a fixed comoving separation freezes in to a nonzero constant value at late times.  The usual result in QFT is for it to go to zero as the physical
separation increases.

Combining Eq.~(\ref{sumthetafinalseries}) and Eq.~(\ref{Cpower}) in Eq.~(\ref{sumtheta}) gives the correlation function as a series expansion
\be
\hspace{-0.9cm}\langle\Omega|{\bar\varphi}_0(t,\vec{x}) {\bar\varphi}_0(t'\!,\vec{x}\,')|\Omega\rangle\!\!&\simeq&\!\!\frac{H^{D-2}}{2^{D-1}\pi^{{D}/{2}}}
\frac{\Gamma\!\left(\!D\!-\!1\!\right)}{\Gamma\!\left(\!\frac{D}{2}\right)}\!\!\left[\mathcal{C}(\!\Delta x)\!+\!\ln(a(t'))\!+\!\!\sum_{n=1}^\infty\!
                 \frac{(\!-1)^{n}(a(t')H\!\Delta x)^{2n}}{2n\,(2n\!+\!\!1)!}\right] \label{VEVgeneral}\\
                 \!\!&\simeq&\!\!\frac{H^{D-2}}{2^{D-1}\pi^{{D}/{2}}}
\frac{\Gamma\!\left(\!D\!-\!1\!\right)}{\Gamma\!\left(\!\frac{D}{2}\right)}\!\!\left[\ln(a(t'))\!+\!\!\sum_{n=1}^\infty\!
                 \frac{(\!-1)^{n}(\!H\!\Delta x)^{2n}}{2n\,(2n\!+\!\!1)!}\left(a^{2n}(t')\!-\!1\right)\right] \label{VEVgeneralpowr}\; .
\ee
Equation~(\ref{expphisq}) (or Eq.~(\ref{VEVgeneralpowr})) can be used to interpret the correlator at a {\it fixed physical distance}. As in Ref.~\cite{W0}, we may choose $\Delta x$ so as to keep the physical distance $a(t')\Delta x$ a constant fraction $K$ of the Hubble length,
\beeq
\Delta x\!=\!\frac{K}{Ha(t')}\; ,\label{DeltaX}
\eneq
and use Eq.~(\ref{ciminussin}) in Eq.~(\ref{expphisq}) to obtain
\be
&&\hspace{-1.3cm}\langle\Omega|{\bar\varphi}_0(t,\vec{x}) {\bar\varphi}_0(t'\!,\vec{x}\,')|\Omega\rangle\simeq\!\frac{H^{D-2}}{2^{D-1}\pi^{{D}/{2}}}
\frac{\Gamma\!\left(\!D\!-\!1\right)}{\Gamma\!\left(\!\frac{D}{2}\right)}\Big[\!\ln(a(t'))
\!-\!\ln(\!K)\!+\!{\rm ci}(\!K)
\!-\!\frac{\sin(\!K)}{K}\!-\!\gamma\!+\!1\nonumber\\
&&\hspace{9.3cm}-\!\sum_{n=1}^\infty\!
                 \frac{(-1)^{n}K^{2n}a^{-2n}(t')}{2n\,(2n\!+\!\!1)!}\Big]\label{corrK0}\\
&&\hspace{2.9cm}\simeq\!\frac{H^{D-2}}{2^{D-1}\pi^{{D}/{2}}}
\frac{\Gamma\!\left(\!D\!-\!1\right)}{\Gamma\!\left(\!\frac{D}{2}\right)}\!\!\left[\ln(a(t'))\!+\!\!\sum_{n=1}^\infty\!
                 \frac{(-1)^{n}K^{2n}}{2n\,(2n\!+\!\!1)!}\left(1\!-\!a^{-2n}(t')\right)\right] \; .\label{corrK}
\ee
Notice that the time dependent part in the infinite sum of Eq.~(\ref{corrK0}) (or Eq.~(\ref{corrK})) decays rapidly as $\mathcal{O}(a^{-2}(t'))$. Since $\ln(a(t'))\!=\!Ht'$, we infer that the correlator at fixed physical distance grows linearly with comoving time. This is because more and more particles are created, which increases
the local field strength. At this point, let us note that the equal space limit of Eq.~(\ref{VEVgeneralpowr}) yields\beeq
\langle\Omega|{\bar\varphi}_0(t,\vec{x}) {\bar\varphi}_0(t'\!,\vec{x})|\Omega\rangle\!=\!\lim_{\vec{x}\,'\rightarrow \vec{x}}\langle\Omega|{\bar\varphi}_0(t,\vec{x}) {\bar\varphi}_0(t'\!,\vec{x}\,')|\Omega\rangle\simeq\!\frac{H^{D-2}}{2^{D-1}\pi^{{D}/{2}}}
\frac{\Gamma\!\left(\!D\!-\!1\right)}{\Gamma\!\left(\!\frac{D}{2}\right)}\ln(a(t'))\; ,\label{equalspaceVEV}
\eneq
and taking the equal time limit of Eq.~(\ref{equalspaceVEV}) gives
\beeq
\langle\Omega|\bar{\varphi}^2_0(t,\vec{x})|\Omega\rangle\!=\!\lim_{t'\rightarrow t}\langle\Omega|{\bar\varphi}_0(t,\vec{x}) {\bar\varphi}_0(t'\!,\vec{x})|\Omega\rangle\simeq\!\frac{H^{D-2}}{2^{D-1}\pi^{{D}/{2}}}
\frac{\Gamma\!\left(\!D\!-\!1\right)}{\Gamma\!\left(\!\frac{D}{2}\right)}\ln(a(t))\; .\label{phisquare}
\eneq

The stochastic analog of the two-point correlation function for the $f_{\!\bar\varphi_0}(t,\vec{x})$ will be needed in Sec.~\ref{subsec:check}. For the sake of completeness, let us compute it here, in the context of QFT, as well.
Equations~(\ref{aadagger}) and (\ref{stoksour}) imply that
\be
\langle\Omega|f_{\!{\bar{\varphi}}_0}(t,\vec{x}) f_{\!{\bar{\varphi}}_0}(t'\!,\vec{x}\,') |\Omega\rangle\!=\!\frac{\Gamma^2(\!D\!-\!1)}{\Gamma^2(\!\frac{D}{2})}
\frac{H^{2D-1}}{2^{D-1}}\!\sum_{\vec{n}\neq 0}\!\sum_{\vec{m}\neq 0}\!\frac{\delta_{Ha(t_n),\, k}}{k^{{(D-3)}/{2}}}\,\frac{\delta_{Ha(t'_m),\, k\,'}}{k'^{{(D-3)}/{2}}}e^{i\vec{k}\cdot\vec{x}}e^{-i\vec{k}'\cdot\vec{x}\,'}\delta_{\vec{n},\,\vec{m}}\; ,
\ee
where $t'\!\!\leq\!t$, $\vec{x}\,'\!\neq\!\vec{x}$ and $\vec{k}'\!=\!2\pi H\vec{m}$. When $\vec{m}\!=\!\vec{n}$ we have $\vec{k}'\!=\!\vec{k}$, therefore,
\be
\langle\Omega|f_{\!{\bar{\varphi}}_0}(t,\vec{x}) f_{\!{\bar{\varphi}}_0}(t'\!,\vec{x}\,') |\Omega\rangle\!\!&=&\!\!\frac{\Gamma^2(\!D\!-\!1)}{\Gamma^2(\!\frac{D}{2})}
\frac{H^{D+2}}{2^{2D-4}\pi^{D-3}}\!\sum_{\vec{n}\neq 0}\frac{\delta_{Ha(t_n),\, Ha(t'_n)}}{n^{D-3}}\,e^{i2\pi H\vec{n}\cdot(\vec{x}-\vec{x}\,')}
\; .
\label{whitenoise}\ee We make the integral approximation to the discrete mode sum in Eq.~(\ref{whitenoise}) to obtain
\be
\hspace{-0.8cm}\sum_{\vec{n}\neq 0}\frac{\delta_{Ha(t_n),\, Ha(t'_n)}}{n^{D-3}}e^{i2\pi H\vec{n}\cdot(\vec{x}-\vec{x}\,')}\!\!&\simeq&\!\!\!\!\int\!\! d\Omega_{D-1}\!\!\int_0^\infty\!\!\!\frac{d k\, k}{\left(2\pi H\right)^2}\,\delta(\!Ha(t)\!-\!k)\,\delta(\!Ha(t')\!-\!k)e^{ik \Delta x \cos{(\theta)}}\, .\label{sumint}
\ee
Evaluating the angular integrations on the right side of Eq.~(\ref{sumint}) yields\be
\hspace{-0.8cm}\frac{2^{D}\pi^{{D}/{2}}}{(2\pi)^3H^2}\frac{\Gamma\left(\!\frac{D}{2}\right)}{\Gamma(\!D\!-\!1)}
\int_0^\infty\!\!\!dk\,\delta(\!Ha(t)\!-\!k)\,\delta(\!Ha(t')\!-\!k)\frac{\sin(k\Delta x)}{\Delta x}\, .\label{intoverk}
\ee
Then we perform the remaining integral over $k$ in Eq.~(\ref{intoverk}),
\be
\int_0^\infty\!\!\!dk\,\delta(\!Ha(t)\!-\!k)\delta(\!Ha(t')\!-\!k)\frac{\sin(k\Delta x)}{\Delta x}
\!&=&\!\frac{\sin(a(t')H\!\Delta x)}{\Delta x}\delta\!\left(Ha(t)\!-\!Ha(t')\right)\nonumber\\
\!&=&\!\frac{\sin(a(t')H\!\Delta x)}{a(t')H\!\Delta x}\frac{\delta(t\!-\!t')}{H}\; ,\label{intdelta}
\ee
where we used $\delta\!\left(\!Ha(t)\!-\!Ha(t')\right)\!=\!\frac{\delta(t-t')}{H\dot{a}(t')}\!=\!\frac{\delta(t-t')}{H^2a(t')}$. Using Eqs.~(\ref{intoverk}) and (\ref{intdelta}) in Eq.~(\ref{sumint}), we find
\be
\sum_{\vec{n}\neq 0}\frac{\delta_{Ha(t_n)\, ,\, Ha(t'_n)}}{n^{D-3}}e^{i2\pi H\vec{n}\cdot(\vec{x}-\vec{x}\,')}\!\simeq\!\frac{2^D\pi^{D/2}}{(2\pi H)^3}\frac{\Gamma\!\left(\!\frac{D}{2}\right)}{\Gamma\!\left(\!D\!-\!1\right)}\frac{\sin(a(t')H\!\Delta x)}{a(t')H\!\Delta x}\delta(t\!-\!t')\; .\label{sum}
\ee
Finally, inserting Eq.~(\ref{sum}) into Eq.~(\ref{whitenoise}) gives the correlation function we wanted to compute,
\beeq
\langle\Omega|f_{{\bar{\varphi}}_0}(t,\vec{x}) f_{{\bar{\varphi}}_0}(t'\!,\vec{x}\,') |\Omega\rangle\!\simeq\!\frac{H^{D-1}}{2^{D-1}\pi^{{D}/{2}}}\frac{\Gamma(\!D\!-\!1)}{\Gamma(\frac{D}{2})}
\frac{\sin(a(t')H\!\Delta x)}{a(t')H\!\Delta x}
\delta(t\!-\!t')\; .\label{expect}
\eneq
The equal space limit of Eq.~(\ref{expect}) yields
\beeq
\langle\Omega|f_{{\bar{\varphi}}_0}(t,\vec{x}) f_{{\bar{\varphi}}_0}(t'\!,\vec{x}) |\Omega\rangle\!\simeq\!\frac{H^{D-1}}{2^{D-1}\pi^{{D}/{2}}}\frac{\Gamma(\!D\!-\!1)}{\Gamma(\frac{D}{2})}
\delta(t\!-\!t')\; .\label{equalspacef0f0}
\eneq
This brings our discussion on the free theory to an end. In the next section, we study the interacting theory with a quartic self-interaction potential.

\subsection{Interacting Theory}
\label{subsect:int}

The solution of the equation of motion for a MMC scalar with a self-interaction potential
$V(\varphi)$ is given in Eq.~(\ref{fullfield}) as
\beeq
\varphi(t,\vec{x}) \!=\! \varphi_0(t,\vec{x})
\!-\!\!\int_0^t \!\!dt' a^{D-1}(t')\!\!\int\! d^{D-1}x'G(t,
\vec{x};t'\!, \vec{x}\,')\frac{V'(\varphi)(t'\!,\vec{x}\,')}{1\!+\!\delta Z} \; ,
\nonumber
\eneq
where the Green's function, defined in Eq.~(\ref{greeneqn}), is \cite{W1}
\be
G(t,
\vec{x};t'\!, \vec{x}\,')\!\!&\equiv&\!\!i \Theta(t\!-\!t')\Bigl[\varphi_0(t,
\vec{x}),\varphi_0(t'\!, \vec{x}\,')\Bigr]\nonumber\\
\!\!&=&\!\!i\Theta(t\!-\!t') \!\!\int\!\! \frac{d^{D-1}k}{(2
\pi)^{D-1}} \, e^{i \vec{k} \cdot (\vec{x}-\vec{x}'\!)} \Bigl\{ u(t,k) u^*(t'\!,k)
\!-\!u^*(t,k) u(t'\!,k) \Bigr\} \; .\label{Greenexpanded}
\ee
To get the IR truncated full field $\bar{\varphi}(t,\vec{x})$, we need the leading order IR limit of the $G(t,\vec{x};t'\!, \vec{x}\,')$. Combining Eqs.~(\ref{udef}), (\ref{HankelinBessel}) and (\ref{Neumann}) we find
\be
&&\hspace{-1.2cm}u(t,k) u^*(t'\!,k)\!-\!u^*(t,k) u(t',k)\!=\!i\frac{\pi}{2H}\frac{\csc\!\left(\!{\frac{(\!D-1)\pi}{2}}\!\right)}
{a^{\frac{D-1}{2}}(t)\,a^{\frac{D-1}{2}}(t')}\Bigg[{J}_{\frac{D-1}{2}}\!(z)\,
{J}_{-\left(\!\frac{D-1}{2}\!\right)}\!(z')\!-\!(z\leftrightarrow z')\Bigg]\; ,\\
&&\hspace{2.85cm}\longrightarrow\!\frac{i}{H}\Biggl\{\!\frac{1}{D\!-\!1}\!+\!\mathcal{O}\Bigl(\frac{k^2}{H^2 a^2(t)},
\frac{k^2}{H^2 a^{2}(t')}\Bigr)\!\Biggr\}\!\Biggl[\frac{1}{a^{D-1}(t)}-\frac{1}{a^{D-1}(t')}\Biggr]\; .\label{uulimit}
\ee
Equations~(\ref{BesselSeries}) and (\ref{gammareflection}) have been used in obtaining the limit in the last line. Hence, plugging Eq.~(\ref{uulimit}) into Eq.~(\ref{Greenexpanded}), retaining only the leading terms, gives \cite{Wstocqgrav, Wstocsqed, W3}
\beeq
G(t, \vec{x} ; t',\vec{x}\,')
\longrightarrow
\frac{1}{(\!D\!-\!1)H}\,\Theta(t\!-\!t')\,\delta^{D-1}(\vec{x}\!-\!\vec{x}\,')
\Biggl[\frac1{a^{D-1}(t')}\!-\!\frac{1}{a^{D-1}(t)}\Biggr]
\; . \label{limitGREEN}
\eneq
When this limit is inserted into Eq.~(\ref{fullfield}) it is multiplied by the integration measure $a^{D-1}(t')$. Then, the first term in the square brackets, ${a^{D-1}(t')}/{a^{D-1}(t')}\!=\!1$, contributes over the whole range of the integration. The second term proportional to ${a^{D-1}(t')}/{a^{D-1}(t)}$, on the other hand, contributes significantly only for $t'\sim t$, and hence, negligible in the leading logarithm approximation we consider. Thus, Eq.~(\ref{fullfield}) yields the IR truncated full field as\beeq
\bar\varphi(t, \vec{x})\!=\!\bar\varphi_0(t, \vec{x})\!-\!\frac{1}{(\!D\!-\!1)H}\!\!\int_0^t\!\!dt' \, \frac{V'(\bar\varphi(t'\!, \vec{x}))}{1\!+\!\delta Z}\; .\label{fieldinfra}
\eneq We consider a scalar field with quartic self-interaction, hence
\beeq
V(\varphi) \!=\! \frac{1}{2} \delta m^2 \varphi^2 \!+\! \frac{1}{4!} (\lambda \!+\! \delta \lambda) \varphi^4 \; ,
\eneq
where $\lambda$ is the coupling constant, $\delta \lambda$ and $\delta m^2$ are the coupling constant and the mass squared counterterms, respectively. Counterterms cannot contribute in leading logarithm order. To see this, let us first note that $\delta Z \!\sim\! \mathcal{O}(\lambda^2)$, $\delta m^2 \!\sim\! \mathcal{O}(\lambda)$ and $\delta \lambda \!\sim\! \mathcal{O}(\lambda^2)$ in the model \cite{BOW}. In the $\frac{1}{4!} (\lambda\!+\!\delta \lambda) \varphi^4$ term, $\lambda$ and $\delta\lambda$ multiply the same $\varphi^4$. Therefore, contributions involving $\frac{\lambda}{4!}\varphi^4$ and contributions involving $\frac{\delta\lambda}{4!}\varphi^4$ must have the same IR logarithm structure. Because the latter are suppressed by at least one extra factor of $\lambda$, with no more IR logarithm, they cannot be in leading order. Next, let us compare the $\frac{\delta m^2}{2}\varphi^2$ and $\frac{\lambda}{4!}\varphi^4$ terms. Although, $\delta m^2\!\sim\!\lambda$, it multiplies $\varphi^2$. Therefore, at a given order in $\lambda$, contributions involving the former cannot produce as high order IR logarithms as contributions involving the latter. Finally, the field strength counterterm $\delta Z$ appears in the field equations in the form $V'(\varphi)/(1\!+\!\delta Z)\!=\!V'(\varphi)\left[1\!-\!\delta Z\!+\!(\delta Z)^2\!-\!\cdots\right]$ where $\delta Z \!\sim\! \mathcal{O}(\lambda^2)$. Consequently, exactly the same leading logarithm contributions is obtained from the simplified field equation without the counterterms,
\beeq
\bar\varphi(t, \vec{x})\!=\!\bar\varphi_0(t, \vec{x})\!-\!\frac{\lambda}{6(\!D\!-\!1)H}\!\!\int_0^t\!\!dt'' \,\bar\varphi^3(t''\!, \vec{x})\; .\label{fullinfrafotphi4}
\eneq
Iterating Eq.~(\ref{fullinfrafotphi4}) twice, we find the IR truncated full field in terms of $\bar\varphi_0(t,\vec{x})$ as
\be
&&\hspace{-1cm}\bar\varphi(t, \vec{x})\!=\!\bar\varphi_0(t,\vec{x})\!-\!\frac{\l}{6(\!D\!-\!1)H}\!\!\int_0^t\!dt'
\bar\varphi^3_0(t'\!,\vec{x})\!+\!\frac{\l^2}{12(\!D\!-\!1)^2H^2}\!\!\int_0^t\!
dt'\bar\varphi^2_0(t'\!,\vec{x})\!\!\int_0^{t'}\!\!dt''\bar\varphi^3_0(t''\!,\vec{x})\nonumber\\
&&\hspace{1.9cm}-\frac{\l^3}{24(\!D\!-\!1)^3H^3}\Bigg\{\!\!\int_0^t\!
dt'\bar\varphi^2_0(t'\!,\vec{x})\!\!\int_0^{t'}
\!\!dt''\bar\varphi^2_0(t''\!,\vec{x})\!\!\int_0^{t''}
\!\!\!dt'''\bar\varphi^3_0(t'''\!,\vec{x})\nonumber\\
&&\hspace{5.8cm}+\frac{1}{3}\!\int_0^t\!
dt'\bar\varphi_0(t'\!,\vec{x})\!\left[\int_0^{t'}\!\!dt''\bar\varphi^3_0(t''\!,\vec{x})\right]^2\!\Bigg\}
\!+\!\mathcal{O}(\l^4)\; .\label{freefieldexpfulf}\ee
Using Eq.~(\ref{freefieldexpfulf}), we can compute the two-point correlation function of the IR truncated full field. Up to $\mathcal{O}(\l^3)$ we have
\be
&&\hspace{-1.4cm}\langle\Omega|
\bar{\varphi}(t,\vec{x})\bar{\varphi}(t'\!,\vec{x}\,')|\Omega\rangle\!=\!\langle\Omega|
\bar{\varphi}_0(t,\vec{x})\bar{\varphi}_0(t'\!,\vec{x}\,')|\Omega\rangle\nonumber\\
&&\hspace{-1.3cm}-\frac{\lambda}{6(\!D\!-\!1)H}\!\!\left[\langle\Omega|
\bar{\varphi}_0(t,\vec{x})\!\!\int_0^{t'}\!\!d\tilde{t}\,{\bar{\varphi}}^3_0(\tilde{t},\vec{x}\,')
|\Omega\rangle\!+\!\langle\Omega|\!\!
\int_0^{t}\!dt''{\bar{\varphi}}^3_0(t''\!,\vec{x})
\bar{\varphi}_0(t'\!,\vec{x}\,')|\Omega\rangle\right]\nonumber\\
&&\hspace{-1.3cm}+\frac{\lambda^2}{12(\!D\!-\!1)^2H^2}\!\Bigg[\langle\Omega|
\bar{\varphi}_0(t,\vec{x})\!\!\int_0^{t'}\!\!d\tilde{t}\,{\bar{\varphi}}^2_0(\tilde{t},\vec{x}\,')
\!\!\int_0^{\tilde{t}}\!d\tilde{\tilde{t}}\, {\bar{\varphi}}^3_0(\tilde{\tilde{t}},\vec{x}\,')
|\Omega\rangle\nonumber\\
&&\hspace{-1.3cm}+\langle\Omega|\!\!
\int_0^{t}\!dt''{\bar{\varphi}}^2_0(t''\!,\vec{x})\!\!
\int_0^{t''}\!\!\!dt'''{\bar{\varphi}}^3_0(t'''\!,\vec{x})
\bar{\varphi}_0(t'\!,\vec{x}\,')
|\Omega\rangle\!+\!\frac{1}{3}\langle\Omega|\!\!
\int_0^{t}\!dt''{\bar{\varphi}}^3_0(t''\!,\vec{x})\!\!
\int_0^{t'}\!\!d\tilde{t}\,{\bar{\varphi}}^3_0(\tilde{t},\vec{x}\,')
|\Omega\rangle\Bigg]\; ,\label{expectequaltime}
\ee
where $0\!\leq\!t'''\!\leq\!t''\!\leq\!t$, $0\!\leq\!\tilde{\tilde{t}}\!\leq\!\tilde{t}\!\leq\!t'$ and $t'\!\leq\!t$. The perturbation theory breaks down when $\ln(a(t))\!=\!Ht\!\sim\!1/\sqrt{\lambda}$ \cite{OW1}. So, for $\lambda\!\ll\!1$, one can have a long period of time during which the perturbation theory and hence Eq.~(\ref{expectequaltime}) are valid. In the next three sections, we compute the tree, one and two-loop contributions in correlation function~(\ref{expectequaltime}).

\subsubsection{Tree-order contribution}
\label{subsec:tree}
The tree-order contribution in the two-point correlation function of the IR truncated full field is just the two-point correlation function of the IR truncated free field which is obtained in Eqs.~(\ref{expphisq}) and (\ref{VEVgeneralpowr}) as
\be
&&\hspace{-1cm}\langle\Omega|
\bar{\varphi}(t,\vec{x})\bar{\varphi}(t'\!,\vec{x}\,')|\Omega\rangle_{\rm tree}\!=\!\langle\Omega|{\bar\varphi}_0(t,\vec{x}) {\bar\varphi}_0(t'\!,\vec{x}\,')|\Omega\rangle\nonumber\\
&&\hspace{0cm}\simeq\!\frac{H^{D-2}}{2^{D-1}\pi^{{D}/{2}}}
\frac{\Gamma\!\left(\!D\!-\!1\right)}{\Gamma\!\left(\!\frac{D}{2}\right)}\!\!\left[{\rm ci}\!\left(a(t')H\!\Delta x\right)\!-\!\frac{\sin\left(a(t')H\!\Delta x\right)}{a(t')H\!\Delta x}\!-\!{\rm ci}(\!H\!\Delta x)\!+\!\frac{\sin(\!H\!\Delta x)}{H\!\Delta x}\right]\\
&&\hspace{0cm}\simeq\!\frac{H^{D-2}}{2^{D-1}\pi^{{D}/{2}}}
\frac{\Gamma\!\left(\!D\!-\!1\!\right)}{\Gamma\!\left(\!\frac{D}{2}\right)}\!\!\left[\ln(a(t'))\!+\!\!\sum_{n=1}^\infty\!
                 \frac{(\!-1)^{n}(H\!\Delta x)^{2n}}{2n\,(2n\!+\!\!1)!}\left(a^{2n}(t')\!-\!1\right)\right]\; .\label{treeorder}
\ee
Quantum correcting this classical result, however, requires some work. In the following two sections we carry out the calculations for the one and two-loop contributions, respectively.

\subsubsection{One-loop contribution}
\label{subsec:1loop}

Computing the one-loop contribution to the tree-order two-point correlation function in Eq.~(\ref{expectequaltime}) involves evaluations of two VEVs. The first one is
\be
\langle\Omega|
\bar{\varphi}_0(t,\vec{x})\!\!\int_0^{t'}\!\!\!d\tilde{t}\,{\bar{\varphi}}^3_0(\tilde{t},\vec{x}\,')
|\Omega\rangle\!=\!\!\int_0^{t'}\!\!\!d\tilde{t}\,\langle\Omega|
\bar{\varphi}_0(t,\vec{x})\, {\bar{\varphi}}^3_0(\tilde{t},\vec{x}\,')|\Omega\rangle \; .\label{phiphicubeequaltime}
\ee
It can be calculated either by directly inserting free field (\ref{whitef}) for each power of $\bar{\varphi}_0$ in Eq.~(\ref{phiphicubeequaltime}) and using the commutator algebra~(\ref{aadagger}) or, equivalently, by {\it topologically inequivalent field-pairings} which yield\be
&&\hspace{-1.3cm}\int_0^{t'}\!\!d\tilde{t}\,\langle\Omega|
\bar{\varphi}_0(t,\vec{x})\, {\bar{\varphi}}^3_0(\tilde{t},\vec{x}\,')|\Omega\rangle\!=\!\!\int_0^{t'}\!\!d\tilde{t}\,3\cdot\!1\langle\Omega|
\bar{\varphi}_0(t,\vec{x})\, \bar{\varphi}_0(\tilde{t},\vec{x}\,')|\Omega\rangle
\langle\Omega|{\bar{\varphi}}^2_0(\tilde{t},\vec{x}\,')|\Omega\rangle \nonumber\\
&&\hspace{-0.6cm}=\!\frac{H^{2D-4}}{2^{2D-2}\pi^{D}}
\frac{\Gamma^2\!\left(\!D\!-\!1\right)}{\Gamma^2\!\left(\!\frac{D}{2}\right)}
3\!\!\int_0^{t'}\!\!d\tilde{t}\!\left[\mathcal{C}(\!\Delta x)\!+\!\ln(a(\tilde{t}))\!+\!\sum_{n=1}^\infty
                 \frac{(-1)^{n}(a(\tilde{t})H\!\Delta x)^{2n}}{2n\,(2n\!+\!1)!}\right]\!\ln(a(\tilde{t}))\; ,\label{pairequaltime}
\ee
where we used Eq.~(\ref{VEVgeneral}), the fact that $\tilde{t}\leq t$ and Eq.~(\ref{phisquare}). Evaluating the integral in Eq.~(\ref{pairequaltime}) we obtain
\be
&&\hspace{-1cm}\int_0^{t'}\!\!d\tilde{t}\,\langle\Omega|
\bar{\varphi}_0(t,\vec{x})\, {\bar{\varphi}}^3_0(\tilde{t},\vec{x}\,')|\Omega\rangle=\frac{H^{2D-5}}{2^{2D-1}\pi^{D}}
\frac{\Gamma^2\!\left(\!D\!-\!1\right)}{\Gamma^2\!\left(\!\frac{D}{2}\right)}3\Bigg\{\!\frac{2\!\ln^3(a(t'))}{3}
\!+\!\mathcal{C}(\!\Delta x)\!\ln^2(a(t'))\nonumber\\
&&\hspace{4cm}+\frac{1}{2}\!\sum_{n=1}^\infty
\!\frac{(-1)^{n}(\!H\!\Delta x)^{2n}}{n^2(2n\!+\!1)!}\!\left[a^{2n}(t')\!\left(\!\ln(a(t'))\!-\!\frac{1}{2n}\!\right)\!\!+\!\frac{1}{2n}\right]\!\!\Bigg\}
\; .\label{1loop1result}
\ee

The second VEV which contributes at one-loop order in Eq.~(\ref{expectequaltime}) can be calculated similarly,
\be
\langle\Omega|\!\!
\int_0^{t}\!\!dt''{\bar{\varphi}}^3_0(t''\!,\vec{x})
\bar{\varphi}_0(t'\!,\vec{x}\,')|\Omega\rangle\!=\!\int_0^{t}\!\!dt'' \,3\cdot\!1\langle\Omega|
\bar{\varphi}_0(t''\!,\vec{x})\bar{\varphi}_0(t'\!,\vec{x}\,')|\Omega\rangle
\langle\Omega|{\bar{\varphi}}^2_0(t''\!,\vec{x})|\Omega\rangle\nonumber\\
\!=\!\frac{H^{D-2}}{2^{D-1}\pi^{D/2}}
\frac{\Gamma\!\left(\!D\!-\!1\right)}{\Gamma\!\left(\!\frac{D}{2}\right)}3\!\int_0^{t}\!\!dt'' \langle\Omega|
\bar{\varphi}_0(t''\!,\vec{x})\bar{\varphi}_0(t'\!,\vec{x}\,')|\Omega\rangle\!\ln(a(t''))\; ,\label{1loopsecond}
\ee
where we used Eq.~(\ref{phisquare}) in the last equality. We have $0\leq\!t''\!\leq\!t$ and $t'\!\leq\!t$. Therefore, in order to evaluate the remaining VEV in the integrand we break up the integral into two as $\int_0^{t}\!dt''\!=\!\int_0^{t'}\!dt''\!+\!\int_{t'}^t\!dt''$. In the first integral $t''\!\leq\!t'$, whereas in the second $t'\!\leq\!t''$. Hence, appropriate use of Eq.~(\ref{VEVgeneral}) in  Eq.~(\ref{1loopsecond}) gives
\be
&&\hspace{-1cm}\langle\Omega|\!\!
\int_0^{t}\!\!dt''{\bar{\varphi}}^3_0(t''\!,\vec{x})
\bar{\varphi}_0(t'\!,\vec{x}\,')|\Omega\rangle\nonumber\\
&&\hspace{-0.9cm}=\!\frac{H^{2D-4}}{2^{2D-2}\pi^{D}}
\frac{\Gamma^2\!\left(\!D\!-\!1\right)}{\Gamma^2\!\left(\!\frac{D}{2}\right)}3
\Bigg\{\!\mathcal{C}(\!\Delta x)\!\!\int_0^{t}\!dt''\!\ln(a(t''))\!+\!\!\int_0^{t'}\!\!dt''\!\ln^2(a(t''))
\!+\!\ln(a(t'))\!\!\int_{t'}^{t}\!dt''\!\ln(a(t''))\nonumber\\
&&\hspace{1cm}+\frac{1}{2}\!\sum_{n=1}^\infty
\!\frac{(-1)^{n}(H\!\Delta x)^{2n}}{n\,(2n\!+\!1)!}\!\left[\int_0^{t'}\!\!dt''a^{2n}(t'')\!\ln(a(t''))
\!+\!a^{2n}(t')\!\!\int_{t'}^{t}\!dt''\ln(a(t''))\right]\!\!\Bigg\}\; .\label{quant1loopsecnd}
\ee
We evaluate the integrals and find
\be
&&\hspace{-0.5cm}\langle\Omega|\!\!
\int_0^{t}\!dt''{\bar{\varphi}}^3_0(t''\!,\vec{x})
\bar{\varphi}_0(t',\vec{x}\,')|\Omega\rangle\!=\!\frac{H^{2D-5}}{2^{2D-1}\pi^{D}}
\frac{\Gamma^2\!\left(\!D\!-\!1\right)}{\Gamma^2\!\left(\!\frac{D}{2}\right)}3
\Bigg\{\!\Big[\mathcal{C}(\!\Delta x)\!+\!\ln(a(t'))\Big]\!\ln^2(a(t))\!-\!\frac{\ln^3(a(t'))}{3}\nonumber\\
&&\hspace{0.5cm}+\frac{1}{2}\!\sum_{n=1}^\infty
\!\frac{(-1)^{n}(\!H\!\Delta x)^{2n}}{n(2n\!+\!1)!}\!\left[a^{2n}(t')\!\left(\!\ln^2(a(t))\!-\!\ln^2(a(t'))
\!+\!\frac{\ln(a(t'))}{n}\!-\!\frac{1}{2n^2}\!\right)\!\!+\!\frac{1}{2n^2}\right]\!\Bigg\}\; .
\label{1loop2result}
\ee
Combining Eqs.~(\ref{Cpower}), (\ref{1loop1result}) and (\ref{1loop2result}) in Eq.~(\ref{expectequaltime}) yields the following total $\mathcal{O}(\lambda)$-correction\be
&&\hspace{-0.4cm}\langle\Omega|
\bar{\varphi}(t,\vec{x})\bar{\varphi}(t'\!,\vec{x}\,')|\Omega\rangle_{\rm 1-loop}\!\simeq\!-\frac
{\lambda}{(\!D\!-\!1)}\frac{H^{2D-6}}{2^{2D}\pi^{D}}
\frac{\Gamma^2\!\left(\!D\!-\!1\right)}{\Gamma^2\!\left(\!\frac{D}{2}\right)}
\Bigg\{\frac{\ln^3(a(t'))}{3}\!+\!\ln(a(t'))\!\ln^2(a(t))\nonumber\\
&&\hspace{4cm}
\!+\frac{1}{2}\!\sum_{n=1}^\infty
\!\frac{(\!-1)^{n}(\!H\!\Delta x)^{2n}}{n(2n\!+\!1)!}\Bigg\{a^{2n}(t')\!\left[\ln^2(a(t))\!-\!\ln^2(a(t'))
\!+\!\frac{2\!\ln(a(t'))}{n}\!-\!\frac{1}{n^2}\right]\nonumber\\
&&\hspace{9.5cm}\!-\!\ln^2(a(t))\!-\!\ln^2(a(t'))\!+\!\frac{1}{n^2}\!\Bigg\}\Bigg\}\; .\label{1loopequaltimefinal}
\ee
As in Eq.~(\ref{expphisq}), to infer the result at a fixed comoving separation at late times during inflation we must fix $\Delta x$ and then take $t'$ large. (Recall $t\!\geq\!t'$.) The asymptotic form of the one-loop correlator is obtained as
\be
&&\hspace{-1cm}\langle\Omega|
\bar{\varphi}(t,\vec{x})\bar{\varphi}(t'\!,\vec{x}\,')|\Omega\rangle_{\rm 1-loop}\!\simeq\!-\frac
{\lambda}{(\!D\!-\!1)}\frac{H^{2D-6}}{2^{2D}\pi^{D}}
\frac{\Gamma^2\!\left(\!D\!-\!1\right)}{\Gamma^2\!\left(\!\frac{D}{2}\right)}
\Bigg\{\frac{4\!\ln^3(a(t'))}{3}\nonumber\\
&&\hspace{-1cm}-\!\left[{\rm ci}(\!H\!\Delta x)\!-\!\frac{\sin\left(\!H\!\Delta x\right)}{H\!\Delta x}\right]\!\Big[\!\ln^2(a(t))\!+\!\ln^2(a(t'))\!\Big]\!+\!2\Big[\!\ln\!\left(\!H\!\Delta x\right)\!+\!\gamma\!-\!1\Big]\!\ln^2(a(t'))\nonumber\\
&&\hspace{-1cm}-\frac{\alpha^2}{6}\,
{}_{3}\mathcal{F}_{4}\!\left(\!\!1, 1, 1; 2, 2, 2, \frac52; -\frac{\alpha^2}{4}\!\right)\!\ln(a(t'))
\!+\!\frac{\alpha^2}{12}\,
{}_{4}\mathcal{F}_{5}\!\left(\!\!1, 1, 1, 1; 2, 2, 2, 2, \frac52; -\frac{\alpha^2}{4}\!\right)\nonumber\\
&&\hspace{5cm}-\frac{H^2\!\Delta x^2}{12}\,
{}_{4}\mathcal{F}_{5}\!\left(\!\!1, 1, 1, 1; 2, 2, 2, 2, \frac52; -\frac{H^2\!\Delta x^2}{4}\!\right)\!\Bigg\}\; ,
\label{asym1loop}
\ee
where we used the large argument expansion~(\ref{ciasym}) of the ${\rm ci}(\alpha)$ function with $\alpha\!=\!a(t')H\!\Delta x$. The ${}_{m}\mathcal{F}_{n}$'s are generalized hypergeometric functions. Equation~(\ref{asym1loop}) implies that one loop correlator grows negatively for a fixed comoving separation during inflation.

To obtain the one-loop correlator for a fixed physical distance, we choose $a(t')\Delta x$ a constant fraction $K$ of the Hubble length as in Eq.~(\ref{DeltaX}),
\be
&&\hspace{-1cm}\langle\Omega|
\bar{\varphi}(t,\vec{x})\bar{\varphi}(t'\!,\vec{x}\,')|\Omega\rangle_{\rm 1-loop}\!\simeq\!-\frac
{\lambda}{(\!D\!-\!1)}\frac{H^{2D-6}}{2^{2D}\pi^{D}}
\frac{\Gamma^2\!\left(\!D\!-\!1\right)}{\Gamma^2\!\left(\!\frac{D}{2}\right)}
\Bigg\{\!\!-\!\frac{2\!\ln^3(a(t'))}{3}\nonumber\\
&&\hspace{-1cm}\!+\!\left\{{\rm ci}(\!K)\!-\!\frac{\sin(\!K)}{K}\!-\!\left[{\rm ci}(\!Ka^{-1}(t'))\!-\!\frac{\sin(\!Ka^{-1}(t'))}{Ka^{-1}(t')}\right] \right\}\ln^2(a(t))\nonumber\\
&&\hspace{-1cm}\!-\!\left\{{\rm ci}(\!K)\!-\!\frac{\sin(\!K)}{K}\!-\!2\ln(\!K)\!+\!\left[{\rm ci}(\!Ka^{-1}(t'))\!-\!\frac{\sin(\!Ka^{-1}(t'))}{Ka^{-1}(t')}\right]\!+\!2\!-\!2\gamma\right\}\ln^2(a(t'))\nonumber\\
&&\hspace{-1cm}-\frac{K^2}{6}\,
{}_{3}\mathcal{F}_{4}\!\left(\!\!1, 1, 1; 2, 2, 2, \frac52; -\frac{K^2}{4}\!\right)\!\ln(a(t'))
\!+\!\frac{K^2}{12}\,
{}_{4}\mathcal{F}_{5}\!\left(\!\!1, 1, 1, 1; 2, 2, 2, 2, \frac52; -\frac{K^2}{4}\!\right)\nonumber\\
&&\hspace{5cm}-\frac{K^2}{12a^2(t')}\,
{}_{4}\mathcal{F}_{5}\!\left(\!\!1, 1, 1, 1; 2, 2, 2, 2, \frac52; -\frac{K^2}{4a^2(t')}\!\right)\!\!\Bigg\}\; .
\ee
We employ Eq.~(\ref{ciminussin}) in this analytic form of the one-loop correlator and get
\be
&&\hspace{-1cm}\langle\Omega|
\bar{\varphi}(t,\vec{x})\bar{\varphi}(t'\!,\vec{x}\,')|\Omega\rangle_{\rm 1-loop}\!\simeq\!-\frac
{\lambda}{(\!D\!-\!1)}\frac{H^{2D-6}}{2^{2D}\pi^{D}}
\frac{\Gamma^2\!\left(\!D\!-\!1\right)}{\Gamma^2\!\left(\!\frac{D}{2}\right)}
\Bigg\{\frac{\ln^3(a(t'))}{3}\!+\!\ln(a(t'))\!\ln^2(a(t))\nonumber\\
&&\hspace{-1cm}+\!\left[\ln^2(a(t))\!-\!\ln^2(a(t'))\right]\!\sum_{n=1}^\infty
\!\frac{(\!-1)^{n}\!K^{2n}}{2n(2n\!+\!\!1)!}
\!+\!\ln(a(t'))\!\sum_{n=1}^\infty
\!\frac{(\!-1)^{n}\!K^{2n}}{n^2(2n\!+\!\!1)!}-\!\sum_{n=1}^\infty
\!\frac{(\!-1)^{n}\!K^{2n}}{2n^3(2n\!+\!\!1)!}\nonumber\\
&&\hspace{1.5cm}-\!\left[\ln^2(a(t))\!+\!\ln^2(a(t'))\right]\!\sum_{n=1}^\infty
\!\frac{(\!-1)^{n}\!K^{2n}a^{-2n}(t')}{2n(2n\!+\!\!1)!}\!+\!\sum_{n=1}^\infty
\!\frac{(\!-1)^{n}\!K^{2n}a^{-2n}(t')}{2n^3(2n\!+\!\!1)!}\Bigg\}\; ,\label{1loopfixeddist}
\ee
where the terms in the last line decay exponentially during inflation. Equation~(\ref{1loopfixeddist}) tells us that the one-loop correlator at fixed physical distance grows negatively during inflation.

In the next section, we compute the two-loop contribution to the correlation function of the IR truncated scalar field.
\subsubsection{Two-loop contribution}
\label{subsec:2loop}
Computing the two-loop contribution in two-point correlation function~(\ref{expectequaltime}) involves evaluations of three VEVs, of which the first is
\be
&&\hspace{-1.5cm}\langle\Omega|
\bar{\varphi}_0(t,\vec{x})\!\!\int_0^{t'}\!\!\!\!d\tilde{t}\,{\bar{\varphi}}^2_0(\tilde{t}\!,\vec{x}\,')
\!\!\int_0^{\tilde{t}}\!\!d\tilde{\tilde{t}}\, {\bar{\varphi}}^3_0(\tilde{\tilde{t}},\vec{x}\,')
|\Omega\rangle\!=\!\int_0^{t'}\!\!\!\!d\tilde{t}\!\!\int_0^{\tilde{t}}\!\!\!d\tilde{\tilde{t}}\Bigg\{\nonumber\\
&&\hspace{0cm}2\!\cdot\!3\!\cdot\!1\,\langle\Omega|
\bar{\varphi}_0(t,\vec{x})\bar{\varphi}_0(\tilde{t},\vec{x}\,')|\Omega\rangle\langle\Omega|
\bar{\varphi}_0(\tilde{t},\vec{x}\,')\bar{\varphi}_0(\tilde{\tilde{t}}\!,\vec{x}\,')|\Omega\rangle\langle\Omega|
\bar{\varphi}^2_0(\tilde{\tilde{t}},\vec{x}\,')|\Omega\rangle\nonumber\\
&&\hspace{1.5cm}3\!\cdot\!1\!\cdot\!1\,\langle\Omega|
\bar{\varphi}_0(t,\vec{x})\bar{\varphi}_0(\tilde{\tilde{t}},\vec{x}\,')|\Omega\rangle\langle\Omega|
\bar{\varphi}^2_0(\tilde{t},\vec{x}\,')|\Omega\rangle\langle\Omega|
\bar{\varphi}^2_0(\tilde{\tilde{t}},\vec{x}\,')|\Omega\rangle\nonumber\\
&&\hspace{3cm}+3\!\cdot\!2\!\cdot\!1\,\langle\Omega|
\bar{\varphi}_0(t,\vec{x})\bar{\varphi}_0(\tilde{\tilde{t}},\vec{x}\,')|\Omega\rangle\!\left[\langle\Omega|
\bar{\varphi}_0(\tilde{t},\vec{x}\,')\bar{\varphi}_0(\tilde{\tilde{t}}\!,\vec{x}\,')|\Omega\rangle\right]^2
\!\Bigg\}\; ,\label{twoloopfirstexpandequaltime}
\ee
where we have $0\!\leq\!\tilde{\tilde{t}}\!\leq\!\tilde{t}\!\leq\!t'$ and $t'\leq t$. This time ordering makes it easy to read off the VEVs in the integrand from Eqs.~(\ref{VEVgeneral}), (\ref{equalspaceVEV}) and (\ref{phisquare}). Substituting the outcomes yields\be
&&\hspace{-1.2cm}\langle\Omega|
\bar{\varphi}_0(t,\vec{x})\!\!\int_0^{t'}\!\!\!d\tilde{t}\,{\bar{\varphi}}^2_0(\tilde{t}\!,\vec{x}\,')
\!\!\int_0^{\tilde{t}}\!\!\!d\tilde{\tilde{t}}\, {\bar{\varphi}}^3_0(\tilde{\tilde{t}},\vec{x}\,')
|\Omega\rangle\!=\!\frac{H^{3D-6}}{2^{3D-3}\pi^{3D/2}}\frac{\Gamma^3(\!D\!-\!1)}{\Gamma^3(\!\frac{D}{2})}\Bigg\{\nonumber\\
&&\hspace{-0.75cm}\int_0^{t'}\!\!\!\!d\tilde{t}\!\!\int_0^{\tilde{t}}\!\!\!d\tilde{\tilde{t}}\,
2\!\cdot\!3\!\cdot\!1\!\!\left[\mathcal{C}(\!\Delta x)\!+\!\ln(a(\tilde{t}))\!+\!\!\sum_{n=1}^\infty\!
                 \frac{(-1)^{n}(a(\tilde{t})H\!\Delta x)^{2n}}{2n(2n\!+\!1)!}\right]\!\!\ln^2(a(\tilde{\tilde{t}}))\nonumber\\
&&\hspace{-1cm}+\!\!\int_0^{t'}\!\!\!\!d\tilde{t}\!\!\int_0^{\tilde{t}}\!\!\!d\tilde{\tilde{t}}\,
3\!\cdot\!1\!\cdot\!1\!\!\left[\mathcal{C}(\!\Delta x)\!+\!\ln(a(\tilde{\tilde{t}}))\!+\!\!\sum_{n=1}^\infty\!
                 \frac{(-1)^{n}(a(\tilde{\tilde{t}})H\!\Delta x)^{2n}}{2n(2n\!+\!1)!}\right]\!\!\ln(a(\tilde{t}))\!\ln(a(\tilde{\tilde{t}}))\nonumber\\
&&\hspace{-1cm}+\!\!\int_0^{t'}\!\!\!\!d\tilde{t}\!\!\int_0^{\tilde{t}}\!\!\!d\tilde{\tilde{t}}\,
3\!\cdot\!2\!\cdot\!1\!\!\left[\mathcal{C}(\!\Delta x)\!+\!\ln(a(\tilde{\tilde{t}}))\!+\!\!\sum_{n=1}^\infty\!
                 \frac{(-1)^{n}(a(\tilde{\tilde{t}})H\!\Delta x)^{2n}}{2n(2n\!+\!1)!}\right]\!\!\ln^2(a(\tilde{\tilde{t}}))\Bigg\}\; .\label{2loop1stequaltime}
\ee
The results of the integrals in Eq.~(\ref{2loop1stequaltime}) are given in Appendix \ref{App:VEV1}.
Using Eqs.~(\ref{2loop1st1stintequaltime}), (\ref{2loop1st2ndintequaltime}) and (\ref{2loop1st3rdintequaltime}) in Eq.~(\ref{2loop1stequaltime}) we obtain\be
&&\hspace{-1.2cm}\langle\Omega|
\bar{\varphi}_0(t,\vec{x})\!\!\!\int_0^{t'}\!\!\!\!d\tilde{t}\,{\bar{\varphi}}^2_0(\tilde{t},\vec{x}\,')
\!\!\int_0^{\tilde{t}}\!\!\!d\tilde{\tilde{t}}\, {\bar{\varphi}}^3_0(\tilde{\tilde{t}},\vec{x}\,')
|\Omega\rangle\nonumber\\
&&\hspace{-1.2cm}=\!\frac{H^{3D-8}}{2^{3D-2}\pi^{3D/2}}\frac{\Gamma^3(\!D\!-\!1)}{\Gamma^3(\!\frac{D}{2})}\Bigg\{\!
\frac{9\!\ln^5(a(t'))}{5}\!+\!\frac{11\mathcal{C}(\!\Delta x)}{4}\!\ln^4(a(t'))\!+\!\!\sum_{n=1}^\infty
\!\frac{(-1)^{n}(\!H\!\Delta x)^{2n}}{n^2(2n\!+\!1)!}\nonumber\\
&&\hspace{-1.2cm}\times\!\Bigg\{\!a^{2n}(t')\!\!\left[\!\ln^3(a(t'))
\!+\!\frac{3\!\ln^2(a(t'))}{4n}\!-\!\frac{21\!\ln(a(t'))}{8n^2}\!+\!\frac{33}{16n^3}\!\right]
\!\!+\!\frac{3\!\ln^2(a(t'))}{8n}\!-\!\frac{3\!\ln(a(t'))}{2n^2}\!-\!\frac{33}{16n^3}
\!\Bigg\}\!\!\Bigg\}\; .\label{2loop1stequaltimeresult}
\ee

The second VEV which contributes at two-loop order in correlation function~(\ref{expectequaltime}) is evaluated similarly,
\be
&&\hspace{-1cm}\langle\Omega|\!\!
\int_0^{t}\!\!\!dt''{\bar{\varphi}}^2_0(t''\!,\vec{x})\!\!
\int_0^{t''}\!\!\!\!dt'''{\bar{\varphi}}^3_0(t'''\!,\vec{x})
\bar{\varphi}_0(t'\!,\vec{x}\,')
|\Omega\rangle\!=\!\int_0^{t}\!\!\!dt''\!\!\int_0^{t''}\!\!\!\!dt'''\Bigg\{\nonumber\\
&&\hspace{0cm}2\!\cdot\!3\!\cdot\!1\,\langle\Omega|
\bar{\varphi}_0(t''\!,\vec{x})\bar{\varphi}_0(t'\!,\vec{x}\,')|\Omega\rangle\,\langle\Omega|
\bar{\varphi}_0(t''\!,\vec{x})\bar{\varphi}_0(t'''\!,\vec{x})|\Omega\rangle\,\langle\Omega|
\bar{\varphi}^2_0(t'''\!,\vec{x})|\Omega\rangle\nonumber\\
&&\hspace{1cm}+1\!\cdot\!3\!\cdot\!1\,\langle\Omega|
\bar{\varphi}^2_0(t''\!,\vec{x})|\Omega\rangle\,\langle\Omega|
\bar{\varphi}_0(t'''\!,\vec{x})\bar{\varphi}_0(t'\!,\vec{x}\,')|\Omega\rangle\,\langle\Omega|
\bar{\varphi}^2_0(t'''\!,\vec{x})|\Omega\rangle\nonumber\\
&&\hspace{2cm}+3\!\cdot\!2\!\cdot\!1\Big[\langle\Omega|
\bar{\varphi}_0(t''\!,\vec{x})\bar{\varphi}_0(t'''\!,\vec{x})|\Omega\rangle\Big]^2\,\langle\Omega|
\bar{\varphi}_0(t'''\!,\vec{x})\bar{\varphi}_0(t'\!,\vec{x}\,')|\Omega\rangle
\!\Bigg\}\; ,\label{quant2loopsecnd}
\ee
where $0\!\leq\!t'''\!\leq\!t''\!\leq\!t$ and $t'\!\leq\!t$. Employing Eqs.~(\ref{equalspaceVEV}) and (\ref{phisquare}) in Eq.~(\ref{quant2loopsecnd}) we get
\be
&&\hspace{-2cm}\langle\Omega|\!\!
\int_0^{t}\!\!\!dt''{\bar{\varphi}}^2_0(t''\!,\vec{x})\!\!
\int_0^{t''}\!\!\!\!dt'''{\bar{\varphi}}^3_0(t'''\!,\vec{x})
\bar{\varphi}_0(t'\!,\vec{x}\,')
|\Omega\rangle\!=\!\frac{H^{2D-4}}{2^{2D-2}\pi^{D}}\frac{\Gamma^2(\!D\!-\!1)}{\Gamma^2(\!\frac{D}{2})}
\Bigg\{\nonumber\\
&&\hspace{1cm}\int_0^{t}\!\!\!dt''\!\!\int_0^{t''}\!\!\!\!dt''' 2\!\cdot\!3\!\cdot\!1\langle\Omega|
\bar{\varphi}_0(t''\!,\vec{x})\bar{\varphi}_0(t'\!,\vec{x}\,')|\Omega\rangle\!\ln^2(a(t'''))\nonumber\\
&&\hspace{1cm}+\!\!\int_0^{t}\!\!\!dt''\!\!\int_0^{t''}\!\!\!\!dt''' 1\!\cdot\!3\!\cdot\!1\langle\Omega|
\bar{\varphi}_0(t'''\!,\vec{x})\bar{\varphi}_0(t'\!,\vec{x}\,')|\Omega\rangle\!\ln(a(t''))\!\ln(a(t'''))\nonumber\\
&&\hspace{1cm}+\!\!\int_0^{t}\!\!\!dt''\!\!\int_0^{t''}\!\!\!\!dt''' 3\!\cdot\!2\!\cdot\!1\langle\Omega|
\bar{\varphi}_0(t'''\!,\vec{x})\bar{\varphi}_0(t'\!,\vec{x}\,')|\Omega\rangle\!\ln^2(a(t'''))
\!\Bigg\}\; .\label{twoloopsecondexpandequaltime}
\ee
Because the ordering between $t'$ and $t''$ and between $t'$ and $t'''$ are unknown in the double integrals, we cannot just read off the VEVs from Eq.~(\ref{VEVgeneral}). We can, however, decompose the double integrals so that a definite ordering between the time parameters exists. That makes the evaluation of the integrals, using Eq.~(\ref{VEVgeneral}), possible; see Appendix \ref{App:VEV2}. Substituting Eqs.~(\ref{firstint}), (\ref{secondint}) and (\ref{thirdint}) into Eq.~(\ref{twoloopsecondexpandequaltime}), we find
\be
&&\hspace{-0.3cm}\langle\Omega|\!\!
\int_0^{t}\!\!dt''{\bar{\varphi}}^2_0(t''\!,\vec{x})\!\!
\int_0^{t''}\!\!\!\!dt'''{\bar{\varphi}}^3_0(t'''\!,\vec{x})
\bar{\varphi}_0(t'\!,\vec{x})
|\Omega\rangle\!=\!\frac{H^{3D-8}}{2^{3D-2}\pi^{3D/2}}\frac{\Gamma^3(\!D\!-\!1)}{\Gamma^3(\!\frac{D}{2})}
\Bigg\{\!\frac{11\!\ln^5(a(t'))}{20}\nonumber\\
&&\hspace{-0.3cm}-\ln^4(a(t'))\!\ln(a(t))\!-\!\frac{\ln^3(a(t'))}{2}\!\ln^2(a(t))
\!+\!\frac{11\!\left[\ln(a(t'))\!+\!\mathcal{C}(\!\Delta x)\right]}{4}\!\ln^4(a(t))\nonumber\\
&&\hspace{-0.3cm}+\!\sum_{n=1}^\infty
\!\frac{(-1)^{n}(\!H\!\Delta x)^{2n}}{n\,(2n\!+\!1)!}\Bigg\{\!a^{2n}(t')\!\Bigg[2\!\ln^5(a(t'))\!-\!2\!\ln^4(a(t'))\!\ln(a(t))\!-\!\frac{5\!\ln^4(a(t'))}{8}
\!+\!\frac{11\!\ln^4(a(t))}{8}\nonumber\\
&&\hspace{-0.3cm}-\frac{3\!\ln^2(a(t'))}{4}\ln^2(a(t))
\!-\!\frac{1}{n}\!\left(\!\frac{11\!\ln^3(a(t'))}{4}\!-\!3\!\ln^2(a(t'))\!\ln(a(t))\!-\!\frac{3\!\ln(a(t'))}{4}\!\ln^2(a(t))\!\right)\nonumber\\
&&\hspace{-0.3cm}+\frac{1}{n^2}\!\left(\!\frac{33\!\ln^2(a(t'))}{8}
\!-\!3\!\ln(a(t'))\!\ln(a(t))\!-\!\frac{3\!\ln^2(a(t))}{8}\!\right)
\!-\!\frac{1}{n^3}\!\left(\!\frac{33\!\ln(a(t'))}{8}\!-\!\frac{3\!\ln(a(t))}{2}\right)\!+\!\frac{33}{16n^4}\Bigg]\nonumber\\
&&\hspace{8cm}+\frac{3\!\ln^2(a(t))}{8n^2}\!-\!\frac{3\!\ln(a(t))}{2n^3}\!-\!\frac{33}{16n^4}\!\Bigg\}\!\Bigg\}\; .\label{2loop2ndresult}
\ee

Finally, the third VEV which contributes at two-loop order in correlation function~(\ref{expectequaltime})~is
\be
&&\hspace{0cm}\langle\Omega|\!\!
\int_0^{t}\!\!dt''{\bar{\varphi}}^3_0(t''\!,\vec{x})\!\!
\int_0^{t'}\!\!\!d\tilde{t}\,{\bar{\varphi}}^3_0(\tilde{t},\vec{x}\,')
|\Omega\rangle\!=\!\int_0^{t}\!\!dt''\!\!\int_0^{t'}\!\!\!d\tilde{t}\Bigg\{\nonumber\\
&&\hspace{2cm}3\!\cdot\!1\!\cdot\!1\,\langle\Omega|
\bar{\varphi}_0(t''\!,\vec{x})\bar{\varphi}_0(\tilde{t},\vec{x}\,')|\Omega\rangle\,\langle\Omega|
\bar{\varphi}^2_0(t''\!,\vec{x})|\Omega\rangle\,\langle\Omega|
\bar{\varphi}^2_0(\tilde{t}\!,\vec{x}\,')|\Omega\rangle\nonumber\\
&&\hspace{3cm}+2\!\cdot\!3\!\cdot\!1\,\langle\Omega|
\bar{\varphi}^2_0(t''\!,\vec{x})|\Omega\rangle\,\langle\Omega|
\bar{\varphi}_0(t''\!,\vec{x})\bar{\varphi}_0(\tilde{t},\vec{x}\,')|\Omega\rangle\,\langle\Omega|
\bar{\varphi}^2_0(\tilde{t}\!,\vec{x}\,')|\Omega\rangle\nonumber\\
&&\hspace{4cm}3\!\cdot\!2\!\cdot\!1\Big[\langle\Omega|
\bar{\varphi}_0(t''\!,\vec{x})\bar{\varphi}_0(\tilde{t},\vec{x}\,')|\Omega\rangle\Big]^3
\Bigg\}\; ,\label{2loop3rd}
\ee
where $0\!\leq\!t''\!\leq\!t$, $0\!\leq\!\tilde{t}\!\leq\!t'$, and $t'\!\leq\!t$. The integrand on the right of Eq.~(\ref{2loop3rd}) consists of three terms that are added together. Notice that the second term is twice the first term. Using Eq.~(\ref{phisquare}) in Eq.~(\ref{2loop3rd}) we get
\be
&&\hspace{-1.5cm}\langle\Omega|\!
\int_0^{t}\!\!dt''{\bar{\varphi}}^3_0(t''\!,\vec{x})\!\!
\int_0^{t'}\!\!\!d\tilde{t}\,{\bar{\varphi}}^3_0(\tilde{t},\vec{x}\,')
|\Omega\rangle\nonumber\\
&&\hspace{-1.4cm}=\!\!\int_0^{t}\!\!dt''\!\!\int_0^{t'}\!\!\!d\tilde{t}\,\Bigg\{9\langle\Omega|
\bar{\varphi}_0(t''\!,\vec{x})\bar{\varphi}_0(\tilde{t},\vec{x}\,')|\Omega\rangle\frac{H^{2D-4}}{2^{2D-2}\pi^D}\frac{\Gamma^2(\!D\!-\!1)}{\Gamma^2(\!\frac{D}{2})}\!
\ln(a(t''))\!\ln(a(\tilde{t}))\nonumber\\
&&\hspace{7cm}+6\Big[\langle\Omega|
\bar{\varphi}_0(t''\!,\vec{x})\bar{\varphi}_0(\tilde{t},\vec{x}\,')|\Omega\rangle\Big]^3
\Bigg\}\; .\label{2loop3rdmidequaltime}
\ee
As in the evaluation of the second VEV given in Eq.~(\ref{quant2loopsecnd}), we need an ordering of the time parameters $t''$ and $\tilde{t}$ to evaluate the expectation values in the integrand on the right of Eq.~(\ref{2loop3rdmidequaltime}). Therefore, we similarly break up the double integral into two parts: $\int_0^{t}\!dt''\!\!\int_0^{t'}\!d\tilde{t}\!=\!\int_0^{\tilde{t}}\!dt''\!\!\int_0^{t'}\!d\tilde{t}
\!+\!\int_{\tilde{t}}^{t}\!dt''\!\!\int_0^{t'}\!d\tilde{t}$ where in the first part on the right $t''\!\leq\!\tilde{t}$, whereas $\tilde{t}\!\leq\!t''$ in the second. Hence, employing Eq.~(\ref{VEVgeneral}) appropriately, we obtain\be
&&\hspace{-1cm}\langle\Omega|\!\!
\int_0^{t}\!\!dt''{\bar{\varphi}}^3_0(t''\!,\vec{x})\!\!
\int_0^{t'}\!\!\!d\tilde{t}\,{\bar{\varphi}}^3_0(\tilde{t},\vec{x}\,')
|\Omega\rangle\!=\!\frac{H^{3D-6}}{2^{3D-3}\pi^{3D/2}}\frac{\Gamma^3(\!D\!-\!1)}{\Gamma^3(\!\frac{D}{2})}
\Bigg\{\nonumber\\
&&\hspace{0cm}\int_0^{\tilde{t}}\!\!dt''\!\!\int_0^{t'}\!\!\!d\tilde{t}\Bigg\{9\!\left[\mathcal{C}(\!\Delta x)\!+\!\ln(a(t''))\!+\!\!\sum_{n=1}^\infty
\!\frac{(-1)^{n}(a(t'')H\!\Delta x)^{2n}}{2n\,(2n\!+\!1)!}\!\right]\!\!\ln(a(t''))\ln(a(\tilde{t}))\nonumber\\
&&\hspace{3cm}+6\!\left[\mathcal{C}(\!\Delta x)\!+\!\ln(a(t''))\!+\!\!\sum_{n=1}^\infty
\!\frac{(-1)^{n}(a(t'')H\!\Delta x)^{2n}}{2n\,(2n\!+\!1)!}\right]^3\!\Bigg\}\nonumber\\
&&\hspace{-0.4cm}+\!\!\int_{\tilde{t}}^{t}\!\!dt''\!\!\int_0^{t'}\!\!\!d\tilde{t}\Bigg\{9\!\left[\mathcal{C}(\!\Delta x)\!+\!\ln(a(\tilde{t}))\!+\!\!\sum_{n=1}^\infty
\!\frac{(-1)^{n}(a(\tilde{t})H\!\Delta x)^{2n}}{2n\,(2n\!+\!1)!}\!\right]\!\!\ln(a(t''))\ln(a(\tilde{t}))\nonumber\\
&&\hspace{3cm}+6\!\left[\mathcal{C}(\!\Delta x)\!+\!\ln(a(\tilde{t}))\!+\!\!\sum_{n=1}^\infty
\!\frac{(-1)^{n}(a(\tilde{t})H\!\Delta x)^{2n}}{2n\,(2n\!+\!1)!}\!\right]^3\!\Bigg\}
\!\Bigg\}\; .\label{2loop3rdaftmidequaltime}
\ee
Now, we need to evaluate the four double integrals in Eq.~(\ref{2loop3rdaftmidequaltime}) to get the VEV. Results of the integrals are given in Appendix \ref{App:VEV3}. Adding up Eqs.~(\ref{2loop3rdint1}), (\ref{2loop3rdint2}), (\ref{2loop3rdint3}) and (\ref{2loop3rdint4}) and substituting the sum into Eq.~(\ref{2loop3rdaftmidequaltime}), we find
\be
&&\hspace{-0.4cm}\langle\Omega|\!\!
\int_0^{t}\!\!dt''{\bar{\varphi}}^3_0(t''\!,\vec{x})\!\!\!
\int_0^{t'}\!\!\!d\tilde{t}\,{\bar{\varphi}}^3_0(\tilde{t},\vec{x}\,')
|\Omega\rangle\!=\!\frac{H^{3D-8}}{2^{3D-2}\pi^{3D/2}}\frac{\Gamma^3(\!D\!-\!1)}{\Gamma^3(\!\frac{D}{2})}3
\Bigg\{\!\!\!-\!\frac{4}{5}\ln^5(a(t'))\!+\!\ln^4(a(t'))\!\ln(a(t))\nonumber\\
&&\hspace{-0.4cm}+\!\ln^3(a(t'))\!\ln^2(a(t))\!-\!2\mathcal{C}(\!\Delta x)\!\!\left[\ln^4(a(t'))\!-\!2\!\ln^3(a(t'))\!\ln(a(t))\!-\!\frac{3\!\ln^2(a(t'))}{4}\!\ln^2(a(t))\right]\nonumber\\
&&\hspace{-0.4cm}-2\mathcal{C}^2(\!\Delta x)\!\left[\ln^3(a(t'))\!-\!3\!\ln^2(a(t'))\!\ln(a(t))\right]\!+\!4\mathcal{C}^3(\!\Delta x)\!\ln(a(t'))\!\ln(a(t))\nonumber\\
&&\hspace{-0.4cm}+3\!\sum_{n=1}^\infty
\!\frac{(-1)^{n}(\!H\!\Delta x)^{2n}}{n^2(2n\!+\!1)!}\Bigg\{\!a^{2n}(t')\!\Bigg\{\!\!\Big[\!\ln(a(t))\!-\!\ln(a(t'))\!\Big]
\!\Bigg[\frac{5\!\ln^2(a(t'))}{4}\!+\!\!\left(\!2\mathcal{C}(\!\Delta x)\!-\!\frac{9}{8n}\!\right)\!\ln(a(t'))\nonumber\\
&&\hspace{-0.4cm}+\frac{\ln(a(t'))}{4}\!\ln(a(t))\!-\!\frac{\ln(a(t))}{8n}\!\Bigg]
\!+\!\frac{3\!\ln^2(a(t'))}{2n}
\!+\!\mathcal{E}(\!\Delta x,\! n)\!\ln(a(t))\!-\!\mathcal{F}(\!\Delta x,\! n)\!\ln(a(t'))\!+\!\mathcal{G}(\!\Delta x,\! n)\!\Bigg\}\nonumber\\
&&\hspace{-0.4cm}-\mathcal{E}(\!\Delta x,\! n)\Big[\!\ln(a(t))\!+\!\ln(a(t'))\!\Big]\!+\!\frac{\ln^2(a(t))}{8n}\!+\!\frac{\ln^2(a(t'))}{8n}
\!-\!\mathcal{G}(\!\Delta x,\! n)\!\Bigg\}\nonumber\\
&&\hspace{-0.4cm}+\frac{3}{2}\!\sum_{p=2}^\infty\!\sum_{n=1}^{p-1}
\!\frac{(-1)^{p}(\!H\!\Delta x)^{2p}}{pn(p\!-\!n)(2n\!+\!\!1)![2(p\!-\!n)\!+\!\!1]!}\Bigg\{\!a^{2p}(t')
\Bigg\{\!\!\Big[\!\ln(a(t))\!-\!\ln(a(t'))\Big]\!
\!\left[\ln(a(t'))\!+\!\mathcal{C}(\!\Delta x)\!-\!\frac{1}{2p}\right]\nonumber\\
&&\hspace{-0.4cm}+\frac{\ln(a(t'))}{p}+\frac{\mathcal{C}(\!\Delta x)}{p}\!-\!\frac{1}{p^2}\!\Bigg\}
\!-\!\left[\mathcal{C}(\!\Delta x)\!-\!\frac{1}{2p}\right]\!\!\Big[\!\ln(a(t))\!+\!\ln(a(t'))\Big]\!-\!\frac{\mathcal{C}(\!\Delta x)}{p}\!+\!\frac{1}{p^2}\Bigg\}\nonumber\\
&&\hspace{-0.4cm}+\frac{1}{4}\!\sum_{q=3}^\infty\!\sum_{p=2}^{q-1}
\!\sum_{n=1}^{p-1}\!\frac{(-1)^{q}(\!H\!\Delta x)^{2q}\!\left[a^{2q}(t')\!\!\left(\!\ln(a(t))\!-\!\ln(a(t'))\!+\!\frac{1}{q}\right)
\!-\!\ln(a(t))\!-\!\ln(a(t'))\!-\!\frac{1}{q}\right]}
{q(q\!-\!p)[2(q\!-\!p)\!+\!\!1]!n(p\!-\!n)(2n\!+\!\!1)![2(p\!-\!n)\!+\!\!1]!}\Bigg\}\; , \label{2loop3rdequaltimeresult}
\ee
where we defined new functions of comoving spatial separation $\Delta x$ and index $n$\be\mathcal{F}(\!\Delta x,\! n)\!\equiv\!\mathcal{C}^2(\!\Delta x)\!-\!\frac{3\mathcal{C}(\!\Delta x)}{n}\!+\!\frac{13}{4n^2}\qquad{\rm and}\qquad
\mathcal{G}(\!\Delta x,\! n)\!\equiv\!\frac{\mathcal{C}^2(\!\Delta x)}{n}\!-\!\frac{2\mathcal{C}(\!\Delta x)}{n^2}\!+\!\frac{15}{8n^3}\; .\ee
Recall that the function $\mathcal{C}(\!\Delta x)$ is expressed as a power series
in Eq.~(\ref{Cpower}). The square and the cube of $\mathcal{C}(\!\Delta x)$ that appear in Eq.~(\ref{2loop3rdequaltimeresult}) can also be expressed as series expansions\be
\mathcal{C}^2(\!\Delta x)\!\!&=&\!\!\!\sum_{p=2}^\infty\sum_{n=1}^{p-1}\frac{(-1)^p(\!H\!\Delta x)^{2p}}
{4n(p\!-\!n)(2n\!+\!1)![2(p\!-\!n)\!+\!\!1]!}\; ,\label{Ckare}\\
\mathcal{C}^3(\!\Delta x)\!\!&=&\!\!-\!\sum_{q=3}^\infty\sum_{p=2}^{q-1}\sum_{n=1}^{p-1}\frac{(-1)^q(\!H\!\Delta x)^{2q}}
{8(q\!-\!p)[2(q\!-\!p)\!+\!\!1]!n(p\!-\!n)(2n\!+\!1)![2(p\!-\!n)\!+\!\!1]!}\; .\label{Ckup}
\ee

We have completed evaluating each of the three VEVs that yields an $\mathcal{O}(\lambda^2)$ correction to tree-order two-point correlation function~(\ref{treeorder}) of the IR truncated scalar full field. Substitution of those VEVs given in Eqs.~(\ref{2loop1stequaltimeresult}), (\ref{2loop2ndresult}) and (\ref{2loop3rdequaltimeresult}) into Eq.~(\ref{expectequaltime}) yields the total two-loop correction as\be
&&\hspace{-0.5cm}\langle\Omega|
\bar{\varphi}(t,\vec{x})\bar{\varphi}(t,\vec{x}\,')|\Omega\rangle_{\rm 2-loop}\!\simeq\!\frac{\lambda^2}{(\!D\!-\!1)^2}\frac{H^{3D-10}}{2^{3D}\pi^{3D/2}}\frac{\Gamma^3(\!D\!-\!1)}
{\Gamma^3(\!\frac{D}{2})}
\Bigg\{\!\frac{31\!\ln^5(a(t'))}{60}\!+\!\frac{\ln^3(a(t'))}{6}\!\ln^2(a(t))\nonumber\\
&&\hspace{-0.5cm}+\frac{11\!\ln(a(t'))}{12}\!\ln^4(a(t))\!+\!\mathcal{C}(\!\Delta x)\!\Bigg[\!\frac{11\!\ln^4(a(t))}{12}\!+\!\frac{\ln^4(a(t'))}{4}\!+\!\frac{4\!\ln^3(a(t'))}{3}\!\ln(a(t))\nonumber\\
&&\hspace{-0.5cm}+\frac{\ln^2(a(t'))}{2}\!\ln^2(a(t))\!\Bigg]\!\!-\!\mathcal{C}^2(\!\Delta x)\!\Bigg[\!\frac{2\!\ln^3(a(t'))}{3}\!-\!2\!\ln^2(a(t'))\!\ln(a(t))\!\Bigg]\!\!+\!\mathcal{C}^3(\!\Delta x)\frac{4\!\ln(a(t'))}{3}\!\ln(a(t))\nonumber\\
&&\hspace{-0.5cm}+\!\!\sum_{n=1}^\infty
\!\frac{(-1)^{n}(\!H\!\Delta x)^{2n}}{n(2n\!+\!1)!}\Bigg\{\!a^{2n}(t')\!\Bigg\{\!2\Big[\!\ln(a(t))\!-\!\ln(a(t'))\Big]\!\Bigg[\!\!-\!\frac{\ln^4(a(t'))}{3}
\!+\!\frac{\ln^2(a(t'))}{n}\nonumber\\
&&\hspace{-0.5cm}+\!\left(\!\frac{\mathcal{C}(\!\Delta x)}{n}\!-\!\frac{1}{n^2}\!\right)\!\!\ln(a(t'))\!+\!\frac{\mathcal{N}(\mathcal{C}, n)}{2}\!\Bigg]\!\!+\!\frac{5}{24}\Big[\!\ln^4(a(t))\!-\!\ln^4(a(t'))\Big]
\!+\!\frac{1}{4}\Big[\!\ln^2(a(t))\!-\!\ln^2(a(t'))\Big]\nonumber\\
&&\hspace{-0.5cm}\times\!\Bigg[\!\ln^2(a(t))\!+\!\frac{2\!\ln(a(t'))}{n}\!-\!\frac{1}{n^2}\!\Bigg]
\!\!+\!\frac{2\!\ln^3(a(t'))}{3n}
\!+\!\frac{2\!\ln^2(a(t'))}{n^2}\!+\!\!\left(\!\frac{2\mathcal{C}(\!\Delta x)}{n^2}\!-\!\frac{9}{2n^3}\!\right)\!\ln(a(t'))\!+\!\mathcal{P}(\mathcal{C}, n)\!\Bigg\}\nonumber\\
&&\hspace{-0.5cm}+\frac{\ln^2(a(t))}{4n^2}\!+\!\frac{\ln^2(a(t'))}{4n^2}\!-\!\mathcal{N}(\mathcal{C}, n)\!\Big[\!\ln(a(t))\!+\!\ln(a(t'))\Big]
\!\!-\!\mathcal{P}(\mathcal{C}, n)\!\Bigg\}\nonumber\\
&&\hspace{-0.5cm}+\frac{1}{2}
\!\sum_{p=2}^\infty\!\sum_{n=1}^{p-1}
\!\frac{(-1)^{p}(\!H\!\Delta x)^{2p}}{pn(p\!-\!n)(2n\!+\!\!1)![2(p\!-\!n)\!+\!\!1]!}\Bigg\{\!a^{2p}(t')
\!\Bigg\{\!\!\Big[\!\ln(a(t))\!-\!\ln(a(t'))\Big]\!\Big[\!\ln(a(t'))\!+\!\mathcal{C}(\!\Delta x)\!-\!\frac{1}{2p}\Big]\nonumber\\
&&\hspace{-0.5cm}+\frac{\ln(a(t'))}{p}\!+\!\frac{\mathcal{C}(\!\Delta x)}{p}\!-\!\frac{1}{p^2}\!\Bigg\}\!-\!\!\left[\mathcal{C}(\!\Delta x)\!-\!\frac{1}{2p}\right]\!\!\Big[\!\ln(a(t))\!+\!\ln(a(t'))\Big]\!\!-\!\frac{\mathcal{C}(\!\Delta x)}{p}\!+\!\frac{1}{p^2}\!\Bigg\}\nonumber\\
&&\hspace{-0.5cm}+\frac{1}{12}\!\sum_{q=3}^\infty\!\sum_{p=2}^{q-1}
\!\sum_{n=1}^{p-1}\!\frac{(-1)^{q}(\!H\!\Delta x)^{2q}\!\left\{\!a^{2q}(t')
\!\left[\ln(a(t))\!-\!\ln(a(t'))\!+\!\frac{1}{q}\right]\!-\!\ln(a(t))\!-\!\ln(a(t'))\!-\!\frac{1}{q}\!\right\}}
{q(q\!-\!p)[2(q\!-\!p)\!+\!1]!n(p-\!n)(2n\!+\!1)![2(p-\!n)\!+\!1]!}\!\Bigg\}\; ,\label{2loopequaltimefinal}
\ee
where we defined
\beeq
\mathcal{N}(\mathcal{C}, n)\!\equiv\!\frac{\mathcal{C}^2(\!\Delta x)}{n}\!-\!\frac{\mathcal{C}(\!\Delta x)}{n^2}\!+\!\frac{1}{n^3} \qquad {\rm and} \qquad \mathcal{P}(\mathcal{C}, n)\!\equiv\!\frac{\mathcal{C}^2(\!\Delta x)}{n^2}\!-\!\frac{2\mathcal{C}(\!\Delta x)}{n^3}\!+\!\frac{13}{4n^4}\; .
\eneq
To get the result for a fixed physical distance, one can choose the comoving separation $\Delta x\!=\!K/Ha(t')$ so that the physical distance $a(t')\Delta x$ remains as a constant fraction $K$ of the Hubble length, as in Eq.~(\ref{DeltaX}).

Adding up tree, one and two-loop correlators~(\ref{treeorder}), (\ref{1loopequaltimefinal}) and (\ref{2loopequaltimefinal})  yields the two-point correlation function of the IR truncated full field at one and two-loop order,
\be
&&\hspace{-1cm}\langle\Omega|
\bar{\varphi}(t,\vec{x})\bar{\varphi}(t'\!,\vec{x}\,')|\Omega\rangle\simeq\langle\Omega|
\bar{\varphi}(t,\vec{x})\bar{\varphi}(t'\!,\vec{x}\,')|\Omega\rangle_{\rm tree}\!+\!\langle\Omega|
\bar{\varphi}(t,\vec{x})\bar{\varphi}(t,\vec{x}\,')|\Omega\rangle_{\rm 1-loop}\nonumber\\
&&\hspace{7.47cm}+\langle\Omega|
\bar{\varphi}(t,\vec{x})\bar{\varphi}(t,\vec{x}\,')|\Omega\rangle_{\rm 2-loop}\; .\label{genelfullexpect}
\ee
(To save space we don't write the result explicitly.) Computation of this correlation function was one of the goals of this paper. We use it, in the next section, to obtain the variance of the IR truncated scalar at one and two-loop order.

\subsection{Fluctuations in Spacetime}
The variation $\Delta \bar{\varphi}$ of the IR truncated scalar $\bar{\varphi}$ is defined as the difference of fields at two events
\beeq
\Delta \bar{\varphi}(t, t' ; \vec{x}, \vec{x}\,') \!\equiv\!
\bar{\varphi}(t, \vec{x})\!-\!\bar{\varphi}(t'\!, \vec{x}\,') \; .
\eneq
Because the VEV of field $\bar{\varphi}$ is zero at any event, the VEV of variation
$\Delta\bar{\varphi}$ also vanishes. However, the variance
\be
&&\hspace{-1.cm}\sigma^2_{\Delta \bar{\varphi}}(t, t' ; \vec{x}, \vec{x}\,')\!\equiv\!\langle\Omega|\!
\left[\Delta\bar{\varphi} \!-\!\langle\Omega|\Delta\bar{\varphi}|\Omega\rangle\right]^2\!|\Omega\rangle\!=\!\langle\Omega|
\left(\Delta\bar{\varphi}\right)^2\!|\Omega\rangle\nonumber\\
&&\hspace{-1cm}=\langle\Omega|{\bar{\varphi}}^2(t, \vec{x})|\Omega\rangle \!-\!\langle\Omega|\bar{\varphi}(t, \vec{x})\bar{\varphi}(t'\!, \vec{x}\,')|\Omega\rangle\!-\!\langle\Omega|\bar{\varphi}(t'\!, \vec{x}\,')\bar{\varphi}(t, \vec{x})|\Omega\rangle \!+\!\langle\Omega|\bar{\varphi}^2(t'\!, \vec{x}\,')|\Omega\rangle\; ,\label{vargeneral}
\ee
is not necessarily zero. The VEVs in the middle of the second line in Eq.~(\ref{vargeneral}) are just the two-point correlation function of the IR truncated scalar obtained in Eq.~(\ref{genelfullexpect}). The remaining VEVs can be obtained from this function by taking appropriate limits:\be
&&\hspace{1cm}\langle\Omega|
\bar{\varphi}^2(t, \vec{x})|\Omega\rangle\!=\!\langle\Omega|
\bar{\varphi}^2(t, \vec{x}\,')|\Omega\rangle\!=\!\frac{H^{D-2}}{2^{D-1}\pi^{{D}/{2}}}
\frac{\Gamma\!\left(\!D\!-\!1\right)}{\Gamma\!\left(\!\frac{D}{2}\right)}\ln(a(t))\nonumber\\
&&\hspace{-1.2cm}-\frac{\lambda}{(\!D\!-\!1)}
\frac{H^{2D-6}}{2^{2D-2}\,3\,\pi^{D}}\frac{\Gamma^2\!\left(\!D\!-\!1\right)}{\Gamma^2\!\left(\!\frac{D}{2}\right)}
\ln^3(a(t))\!+\!\frac{\lambda^2}{(\!D\!-\!1)^2}\frac{H^{3D-10}}{2^{3D-3}\,5\,\pi^{3D/2}}
\frac{\Gamma^3\!\left(\!D\!-\!1\right)}{\Gamma^3\!\left(\!\frac{D}{2}\right)}\ln^5(a(t))
\; ,\label{fieldsquare}
\ee
and \beeq\langle\Omega|\bar{\varphi}^2(t'\!, \vec{x}\,')|\Omega\rangle\!=\!\langle\Omega|\bar{\varphi}^2(t'\!, \vec{x})|\Omega\rangle=\lim_{t\rightarrow t'}\langle\Omega|
\bar{\varphi}^2(t, \vec{x})|\Omega\rangle\; ,\label{fieldprimesquare}
\eneq
which can directly be read off from Eq.~(\ref{fieldsquare}). Thus, using Eqs.~(\ref{genelfullexpect}), (\ref{fieldsquare}) and (\ref{fieldprimesquare}) in Eq.~(\ref{vargeneral}), one obtains the variance $\sigma^2_{\Delta \bar{\varphi}}(t, t' ; \vec{x}, \vec{x}\,')$. In the next two sections, we consider its equal time and equal space limits, respectively.

\subsubsection{Fluctuations in Space}

The spatial variation of the IR truncated scalar is the difference of fields at equal time events
$\Delta \bar{\varphi}(t, \vec{x}, \vec{x}\,')\!=\!
\bar{\varphi}(t, \vec{x})\!-\!\bar{\varphi}(t, \vec{x}\,')$. It derives from the fact that inflationary particle production is a random process, like everything else in quantum mechanics. At some places the
field strength is positive, at others it is negative; at some places the {\it magnitude}
of the field strength increases at others it decreases. Although the {\it expectation value} of the spatial variation $\Delta \bar{\varphi}(t, \vec{x}, \vec{x}\,')$ {\it vanishes},
the spatial variance $\sigma^2_{\Delta \bar{\varphi}}(t, \vec{x}, \vec{x}\,')$, which can be obtained by taking the equal time limit of Eq.~(\ref{vargeneral}), {\it does not}. We employ Eqs.~(\ref{genelfullexpect}) and~(\ref{fieldsquare}) in Eq.~(\ref{vargeneral}), to obtain\be
&&\hspace{-0.6cm}\sigma^2_{\Delta \bar{\varphi}}(t, \vec{x}, \vec{x}\,')\!=\!\langle\Omega|\!
\left[\Delta\bar{\varphi}(t, \vec{x}, \vec{x}\,')\!-\!\langle\Omega|\Delta\bar{\varphi}(t, \vec{x}, \vec{x}\,')|\Omega\rangle\right]^2\!|\Omega\rangle\!=\!\langle\Omega|\!\left[\bar{\varphi}(t, \vec{x})\!-\!\bar{\varphi}(t, \vec{x}\,')\right]^2\!|\Omega\rangle\nonumber\\
&&\hspace{-0.6cm}\simeq\!\frac{H^{D-2}}{2^{D-1}\pi^{{D}/{2}}}
\frac{\Gamma\!\left(\!D\!-\!1\right)}{\Gamma\!\left(\!\frac{D}{2}\right)}\!\sum_{n=1}^\infty
\frac{(-1)^{n+1}(\!H\!\Delta x)^{2n}}{n\,(2n\!+\!\!1)!}\left(a^{2n}(t)\!-\!1\right)\!+\!\frac{\lambda}{(\!D\!-\!1)}\frac{H^{2D-6}}{2^{2D-1}\pi^{D}}
\frac{\Gamma^2\!\left(\!D\!-\!1\right)}{\Gamma^2\!\left(\!\frac{D}{2}\right)}\Bigg\{\nonumber\\
&&\hspace{-0.6cm}\sum_{n=1}^\infty
\frac{(-1)^{n}(\!H\!\Delta x)^{2n}}{n\,(2n\!+\!\!1)!}\!\left[\frac{a^{2n}(t)}{n}\!\ln(a(t))\!-\!\ln^2(a(t))
\!-\!\frac{(a^{2n}(t)\!-\!1)}{2n^2}\right]\!\!\Bigg\}\!-\!\frac{\lambda^2}{(\!D\!-\!1)^2}
\frac{H^{3D-10}}{2^{3D-2}\pi^{3D/2}}\nonumber\\
&&\hspace{-0.6cm}\times\frac{\Gamma^3(\!D\!-\!1)}
{\Gamma^3(\!\frac{D}{2})}
\Bigg\{\!\!\sum_{n=1}^\infty
\!\frac{(-1)^{n}(\!H\!\Delta x)^{2n}}{n\,(2n\!+\!\!1)!}\Bigg\{\!a^{2n}(t)\!\Bigg[\frac{\ln^3(a(t))}{3n}\!+\!\frac{\ln^2(a(t))}{n^2}
\!+\!\left(\!\frac{\mathcal{C}(\!\Delta x)}{n^2}\!-\!\frac{9}{4n^3}\!\right)\!\ln(a(t))\!+\!\frac{\mathcal{P}}{2}\Bigg]\nonumber\\
&&\hspace{-0.6cm}-\frac{3\!\ln^4(a(t))}{4}\!+\!\frac{\ln^2(a(t))}{4n^2}\!-\!\mathcal{N}\!\ln(a(t))
\!-\!\frac{\mathcal{P}}{2}\!\Bigg\}\!+\!\frac{1}{4}
\!\sum_{p=2}^\infty\!\sum_{n=1}^{p-1}
\!\frac{(-1)^{p}(\!H\!\Delta x)^{2p}}{pn(p\!-\!n)(2n\!+\!\!1)![2(p\!-\!n)\!+\!\!1]!}\Bigg\{\nonumber\\
&&\hspace{-0.6cm}\frac{a^{2p}(t)}{p}\!\left[\ln(a(t))\!+\!\mathcal{C}(\!\Delta x)\!-\!\frac{1}{p}\right]\!\!+\!\frac{2p}{3}\!\ln^3(a(t))\!-\!\!\left[2\mathcal{C}(\!\Delta x)\!-\!\frac{1}{p}\right]\!\!\ln(a(t))\!-\!\frac{\mathcal{C}(\!\Delta x)}{p}\!+\!\frac{1}{p^2}\!\Bigg\}\nonumber\\
&&\hspace{0.9cm}+\frac{1}{24}\!\sum_{q=3}^\infty\!\sum_{p=2}^{q-1}
\!\sum_{n=1}^{p-1}\!\frac{(-1)^{q}(\!H\!\Delta x)^{2q}\!\Big[a^{2q}(t)\!-\!2q^2\ln^2(a(t))\!-\!2q\ln(a(t))\!-\!1\Big]}
{q^2(q\!-\!p)[2(q\!-\!p)\!+\!\!1]!n(p\!-\!n)(2n\!+\!\!1)![2(p\!-\!n)\!+\!\!1]!}\Bigg\}\!+\!\mathcal{O}(\lambda^3)\; .\label{spatialvariance}
\ee
Tree order spatial variance in Eq.~(\ref{spatialvariance}),
\be
&&\hspace{-0.9cm}\sigma^2_{{\Delta \bar{\varphi}_{\rm tree}}}\simeq\!\frac{H^{D-2}}{2^{D-2}\pi^{{D}/{2}}}
\frac{\Gamma\!\left(\!D\!-\!1\right)}{\Gamma\!\left(\!\frac{D}{2}\right)}\Bigg\{\!\!\ln(a(t))
\!+\!\frac{\sin\left(a(t)H\!\Delta x\right)}{a(t)H\!\Delta x}\!-\!\frac{\sin(\!H\!\Delta x)}{H\!\Delta x}\nonumber\\
&&\hspace{8.5cm}-{\rm ci}\!\left(a(t)H\!\Delta x\right)\!+{\rm ci}(\!H\!\Delta x)\!\Bigg\}\; ,\label{varfree}
\ee
for a fixed comoving separation (increasing physical distance) asymptotes to
\be
&&\hspace{-0.9cm}\sigma^2_{{\Delta \bar{\varphi}_{\rm tree}}}\rightarrow
\frac{H^{D-2}}{2^{D-2}\pi^{{D}/{2}}}
\frac{\Gamma\!\left(\!D\!-\!1\right)}{\Gamma\!\left(\!\frac{D}{2}\right)}\Bigg\{\!\!\ln(a(t))\!+{\rm ci}(\!H\!\Delta x)
\!-\!\frac{\sin(\!H\!\Delta x)}{H\!\Delta x}\!\Bigg\}\; .\label{varfreeasymp}
\ee
(We used Eq.~(\ref{ciasym}) in Eq.~(\ref{varfree}) to get Eq.~(\ref{varfreeasymp}).) Time dependence of tree order spatial variance implies that the {\it magnitude} of spatial
variation increases with time. This fact is a crucial argument
of some cosmologists against putting any trust in results for
expectation values. They argue that an expectation value, which is measure of average effect, is misleading since the actual effect that would be perceived by any local observer is either an increase or a decrease in the field strength as fluctuations happen. Reference~\cite{W0}, on the other hand, argues that, while there is certainly spatial and temporal variation in the actual effects, one can, under certain circumstances, roughly trust what expectation values are telling us. See also Ref.~\cite{Tegmark} which mentions the problems some cosmologists have with using expectation values and gives conditions under which we can trust them.

To infer our result~(\ref{spatialvariance}) for a fixed physical distance we take comoving spatial separation $\Delta x\!=\!K/Ha(t)$ as in Eq.~(\ref{DeltaX}), use Eq.~(\ref{ciminussin}), and get
\be
&&\hspace{-1.3cm}\sigma^2_{{\Delta \bar{\varphi}_{\rm tree}}}\simeq
\frac{H^{D-2}}{2^{D-2}\pi^{{D}/{2}}}
\frac{\Gamma\!\left(\!D\!-\!1\right)}{\Gamma\!\left(\!\frac{D}{2}\right)}\Bigg\{\!\ln(\!K)
\!+\!\frac{\sin(\!K)}{K}\!-\!{\rm ci}(\!K)\!+\!\gamma\!-\!1\!+\!\sum_{n=1}^\infty\!
                 \frac{(-1)^{n}K^{2n}a^{-2n}(t)}{2n\,(2n\!+\!\!1)!}\Bigg\}\; ,\label{varianceasymp}
\ee
which agrees with the tree order result given in Ref.~\cite{W0} when $D\!=\!4$. Equation~(\ref{varianceasymp}) implies that the tree order variance freezes in to a not especially large value. If the physical distance is taken as (half) the Hubble length, then the quantity in the curly brackets approaches ($0.02$) $0.08$ during inflation. The one-loop variance in Eq.~(\ref{spatialvariance}),\be
&&\hspace{0cm}\sigma^2_{{\Delta \bar{\varphi}_{\rm 1-loop}}}\simeq
-\frac{\lambda}{(\!D\!-\!1)}\frac{H^{2D-6}}{2^{2D-2}\pi^{D}}
\frac{\Gamma^2\!\left(\!D\!-\!1\right)}{\Gamma^2\!\left(\!\frac{D}{2}\right)}
\Bigg\{\!\!\left[{\rm ci}(\!H\!\Delta x)\!-\!\frac{\sin(\!H\!\Delta x)}{H\!\Delta x}\!-\!\ln\!\left(\!H\!\Delta x\right)\!+\!1\!-\!\gamma\right]\!\ln^2(a(t))\nonumber\\
&&\hspace{2.1cm}+\frac{\alpha^2}{12}\,
{}_{3}\mathcal{F}_{4}\!\left(\!\!1, 1, 1; 2, 2, 2, \frac52; -\frac{\alpha^2}{4}\!\right)\!\ln(a(t))
\!-\!\frac{\alpha^2}{24}\,
{}_{4}\mathcal{F}_{5}\!\left(\!\!1, 1, 1, 1; 2, 2, 2, 2, \frac52; -\frac{\alpha^2}{4}\!\right)\nonumber\\
&&\hspace{6cm}+\frac{H^2\!\Delta x^2}{24}\,
{}_{4}\mathcal{F}_{5}\!\left(\!\!1, 1, 1, 1; 2, 2, 2, 2, \frac52; -\frac{H^2\!\Delta x^2}{4}\!\right)\!\!\Bigg\}\; ,
\ee
where $\alpha\!=\!a(t)H\!\Delta x$, for a fixed physical distance $a(t)\Delta x\!=\!K/H$, becomes
\be
&&\hspace{-1.1cm}\sigma^2_{{\Delta \bar{\varphi}_{\rm 1-loop}}}\simeq
-\frac{\lambda}{(\!D\!-\!1)}\frac{H^{2D-6}}{2^{2D-2}\pi^{D}}
\frac{\Gamma^2\!\left(\!D\!-\!1\right)}{\Gamma^2\!\left(\!\frac{D}{2}\right)}
\Bigg\{\!\frac{K^2}{12}\,
{}_{3}\mathcal{F}_{4}\!\left(\!\!1, 1, 1; 2, 2, 2, \frac52; -\frac{K^2}{4}\!\right)\!\ln(a(t))\nonumber\\
&&\hspace{4cm}-\frac{K^2}{24}\,
{}_{4}\mathcal{F}_{5}\!\left(\!\!1, 1, 1, 1; 2, 2, 2, 2, \frac52; -\frac{K^2}{4}\!\right)\!+\!\mathcal{O}(a^{-2}(t))\!\Bigg\}\; ,\label{1loopcorr}
\ee
which grows {\it negatively}. If the physical distance is taken as (half) the Hubble length, then the quantity in the curly brackets approaches ($1.03$) $4.07$ after fifty e-foldings. This means that the spatial variation does not grow as large as tree-order variance~(\ref{varianceasymp}) implies. Hence the VEV describes the behavior of the field better when the loop corrections are included. Next, we study the temporal fluctuations of the scalar field in our model, considering the difference of fields at equal space events.

\subsubsection{Fluctuations in Time}
The temporal variation is the difference of fields at equal space events
$\Delta\bar{\varphi}(t, t'\!, \vec{x}) \equiv
\bar{\varphi}(t, \vec{x})-\!\bar{\varphi}(t'\!,\vec{x})$ whose VEV vanishes.
The temporal variance can be obtained by taking the equal space limit of Eq.~(\ref{vargeneral}). Using Eqs.~(\ref{genelfullexpect}) and (\ref{fieldsquare}), we find
\be
&&\hspace{-0.4cm}\sigma^2_{\Delta \bar{\varphi}}(t, t'\!, \vec{x})\!=\!\langle\Omega|\!
\left[\Delta\bar{\varphi}(t, t'\!, \vec{x})\!-\!\langle\Omega|\Delta\bar{\varphi}(t, t'\!, \vec{x})|\Omega\rangle\right]^2\!|\Omega\rangle\!=\!\langle\Omega|\!\left[\bar{\varphi}(t, \vec{x})\!-\!\bar{\varphi}(t'\!, \vec{x})\right]^2\!|\Omega\rangle\nonumber\\
&&\hspace{-0.4cm}=\!\frac{H^{D-2}}{2^{D-1}\pi^{{D}/{2}}}
\frac{\Gamma\!\left(\!D\!-\!1\right)}{\Gamma\!\left(\!\frac{D}{2}\right)}\Big[\!\ln(a(t))\!-\!\ln(a(t'))\Big]
\!-\!\frac{\lambda}{(\!D\!-\!1)}
\frac{H^{2D-6}}{2^{2D-1}\,\pi^{D}}\frac{\Gamma^2\!\left(\!D\!-\!1\right)}{\Gamma^2\!\left(\!\frac{D}{2}\right)}
\Bigg[\frac{2\!\ln^3(a(t))}{3}\!+\!\frac{\ln^3(a(t'))}{3}\nonumber\\
&&\hspace{-0.4cm}-\!\ln(a(t'))\!\ln^2(a(t))\Bigg]
\!\!+\!\frac{\lambda^2}{(\!D\!-\!1)^2}\frac{H^{3D-10}}{2^{3D}\,\pi^{3D/2}}
\frac{\Gamma^3\!\left(\!D\!-\!1\right)}{\Gamma^3\!\left(\!\frac{D}{2}\right)}
\Bigg[\frac{8\!\ln^5(a(t))}{5}\!+\!\frac{17\!\ln^5(a(t'))}{30}\nonumber\\
&&\hspace{5.5cm}-\frac{\ln^3(a(t'))}{3}\!\ln^2(a(t))
\!-\!\frac{11\!\ln(a(t'))}{6}\!\ln^4(a(t))\Bigg]\!\!+\!\mathcal{O}(\lambda^3)\; .\label{temporalvariance}
\ee
The one-loop correction tells us that the temporal variation does not grow as fast as the tree-order result implies. Thus, the VEV describes the behavior of field better, when the quantum corrections are included. In the next section, we study the IR truncated scalar in the same model applying the SFT instead of the QFT. We will see that it yields the same two-point correlation function and hence the same statistical properties we infer from it.
\section{Stochastic Field Theoretical Analysis}
\label{sec:Stoch}

Although the free field in Eq.~(\ref{expantruncwithu}) involves only the IR modes, it is still a quantum field in the sense that the commutator of the field and its first time derivative does not vanish:
\be
&&\hspace{-0.9cm}\left[\varphi_0(t, \vec{x}) , \dot{\varphi}_0(t, \vec{x}\,')\right]\!=\!H^{D-1}\!\sum_{n\neq0}\!\Theta(\!Ha(t)\!-\!H2\pi n)
\!\left[u(t, k){\dot{u}}^*(t, k)\!-\!u^*(t, k)\dot{u}(t, k) \right]\!e^{i\vec{k}\cdot(\vec{x}-\vec{x}\,')}\label{yuz}\\
&&\hspace{-0.9cm}=\!\frac{i\pi^{-{D}/{2}}}{(a(t)\Delta x)^{D-1}}\frac{\Gamma\!\left(\!\frac{D}{2}\right)}{\Gamma\!\left(\!D\!-\!1\right)}
\Bigg\{\!(\!D\!-\!3)\left(a(t)H\!\Delta x\right)^{D-4}\sin\left(a(t)H\!\Delta x\right)\!+\!\left(\!H\!\Delta x\right)^{D-3}\cos\left(\!H\!\Delta x\right)\nonumber\\
&&\hspace{4cm}-\!\left(a(t)H\!\Delta x\right)^{D-3}\cos\left(a(t)H\!\Delta x\right)\!-\!(\!D\!-\!3)\left(\!H\!\Delta x\right)^{D-4}\sin\left(\!H\!\Delta x\right)\nonumber\\
&&\hspace{4cm}-(\!D\!-\!3)(\!D\!-\!4)\!\!\int_{H\!\Delta x}^{Ha(t)\Delta x}\!\!dy\, y^{D-5}\sin(y) \Bigg\}\; .\label{comlongint}
\ee
See Appendix~\ref{App:commutator} for the derivation of Eq.~(\ref{comlongint}). Recall that we replaced the mode function in Eq.~(\ref{expantruncwithu}) with its constant leading IR limit~(\ref{modeleadingorder}) to obtain the IR truncated free field ${\bar\varphi}_0(t, \vec{x})$ in Eq.~(\ref{whitef}). Thus, Eq.~(\ref{yuz}) implies that the commutator $[{\bar\varphi}_0(t, \vec{x}), {\dot{\bar\varphi}}_0(t, \vec{x}\,')]$ vanishes when the replacement is made. That means that ${\bar\varphi}_0(t, \vec{x})$ behaves like a classical variable ${\bar\phi}_0(t, \vec{x})$ in which the associated annihilation and creation operators are just random variables. Such a field is called stochastic. The full stochastic field $\phi(t, \vec{x})$ in our model obeys
\begin{equation}
\phi(t,\vec{x}) \!=\! \phi_0(t,\vec{x})\!-\!\frac{1}{(\!D\!-\!1)H}\!\!
\int_0^t\!\! dt' V'(\phi(t',\vec{x}))\; , \label{integstarlang}
\end{equation}
first obtained by Starobinsky in Ref.~\cite{Staro}.
Note that Eq.~(\ref{integstarlang}) has the same form as Eq.~(\ref{fieldinfra}), up to the field strength renormalization factor $(1\!+\!\delta Z)^{-1}$.

\subsection{Free Theory}
\label{subsec:stochfree}
As discussed earlier, we take the IR limit of the mode function
and retain only the super-horizon modes ---except the zero-mode, to obtain IR truncated stochastic free~field
\begin{equation}
{\bar\phi}_0(t, \vec{x})\!=\!\sum_{\vec{n} \neq 0} \!\frac{\Gamma(\!D\!-\!1)}{\Gamma(\!\frac{D}{2})}\frac{H^{D-3/2}}{(2 k)^{\frac{D-1}{2}}}
\Theta\Bigl(\!H a(t) \!-\! k\!\Bigr)\!\Bigl[\cos(\vec{k} \!\cdot\! \vec{x})
\alpha_{\vec{n}} \!-\! \sin(\vec{k} \!\cdot\! \vec{x})\beta_{\vec{n}}
\Bigr]\; ,\label{stochfreefield}
\end{equation}
where $\vec{k}\!=\!2\pi H\vec{n}$ as in Eq.~(\ref{whitef}). The annihilation and creation operators $\hat{A}_{\vec{n}}$
and $\hat{A}^\dagger_{\vec{n}}$ in Eq.~(\ref{whitef}) are considered \cite{W0} to be two complex conjugate
stochastic random variables
\beeq
A_{\vec{n}} \equiv \frac{1}{2}\left(\alpha_{\vec{n}} \!+\! i \beta_{\vec{n}}\right)\;\; {\rm and}\;\; A^*_{\vec{n}} \equiv \frac{1}{2}\left(\alpha_{\vec{n}} \!-\! i \beta_{\vec{n}}\right)\; .
\eneq
The real random variables $\alpha_{\vec{n}}$ and $\beta_{\vec{n}}$ are governed by a Gaussian probability distribution function
\beeq
\rho\left(\alpha_{\vec{n}}, \beta_{\vec{n}}\right)
\!=\!\frac{e^{-\frac12 \alpha_{\vec{n}}^2}}{\sqrt{2\pi}} \frac{e^{-\frac12 \beta_{\vec{n}}^2}}{\sqrt{2\pi}} \; ,\label{real}
\eneq
with mean zero and standard deviation one. Note that the factors of $\sqrt{2\pi}$ in the denominators exist for each mode $\vec{n}$.

Now, let us apply the stochastic formalism to compute the two-point correlation function of the IR truncated stochastic  free field
\be
\hspace{-0.5cm}\Big\langle\!\bar{{\phi}}_{0}(t,\vec{x}) \bar{{\phi}}_{0}(t'\!,\vec{x}\,')\!\Big\rangle\!\!&=&\!\prod_{{\vec{q}}\neq 0}\int_{-\infty}^{+\infty}\!\!\!\frac{d\alpha_{\vec{q}}}{\sqrt{2\pi}}e^{-\frac{1}{2}\alpha_{\vec{q}}^2}\!\int_{-\infty}^{+\infty}\!\!\!
\frac{d{\beta_{\vec{q}}}}{\sqrt{2\pi}}e^{-\frac{1}{2}\beta_{\vec{q}}^2}\,\bar{{\phi}}_{0}(t,\vec{x}) \bar{{\phi}}_{0}(t'\!,\vec{x}\,')\nonumber\\
&=&\!\frac{\Gamma^2(\!D\!-\!1)}{\Gamma^2(\!\frac{D}{2})}
\frac{H^{2D-3}}{2^{D-1}}\!\sum_{\vec{n}\neq 0}\!\sum_{\vec{m}\neq 0}\!\frac{\Theta(\!Ha(t)\!-\! k)}{k^{\frac{D-1}{2}}}\,\frac{\Theta(\!Ha(t')\!-\!k')}{k'^{\frac{D-1}{2}}}\nonumber\\
&&\hspace{2cm}\times\delta_{\vec{m},\,\vec{n}}\!\left(\!\cos(\vec{k} \!\cdot\! \vec{x})\cos(\vec{k}' \!\cdot\! \vec{x}\,')\!+\!\sin(\vec{k} \!\cdot\! \vec{x})\sin(\vec{k}' \!\cdot\! \vec{x}\,')\!\right)\; ,
\ee
where $\vec{k}'\!=\!2\pi H\vec{m}$. When $\vec{m}\!=\!\vec{n}$ we have $\vec{k}'\!=\!\vec{k}$. Therefore, we obtain
\be
&&\hspace{-1.5cm}\Big\langle\!\bar{{\phi}}_{0}(t,\vec{x}) \bar{{\phi}}_{0}(t'\!,\vec{x}\,')\!\Big\rangle\!=\!
\frac{\Gamma^2(\!D\!-\!1)}{\Gamma^2(\!\frac{D}{2})}
\frac{H^{D-2}}{2^{2D-2}\pi^{D-1}}\!\sum_{\vec{n}\neq 0}\!\frac{\Theta(\!H\!a(t)\!-\!k)\Theta(\!H\!a(t')\!-\!k)}{n^{D-1}}\!\cos\!{\left(\vec{k}\!\cdot\!(\vec{x}\!-\!\vec{x}\,')\!\right)}\; . \label{whitesum}
\ee
Using the fact that
\be
\cos{\left(\vec{k}\!\cdot\!(\vec{x}\!-\!\vec{x}\,')\!\right)}\!=\!\frac{1}{2}\!\left(e^{ik \Delta x \cos{(\theta)}}\!+\!e^{-ik \Delta x \cos{(\theta)}}\right)\; ,\label{complexexp}
\ee
where $\theta$ is the angle between $\vec{k}\!=\!2\pi H\vec{n}$ and $\Delta\vec{x}\!=\!\vec{x}\!-\!\vec{x}\,'$, and making the integral approximation to the discrete mode sum over $\vec{n}$ as in Eqs.~(\ref{angkeyeqn})-(\ref{logar}) we find
\be
&&\hspace{-1.5cm}\sum_{\vec{n}\neq 0}\frac{\Theta(\!Ha(t)\!-\!k)\,\Theta(\!Ha(t')\!-\!k)}{n^{D-1}}
\cos\!{\left(\!\vec{k}\!\cdot\!(\vec{x}\!-\!\vec{x}\,')\!\right)}\nonumber\\
&&\hspace{-0.5cm}\simeq2^{D-1}\pi^{\frac{D}{2}-1}
\frac{\Gamma\!\left(\!\frac{D}{2}\right)}{\Gamma\!\left(\!D\!-\!1\right)}\!\!\left[{\rm ci}\!\left(a(t')H\!\Delta x\right)\!-\!\frac{\sin\left(a(t')H\!\Delta x\right)}{a(t')H\!\Delta x}\!-\!{\rm ci}(\!H\!\Delta x)\!+\!\frac{\sin(\!H\!\Delta x)}{H\!\Delta x}\right]\label{thetathetacossumone}\\
&&\hspace{-0.5cm}\simeq2^{D-1}\pi^{\frac{D}{2}-1}
\frac{\Gamma\!\left(\!\frac{D}{2}\right)}{\Gamma\!\left(\!D\!-\!1\right)}\!\!\left[\mathcal{C}(\!\Delta x)\!+\!\ln(a(t'))\!+\!\!\sum_{n=1}^\infty\!
                 \frac{(-1)^{n}(a(t')H\!\Delta x)^{2n}}{2n\,(2n\!+\!\!1)!}\right]
\; ,\label{sumthetacos}
\ee
where we used Eqs.~(\ref{ciminussin}) and (\ref{C}) in the last equality. Let us note here that the equal space and equal spacetime limits of Eq.~(\ref{sumthetacos}) yield
\beeq
\sum_{\vec{n}\neq 0}\frac{\Theta(\!Ha(t)\!-\!k)\,\Theta(\!Ha(t')\!-\!k)}{n^{D-1}}\!=\!\sum_{\vec{n}\neq 0}\frac{\Theta(\!Ha(t')\!-\!k)}{n^{D-1}}\!=\!2^{D-1}\pi^{\frac{D}{2}-1}
\frac{\Gamma\!\left(\!\frac{D}{2}\right)}{\Gamma\!\left(\!D\!-\!1\right)}\!\ln(a(t'))\; ,\label{sumsingletheta}
\eneq
where $t'\leq t$. Using Eqs.~(\ref{thetathetacossumone})-(\ref{sumthetacos}) in Eq.~(\ref{whitesum}), we find
\be
&&\hspace{-1.5cm}\Big\langle\!\bar{{\phi}}_{0}(t,\vec{x}) \bar{{\phi}}_{0}(t'\!,\vec{x}\,')\!\Big\rangle\nonumber\\
&&\hspace{-0.5cm}\!\simeq\!\frac{H^{D-2}}{2^{D-1}\pi^{{D}/{2}}}
\frac{\Gamma\!\left(\!D\!-\!1\right)}{\Gamma\!\left(\!\frac{D}{2}\right)}\!\!\left[{\rm ci}\!\left(a(t')H\!\Delta x\right)\!-\!\frac{\sin\!\left(a(t')H\!\Delta x\right)}{a(t')H\!\Delta x}\!-\!{\rm ci}(\!H\!\Delta x)\!+\!\frac{\sin(\!H\!\Delta x)}{H\!\Delta x}\right]\; ,\label{stochtreephizero}\\
&&\hspace{-0.5cm}\!\simeq\!\frac{H^{D-2}}{2^{D-1}\pi^{{D}/{2}}}
\frac{\Gamma\!\left(\!D\!-\!1\right)}{\Gamma\!\left(\!\frac{D}{2}\right)}\!\!\left[\mathcal{C}(\!\Delta x)\!+\!\ln(a(t'))\!+\!\!\sum_{n=1}^\infty\!
                 \frac{(-1)^{n}(a(t')H\!\Delta x)^{2n}}{2n\,(2n\!+\!\!1)!}\right]\; ,\label{stochtreephi}
\ee
respectively. Results~(\ref{stochtreephizero})-(\ref{stochtreephi}) are the same as results~(\ref{expphisq})-(\ref{VEVgeneral}) obtained in the context of QFT. Hence,
\beeq
\Big\langle\!\bar{{\phi}}_{0}(t,\vec{x}) \bar{{\phi}}_{0}(t'\!,\vec{x}\,')\!\Big\rangle\!=\!\langle\Omega|
\bar{\varphi}_0(t,\vec{x})\bar{\varphi}_0(t'\!,\vec{x}\,')|\Omega\rangle\; .\label{stocquanttree}
\eneq

Because it will be needed in Sec.~\ref{subsec:check}, let us compute the two-point correlation function for the first time derivative of the IR truncated stochastic free field, \be
\hspace{-0.35 cm}f_{\!\bar{\phi}_0}(t,\vec{x})\!\equiv\!{\dot{\bar{\phi}}_0}(t,\vec{x})\!=\!\!\sum_{\vec{n} \neq 0} \!\frac{\Gamma(\!D\!-\!1)}{\Gamma(\!\frac{D}{2})}
\frac{H^{D-\frac{1}{2}}}{2^{(D-1)/2}k^{(D-3)/2}}
\delta_{H a(t_n),\,k}\Bigl[\cos(\vec{k} \!\cdot\! \vec{x})
\alpha_{\vec{n}}\!-\!\sin(\vec{k} \!\cdot\! \vec{x}) \beta_{\vec{n}}
\Bigr] \; , \label{} \ee
which is
\be
&&\hspace{-0.5cm}\Big\langle\!
f_{\!\bar{{\phi}}_{0}}(t,\vec{x}) f_{\!\bar{{\phi}}_{0}}(t'\!,\vec{x}\,')\!\Big\rangle\!=\!\prod_{{\vec{q}}\neq 0}\int_{-\infty}^{+\infty}\!\!\!\frac{d\alpha_{\vec{q}}}{\sqrt{2\pi}}e^{-\frac{1}{2}\alpha_{\vec{q}}^2}\!\int_{-\infty}^{+\infty}\!\!\!
\frac{d{\beta_{\vec{q}}}}{\sqrt{2\pi}}e^{-\frac{1}{2}\beta_{\vec{q}}^2} f_{\!{\phi}_{0}}(t,\vec{x}) f_{\!{\phi}_{0}}(t'\!,\vec{x}\,')\nonumber\\
&&\hspace{-0.5cm}\!=\!\frac{\Gamma^2(\!D\!-\!1)}{\Gamma^2(\!\frac{D}{2})}
\frac{H^{2D-1}}{2^{D-1}}\!\sum_{\vec{n}\neq 0}\!\sum_{\vec{m}\neq 0}\!\frac{\delta_{Ha(t_n),\, k}}{k^{(D-3)/2}}\,\frac{\delta_{Ha(t'_m),\, k\,'}}{k'^{(D-3)/2}}\delta_{\vec{m},\,\vec{n}}\!\left(\!\cos(\vec{k} \!\cdot\! \vec{x})\cos(\vec{k}' \!\cdot\! \vec{x}\,')\!+\!\sin(\vec{k} \!\cdot\! \vec{x})\sin(\vec{k}' \!\cdot\! \vec{x}\,')\!\right)\; .\nonumber
\ee
When $\vec{m}\!=\!\vec{n}$ we have $\vec{k}'\!=\!\vec{k}$, hence we obtain
\be
\Big\langle\!
f_{\!{\bar{\phi}}_{0}}(t,\vec{x}) f_{\!{\bar{\phi}}_{0}}(t',\vec{x}\,')\!\Big\rangle
\!=\!\frac{\Gamma^2(\!D\!-\!1)}{\Gamma^2(\!\frac{D}{2})}
\frac{H^{D+2}}{2^{2D-4}\pi^{D-3}}\sum_{\vec{n}\neq 0}\frac{\delta_{Ha(t_n)\, ,\, Ha(t'_n)}}{n^{D-3}}\!\cos{\left(2\pi H\vec{n}\!\cdot\!(\vec{x}\!-\!\vec{x}\,')\right)}
\; .\label{white}
\ee
Employing Eq.~(\ref{complexexp}), making the integral approximation to the sum as in Eqs.~(\ref{sumint})-(\ref{intdelta}), we find
\be
\sum_{\vec{n}\neq 0}\frac{\delta_{Ha(t_n)\, ,\, Ha(t'_n)}}{n^{D-3}}\!\cos{\left(2\pi H\vec{n}\!\cdot\!(\vec{x}\!-\!\vec{x}\,')\right)}\!\simeq\!\frac{2^D\pi^{D/2}}{(2\pi H)^3}\frac{\Gamma\!\left(\!\frac{D}{2}\right)}{\Gamma\!\left(\!D\!-\!1\right)}\frac{\sin(a(t')H\!\Delta x)}{a(t')H\!\Delta x}\delta(t\!-\!t')\;
\; .\label{sumcos}
\ee
This result is the same as the one in Eq.~(\ref{sum}). Thus, substituting Eq.~(\ref{sumcos}) into Eq.~(\ref{white}) we obtain
\beeq
\Big\langle\!
f_{{\bar{\phi}}_{0}}(t,\vec{x}) f_{{\bar{\phi}}_{0}}(t'\!,\vec{x}\,')\!\Big\rangle\simeq\frac{H^{D-1}}{2^{D-1}\pi^{D/2}}
\frac{\Gamma(\!D\!-\!1)}{\Gamma(\!\frac{D}{2})}\frac{\sin(a(t')H\!\Delta x)}{a(t')H\!\Delta x}
\delta(t\!-\!t')\; ,\label{stocwhite}
\eneq
which is the same result obtained in Eq.~(\ref{expect}) applying QFT. The equal space limit of correlation~(\ref{stocwhite}) is the limit given in Eq.~(\ref{equalspacef0f0}), hence $f_{{\bar{\phi}}_{0}}$ is a stochastic white noise. The limit enters the Fokker-Planck equation as a source term in the last section. In the next section, we study the self-interacting theory in the context of SFT.

\subsection{Interacting Theory}
\label{subsec:stochinteract}
Here also we consider $V(\phi)=\frac{\lambda}{4!}\phi^4$ theory in $D$ spacetime dimensions. Equation~(\ref{integstarlang})
implies for the stochastic scalar\be
\phi(t, \vec{x})\!=\!\phi_0(t, \vec{x})\!-\!\frac{\l}{6(\!D\!-\!1)H}
\!\!\int_0^t\!\!dt''\phi^3(t'',\vec{x})\; .
\ee
Iterating this equation twice for the IR truncated field $\bar\phi$ yields
\be
&&\hspace{-1cm}\bar\phi(t, \vec{x})\!=\!\bar\phi_0(t,\vec{x})\!-\!\frac{\l}{6(\!D\!-\!1)H}\!\!\int_0^t\!\!dt'
\bar\phi^3_0(t'\!,\vec{x})\!+\!\frac{\l^2}{12(\!D\!-\!1)^2H^2}\!\!\int_0^t\!\!
dt'\bar\phi^2_0(t'\!,\vec{x})\!\!\int_0^{t'}\!\!dt''\bar\phi^3_0(t''\!,\vec{x})\nonumber\\
&&\hspace{1.85cm}-\frac{\l^3}{24(\!D\!-\!1)^3H^3}\Bigg\{\!\!\int_0^t\!\!
dt'\bar\phi^2_0(t'\!,\vec{x})\!\!\int_0^{t'}
\!\!dt''\bar\phi^2_0(t''\!,\vec{x})\!\!\int_0^{t''}
\!\!\!dt'''\bar\phi^3_0(t'''\!,\vec{x})\nonumber\\
&&\hspace{5.5cm}+\frac{1}{3}\!\int_0^t\!
dt'\bar\phi_0(t'\!,\vec{x})\!\left[\int_0^{t'}\!\!dt''\bar\phi^3_0(t''\!,\vec{x})\right]^2\!\Bigg\}
\!+\!\mathcal{O}(\l^4)\; .\ee
Analogous to Eq.~(\ref{expectequaltime}), the two-point correlation function of the IR truncated stochastic full field is
\be
&&\hspace{-1.5cm}\Big\langle\!
\bar{\phi}(t,\vec{x})\bar{\phi}(t'\!,\vec{x}\,')\!\Big\rangle\!=\!\Big\langle\!
\bar{\phi}_0(t,\vec{x})\bar{\phi}_0(t'\!,\vec{x}\,')\!\Big\rangle\nonumber\\
&&\hspace{-1.5cm}-\frac{\lambda}{6(\!D\!-\!1)H}\!\!\left[\Big\langle\!
\bar{\phi}_0(t,\vec{x})\!\!\int_0^{t'}\!\!\!\!d\tilde{t}\,{\bar{\phi}}^3_0(\tilde{t},\vec{x}\,')\!
\Big\rangle\!+\!\Big\langle\!
\int_0^{t}\!\!\!dt''{\bar{\phi}}^3_0(t''\!,\vec{x})
\bar{\phi}_0(t'\!,\vec{x}\,')\!\Big\rangle\right]\nonumber\\
&&\hspace{-1.5cm}+\frac{\lambda^2}{12(\!D\!-\!1)^2H^2}\!\Bigg[\Big\langle\!
\bar{\phi}_0(t,\vec{x})\!\!\int_0^{t'}\!\!\!d\tilde{t}\,{\bar{\phi}}^2_0(\tilde{t},\vec{x}\,')
\!\!\int_0^{\tilde{t}}\!\!d\tilde{\tilde{t}}\, {\bar{\phi}}^3_0(\tilde{\tilde{t}},\vec{x}\,')
\!\Big\rangle\nonumber\\
&&\hspace{-1.5cm}+\Big\langle\!
\int_0^{t}\!\!dt''{\bar{\phi}}^2_0(t''\!,\vec{x})\!\!
\int_0^{t''}\!\!\!dt'''{\bar{\phi}}^3_0(t'''\!,\vec{x})
\bar{\phi}_0(t'\!,\vec{x}\,')\!
\Big\rangle\!+\!\frac{1}{3}\Big\langle\!
\int_0^{t}\!\!dt''{\bar{\phi}}^3_0(t''\!,\vec{x})\!\!
\int_0^{t'}\!\!\!d\tilde{t}\,{\bar{\phi}}^3_0(\tilde{t},\vec{x}\,')
\!\Big\rangle\Bigg]\!\!+\!\mathcal{O}(\l^3)\; ,\label{expectstoachgenel}
\ee
where $0\!\leq\!t'''\!\leq\!t''\!\leq\!t$, $0\!\leq\!\tilde{\tilde{t}}\!\leq\!\tilde{t}\!\leq\!t'$ and $t'\leq t$.
The first term is the tree-order contribution calculated in Eq.~(\ref{stochtreephi}). The $\mathcal{O}(\lambda)$ contribution involves two expectation values each of which has four free fields multiplied together. Recall that, in the quantum field theoretical analysis of Sec. \ref{sec:Quant}, we reduced the computation to that of the topologically inequivalent pairings of free fields. These pairings originated directly from the commutator algebra $[\hat{A}_{\vec{n}} ,
\hat{A}^{\dagger}_{\vec{m}}]\!=\!\delta_{\vec{n} , \vec{m}}$ of the annihilation and the creation operators and the fact that $\hat{A}_{\vec{n}}|\Omega\rangle\!=\!0$ which exclude any higher order groupings of the free fields. In the stochastic analysis, on the other hand, the Gaussian integrals over the random variables $\alpha_{\vec{n}}$ and $\beta_{\vec{n}}$ do not vanish when an even power of the variables $\alpha_{\vec{n}}$ and $\beta_{\vec{n}}$ are considered. Thus, one needs to exclude those combinations to allow only the pairings of the $\alpha_{\vec{n}}$s and of the $\beta_{\vec{n}}$s. This can be achieved by a projection operator $\mathcal{P}$. Let us now compute the $\mathcal{O}(\lambda)$ contribution in Eq.~(\ref{expectstoachgenel}).

\subsubsection{$\mathcal{O}(\lambda)$ Contribution}
\label{subsubsteqq1loop}
The first of two expectation values which contribute at $\mathcal{O}(\lambda)$ in Eq.~(\ref{expectstoachgenel}) is
\be
\hspace{-1cm}\Big\langle\!
\bar{\phi}_0(t,\vec{x})\!\!\int_0^{t'}\!\!\!d\tilde{t}\,{\bar{\phi}}^3_0(\tilde{t},\vec{x}\,')\!
\Big\rangle\!&=&\!\!\!\int_0^{t'}\!\!\!d\tilde{t}\,\Big\langle\!
\bar{\phi}_0(t,\vec{x}){\bar{\phi}}^3_0(\tilde{t},\vec{x}\,')\!
\Big\rangle\nonumber\\
&=&\!\!\!\int_0^{t'}\!\!\!d\tilde{t}\,\mathcal{P}\prod_{{\vec{q}}\neq 0}\!\int_{-\infty}^{+\infty}\!\!\!\frac{d\alpha_{\vec{q}}}{\sqrt{2\pi}}
e^{-\frac{1}{2}\alpha_{\vec{q}}^2}\!\!\int_{-\infty}^{+\infty}\!\!\!
\frac{d{\beta_{\vec{q}}}}{\sqrt{2\pi}}e^{-\frac{1}{2}\beta_{\vec{q}}^2}\,
\bar{\phi}_0(t,\vec{x}){\bar{\phi}}^3_0(\tilde{t},\vec{x}\,')\; .\label{1loopstoch}
\ee
Inserting free field expansion~(\ref{stochfreefield}), for each $\bar\phi_0$, into Eq.~(\ref{1loopstoch}), evaluating the Gaussian integrals using $\int_{-\infty}^{\infty}\!\frac{dx}{\sqrt{2\pi}} e^{-{x^2}/{2}}\!=\!\int_{-\infty}^{\infty}\!\frac{dx}{\sqrt{2\pi}} x^2e^{-{x^2}/{2}}\!=\!1$ and projecting out quartic combinations of the random variables yields
\be
&&\hspace{-2cm}\Big\langle\!
\bar{\phi}_0(t,\vec{x})\!\!\int_0^{t'}\!\!\!d\tilde{t}\,{\bar{\phi}}^3_0(\tilde{t},\vec{x}\,')\!
\Big\rangle\!=\!\frac{H^{2D-4}}{2^{4D-4}\pi^{2D-2}}
\frac{\Gamma^4\!\left(\!D\!-\!1\right)}{\Gamma^4\!\left(\!\frac{D}{2}\right)}
3\nonumber\\
&&\hspace{0cm}\times\!\!\int_0^{t'}\!\!\!d\tilde{t}\sum_{\vec{n}\neq 0}\frac{\Theta(\!Ha(t)\!-\!k)\,\Theta(\!Ha(\tilde{t})\!-\!k)}{n^{D-1}}
\cos\!{\left(\vec{k}\!\cdot\!(\vec{x}\!-\!\vec{x}\,')\!\right)}\!\sum_{\vec{m}\neq 0}\frac{\Theta(\!Ha(\tilde{t})\!-\!k')}{m^{D-1}}\; ,\label{stoch1loopsecmid}
\ee
where $\tilde{t}\!\leq\!t'\!\leq\!t$. At this point, we employ sums~(\ref{sumthetacos}) and (\ref{sumsingletheta}) in Eq.~(\ref{stoch1loopsecmid}) and find
\be
&&\hspace{-2cm}\Big\langle\!
\bar{\phi}_0(t,\vec{x})\!\!\int_0^{t'}\!\!\!d\tilde{t}\,{\bar{\phi}}^3_0(\tilde{t},\vec{x}\,')\!
\Big\rangle\nonumber\\
&&\hspace{-1cm}\!=\!\frac{H^{2D-4}}{2^{2D-2}\pi^{D}}
\frac{\Gamma^2\!\left(\!D\!-\!1\right)}{\Gamma^2\!\left(\!\frac{D}{2}\right)}
3\!\!\int_0^{t'}\!\!\!d\tilde{t}\!\left[\mathcal{C}(\!\Delta x)\!+\!\ln(a(\tilde{t}))\!+\!\!\sum_{n=1}^\infty
                 \frac{(-1)^{n}(a(\tilde{t})H\!\Delta x)^{2n}}{2n\,(2n\!+\!1)!}\right]\!\!\ln(a(\tilde{t}))\; .
\ee
This result is the same as the one obtained in Eq.~(\ref{pairequaltime}). Performing the remaining temporal integration yields the result given in Eq.~(\ref{1loop1result}) in the context of QFT. Thus,
\be
\Big\langle\!
\bar{\phi}_0(t,\vec{x})\!\!\int_0^{t'}\!\!\!d\tilde{t}\,{\bar{\phi}}^3_0(\tilde{t},\vec{x}\,')\!
\Big\rangle\!=\!\langle\Omega|
\bar{\varphi}_0(t,\vec{x})\!\!\int_0^{t'}\!\!\!d\tilde{t}\,{\bar{\varphi}}^3_0(\tilde{t},\vec{x}\,')
\!|\Omega\rangle \; .\label{stoch1loopfirst}
\ee

The remaining expectation value which contributes at $\mathcal{O}(\lambda)$ in Eq.~(\ref{expectstoachgenel}),
\be
&&\hspace{-1.5cm}\Big\langle\!
\int_0^{t}\!\!dt''{\bar{\phi}}^3_0(t''\!,\vec{x})
\bar{\phi}_0(t'\!,\vec{x}\,')\!\Big\rangle\!=\!\!\int_0^{t}\!\!dt''\,\mathcal{P}\prod_{{\vec{q}}\neq 0}\!\int_{-\infty}^{+\infty}\!\!\!\frac{d\alpha_{\vec{q}}}{\sqrt{2\pi}}e^{-\frac{1}{2}
\alpha_{\vec{q}}^2}\!\!\int_{-\infty}^{+\infty}\!\!\!
\frac{d{\beta_{\vec{q}}}}{\sqrt{2\pi}}e^{-\frac{1}{2}\beta_{\vec{q}}^2}\,
\bar{\phi}^3_0(t''\!,\vec{x}){\bar{\phi}}_0(t'\!,\vec{x}\,')\; ,\label{2ndvev1loopstoch}
\ee
can be computed similarly. For each free field in Eq.~(\ref{2ndvev1loopstoch}) we insert expansion~(\ref{stochfreefield}) and evaluate the Gaussian integrals projecting out quartic combinations of the random variables, to obtain
\be
&&\hspace{-1.5cm}\Big\langle\!
\int_0^{t}\!\!dt''{\bar{\phi}}^3_0(t''\!,\vec{x})
\bar{\phi}_0(t'\!,\vec{x}\,')\!\Big\rangle\!=\!\frac{H^{2D-4}}{2^{4D-4}\pi^{2D-2}}
\frac{\Gamma^4\!\left(\!D\!-\!1\right)}{\Gamma^4\!\left(\!\frac{D}{2}\right)}
3\!\!\int_0^{t}\!\!dt''\!\sum_{\vec{n}\neq 0}\frac{\Theta(\!Ha(t'')\!-\!k)}{n^{D-1}}\nonumber\\
&&\hspace{4.5cm}\times\!\sum_{\vec{m}\neq 0}\frac{\Theta(\!Ha(t'')\!-\!k')\,\Theta(\!Ha(t')\!-\!k')}{m^{D-1}}
\cos\!{\left(\vec{k}'\!\cdot\!(\vec{x}\!-\!\vec{x}\,')\!\right)}\; .\label{phiphi3stoch}
\ee
We then substitute the sum computed in Eq~(\ref{sumsingletheta}) into Eq.~(\ref{phiphi3stoch}),
\be
&&\hspace{-1.4cm}\Big\langle\!
\int_0^{t}\!\!dt''{\bar{\phi}}^3_0(t''\!,\vec{x})
\bar{\phi}_0(t'\!,\vec{x}\,')\!\Big\rangle\nonumber\\
&&\hspace{-1.4cm}=\!\frac{H^{2D-4}}{2^{3D-3}\pi^{3D/2-1}}
\frac{\Gamma^3\!\left(\!D\!-\!1\right)}{\Gamma^3\!\left(\!\frac{D}{2}\right)}
3\!\!\int_0^{t}\!\!dt''\!\ln(a(t''))\nonumber\\
&&\hspace{4cm}\times\!\sum_{\vec{m}\neq 0}\frac{\Theta(\!Ha(t'')\!-\!k')\,\Theta(\!Ha(t')\!-\!k')}{m^{D-1}}
\cos\!{\left(\vec{k}'\!\cdot\!(\vec{x}\!-\!\vec{x}\,')\!\right)}\; .\label{stoch1loopsecnd}
\ee
Because we have $0\leq\!t''\!\leq\!t$ and $t'\!\leq\!t$, we decompose the integral into two parts: $\int_0^{t}\!dt''\!=\!\int_0^{t'}\!dt''\!+\!\int_{t'}^t\!dt''$ to evaluate the remaining sum in the integrand. In the first integral on the right $t''\!\leq\!t'$, whereas in the second $t'\!\leq\!t''$. Employing Eq.~(\ref{complexexp}) in  Eq.~(\ref{stoch1loopsecnd}) appropriately, we get
\be
&&\hspace{-1.4cm}\Big\langle\!
\int_0^{t}\!\!dt''{\bar{\phi}}^3_0(t''\!,\vec{x})
\bar{\phi}_0(t'\!,\vec{x}\,')\!\Big\rangle\nonumber\\
&&\hspace{-1.4cm}=\!\frac{H^{2D-4}}{2^{2D-2}\pi^{D}}
\frac{\Gamma^2\!\left(\!D\!-\!1\right)}{\Gamma^2\!\left(\!\frac{D}{2}\right)}3
\Bigg\{\!\mathcal{C}(\!\Delta x)\!\!\int_0^{t}\!\!dt''\!\ln(a(t''))\!+\!\!\int_0^{t'}\!\!\!dt''\!\ln^2(a(t''))
\!+\!\ln(a(t'))\!\!\int_{t'}^{t}\!\!dt''\!\ln(a(t''))\nonumber\\
&&\hspace{1cm}+\frac{1}{2}\!\sum_{n=1}^\infty
\!\frac{(-1)^{n}(\!H\!\Delta x)^{2n}}{n\,(2n\!+\!1)!}\!\left[\int_0^{t'}\!\!\!dt''a^{2n}(t'')\!\ln(a(t''))
\!+\!a^{2n}(t')\!\!\int_{t'}^{t}\!\!dt''\ln(a(t''))\right]\!\!\Bigg\}\; .
\ee
This result is the same as the one obtained in Eq.~(\ref{quant1loopsecnd}). We then perform the remaining temporal integrations and find the result given in Eq.~(\ref{1loop2result}) in the context of QFT. Thus,
\be
\Big\langle\!
\int_0^{t}\!\!dt''{\bar{\phi}}^3_0(t''\!,\vec{x})
\bar{\phi}_0(t'\!,\vec{x}\,')\!\Big\rangle\!=\!\langle\Omega|\!\!
\int_0^{t}\!\!dt''{\bar{\varphi}}^3_0(t''\!,\vec{x})
\bar{\varphi}_0(t'\!,\vec{x}\,')|\Omega\rangle\;. \label{stoch1lpsec}
\ee
Equations~(\ref{stoch1loopfirst}) and (\ref{stoch1lpsec}) imply the equivalence of the $\mathcal{O}(\lambda)$-stochastic and the one-loop QFT correlators
\be
\Big\langle\!
\bar{\phi}(t,\vec{x})\bar{\phi}(t'\!,\vec{x}\,')\!\Big\rangle_{\mathcal{O}(\lambda)}\!=\!\langle\Omega|
\bar{\varphi}(t,\vec{x})\bar{\varphi}(t'\!,\vec{x}\,')|\Omega\rangle_{\rm 1-loop}\; ,\label{stocquant1lp}
\ee
where the latter was obtained in Eq.~(\ref{1loopequaltimefinal}). In the next section, we show that the same equivalence also holds for the $\mathcal{O}(\lambda^2)$-stochastic and the two-loop QFT correlators.

\subsubsection{$\mathcal{O}(\lambda^2)$ Contribution}
\label{subsubsteqq2loop}
Three expectation values contribute at $\mathcal{O}(\lambda^2)$ in the two-point correlation function of the stochastic scalar full field in Eq.~(\ref{expectstoachgenel}). Each will be computed in this section. Let us start evaluating the first one:
\be
&&\hspace{-1cm}\Big\langle\!
\bar{\phi}_0(t,\vec{x})\!\!\!\int_0^{t'}\!\!\!d\tilde{t}\,{\bar{\phi}}^2_0(\tilde{t},\vec{x}\,')
\!\!\!\int_0^{\tilde{t}}\!\!d\tilde{\tilde{t}}\, {\bar{\phi}}^3_0(\tilde{\tilde{t}},\vec{x}\,')
\!\Big\rangle\nonumber\\
&&\hspace{-1cm}=\!\!\!\int_0^{t'}\!\!\!d\tilde{t}\int_0^{\tilde{t}}\!\!d\tilde{\tilde{t}}
\,\,\mathcal{P}\prod_{{\vec{q}}\neq 0}\!\int_{-\infty}^{+\infty}\!\!\!\frac{d\alpha_{\vec{q}}}{\sqrt{2\pi}}
e^{-\frac{1}{2}\alpha_{\vec{q}}^2}\!\!\int_{-\infty}^{+\infty}\!\!\!
\frac{d{\beta_{\vec{q}}}}{\sqrt{2\pi}}e^{-\frac{1}{2}\beta_{\vec{q}}^2}\,\bar{\phi}_0(t,\vec{x})\,{\bar{\phi}}^2_0(\tilde{t},\vec{x}\,')
\,{\bar{\phi}}^3_0(\tilde{\tilde{t}},\vec{x}\,')\; ,\label{1stvev2loopstoch}
\ee
where the ordering of the time parameters is $0\!\leq\!\tilde{\tilde{t}}\!\leq\!\tilde{t}\!\leq\!t'\!\leq\!t$. We follow the same steps as in the evaluation of  Eq.~(\ref{1loopstoch}) or Eq.~(\ref{2ndvev1loopstoch}). For each free field  we insert expansion~(\ref{stochfreefield}) into Eq.~(\ref{1stvev2loopstoch}) and evaluate the Gaussian integrals over the random variables projecting out quartic and hexic combinations. We find
\be
&&\hspace{-1cm}\Big\langle\!
\bar{\phi}_0(t,\vec{x})\!\!\!\int_0^{t'}\!\!\!d\tilde{t}\,{\bar{\phi}}^2_0(\tilde{t},\vec{x}\,')
\!\!\!\int_0^{\tilde{t}}\!\!d\tilde{\tilde{t}}\, {\bar{\phi}}^3_0(\tilde{\tilde{t}},\vec{x}\,')
\!\Big\rangle\nonumber\\
&&\hspace{-1cm}=\!\frac{H^{3D-6}}{2^{6D-6}\pi^{3D-3}}
\frac{\Gamma^6\!\left(\!D\!-\!1\right)}{\Gamma^6\!\left(\!\frac{D}{2}\right)}
\!\!\int_0^{t'}\!\!\!d\tilde{t}\!\int_0^{\tilde{t}}\!\!d\tilde{\tilde{t}}\Bigg\{6\!\sum_{\vec{n}\neq 0}\frac{\Theta(\!Ha(t)\!-\!k)\,\Theta(\!Ha(\tilde{t})\!-\!k)}{n^{D-1}}
\cos\!{\left(\vec{k}\!\cdot\!(\vec{x}\!-\!\vec{x}\,')\!\right)}\nonumber\\
&&\hspace{1cm}\times\!\!\sum_{\vec{m}\neq 0}\frac{\Theta(\!Ha(\tilde{t})\!-\!k')\Theta(\!Ha(\tilde{\tilde{t}})\!-\!k')}{m^{D-1}}\!\sum_{\vec{r}\neq 0}\frac{\Theta(\!Ha(\tilde{\tilde{t}})\!-\!k'')}{r^{D-1}}\nonumber\\
&&\hspace{-1cm}+3\!\sum_{\vec{n}\neq 0}\frac{\Theta(\!Ha(t)\!-\!k)\,\Theta(\!Ha(\tilde{\tilde{t}})\!-\!k)}{n^{D-1}}
\cos\!{\left(\vec{k}\!\cdot\!(\vec{x}\!-\!\vec{x}\,')\!\right)}\!\!\sum_{\vec{m}\neq 0}\frac{\Theta(\!Ha(\tilde{t})\!-\!k')}{m^{D-1}}\!\sum_{\vec{r}\neq 0}\frac{\Theta(\!Ha(\tilde{\tilde{t}})\!-\!k'')}{r^{D-1}}\nonumber\\
&&\hspace{-1cm}+6\!\sum_{\vec{n}\neq 0}\frac{\Theta(\!Ha(t)\!-\!k)\,\Theta(\!Ha(\tilde{\tilde{t}})\!-\!k)}{n^{D-1}}
\cos\!{\left(\vec{k}\!\cdot\!(\vec{x}\!-\!\vec{x}\,')\!\right)}\nonumber\\
&&\hspace{1cm}\times\!\!\sum_{\vec{m}\neq 0}\frac{\Theta(\!Ha(\tilde{t})\!-\!k')\Theta(\!Ha(\tilde{\tilde{t}})\!-\!k')}{m^{D-1}}\!\sum_{\vec{r}\neq 0}\frac{\Theta(\!Ha(\tilde{t})\!-\!k'')\Theta(\!Ha(\tilde{\tilde{t}})\!-\!k'')}{r^{D-1}}\Bigg\}\; ,\label{1stvev2loopstochwithsums}
\ee
where $k''\!=\!2\pi Hr$. The sums in Eq.~(\ref{1stvev2loopstochwithsums}) are evaluated in Eqs.~(\ref{sumthetacos}) and (\ref{sumsingletheta}). When their outcomes are substituted, the expectation value becomes
\be
&&\hspace{-1cm}\Big\langle\!
\bar{\phi}_0(t,\vec{x})\!\!\!\int_0^{t'}\!\!\!d\tilde{t}\,{\bar{\phi}}^2_0(\tilde{t},\vec{x}\,')
\!\!\!\int_0^{\tilde{t}}\!\!d\tilde{\tilde{t}}\, {\bar{\phi}}^3_0(\tilde{\tilde{t}},\vec{x}\,')
\!\Big\rangle\nonumber\\
&&\hspace{-1cm}=\!\frac{H^{3D-6}}{2^{3D-3}\pi^{3D/2}}
\frac{\Gamma^3\!\left(\!D\!-\!1\right)}{\Gamma^3\!\left(\!\frac{D}{2}\right)}
\!\!\int_0^{t'}\!\!\!d\tilde{t}\!\!\int_0^{\tilde{t}}\!\!d\tilde{\tilde{t}}\Bigg\{\!6\!\left[\mathcal{C}(\!\Delta x)\!+\!\ln(a(\tilde{t}))\!+\!\!\sum_{n=1}^\infty\!
                 \frac{(-1)^{n}(a(\tilde{t})H\!\Delta x)^{2n}}{2n(2n\!+\!1)!}\right]\!\!\ln^2(a(\tilde{\tilde{t}}))\nonumber\\
&&\hspace{0cm}+3\!\left[\mathcal{C}(\!\Delta x)\!+\!\ln(a(\tilde{\tilde{t}}))\!+\!\sum_{n=1}^\infty\!
                 \frac{(-1)^{n}(a(\tilde{\tilde{t}})H\!\Delta x)^{2n}}{2n(2n\!+\!1)!}\right]\!\!\left[\ln(a(\tilde{t}))
                 \ln(a(\tilde{\tilde{t}}))\!+\!2\ln^2(a(\tilde{\tilde{t}}))\right]\!\!\Bigg\}\; .
\ee
This result is the same as the one in Eq.~(\ref{2loop1stequaltime}). We, after integration, obtain the result given in Eq.~(\ref{2loop1stequaltimeresult}). Thus, we conclude that
\be
\Big\langle\!
\bar{\phi}_0(t,\vec{x})\!\!\int_0^{t'}\!\!\!d\tilde{t}\,{\bar{\phi}}^2_0(\tilde{t},\vec{x}\,')
\!\!\int_0^{\tilde{t}}\!\!d\tilde{\tilde{t}}\, {\bar{\phi}}^3_0(\tilde{\tilde{t}},\vec{x}\,')
\!\Big\rangle=\langle\Omega|
\bar{\varphi}_0(t,\vec{x})\!\!\int_0^{t'}\!\!\!d\tilde{t}\,{\bar{\varphi}}^2_0(\tilde{t},\vec{x}\,')
\!\!\int_0^{\tilde{t}}\!\!d\tilde{\tilde{t}}\, {\bar{\varphi}}^3_0(\tilde{\tilde{t}},\vec{x}\,')
|\Omega\rangle \; .\label{eqstqnt1}
\ee

Next, we consider the second expectation value which contributes at $\mathcal{O}(\lambda^2)$ in Eq.~(\ref{expectstoachgenel}),
\be
&&\hspace{-0.5cm}\Big\langle\!
\int_0^{t}\!\!dt''{\bar{\phi}}^2_0(t''\!,\vec{x})\!\!\!
\int_0^{t''}\!\!\!\!dt'''{\bar{\phi}}^3_0(t'''\!,\vec{x})
\bar{\phi}_0(t'\!,\vec{x}\,')\!
\Big\rangle\nonumber\\
&&\hspace{-0.5cm}=\!\!\!\int_0^{t}\!\!dt''\!\!\int_0^{t''}\!\!\!\!dt'''
\,\mathcal{P}\prod_{{\vec{q}}\neq 0}\!\int_{-\infty}^{+\infty}\!\!\!\frac{d\alpha_{\vec{q}}}{\sqrt{2\pi}}
e^{-\frac{1}{2}\alpha_{\vec{q}}^2}\!\!\int_{-\infty}^{+\infty}\!\!\!
\frac{d{\beta_{\vec{q}}}}{\sqrt{2\pi}}e^{-\frac{1}{2}\beta_{\vec{q}}^2}
\,{\bar{\phi}}^2_0(t''\!,\vec{x})\,{\bar{\phi}}^3_0(t'''\!,\vec{x})
\,\bar{\phi}_0(t'\!,\vec{x}\,')\; .
\ee
Proceeding as before, we perform the Gaussian integrals projecting out quartic and hexic combinations of the random variables to obtain
\be
&&\hspace{-0.5cm}\Big\langle\!
\int_0^{t}\!\!dt''{\bar{\phi}}^2_0(t''\!,\vec{x})\!\!\!
\int_0^{t''}\!\!\!\!dt'''{\bar{\phi}}^3_0(t'''\!,\vec{x})
\bar{\phi}_0(t'\!,\vec{x}\,')\!
\Big\rangle\nonumber\\
&&\hspace{-0.5cm}=\!\frac{H^{3D-6}}{2^{6D-6}\pi^{3D-3}}
\frac{\Gamma^6\!\left(\!D\!-\!1\right)}{\Gamma^6\!\left(\!\frac{D}{2}\right)}
\!\!\int_0^t\!\!dt''\!\!\int_0^{t''}\!\!\!\!dt'''\Bigg\{ 6\!\sum_{\vec{n}\neq 0}\frac{\Theta(\!Ha(t'')\!-\!k)\,\Theta(\!Ha(t''')\!-\!k)}{n^{D-1}}\nonumber\\
&&\hspace{-0.5cm}\times\!\sum_{\vec{m}\neq 0}\frac{\Theta(\!Ha(t'')\!-\!k')\,\Theta(\!Ha(t')\!-\!k')}{m^{D-1}}
\cos\!{\left(\vec{k}'\!\cdot\!(\vec{x}\!-\!\vec{x}\,')\!\right)}\!\sum_{\vec{r}\neq 0}\frac{\Theta(\!Ha(t''')\!-\!k'')}{r^{D-1}}\nonumber\\
&&\hspace{-0.5cm}+3\!\sum_{\vec{n}\neq 0}\frac{\Theta(\!Ha(t'')\!-\!k)}{n^{D-1}}
\!\sum_{\vec{m}\neq 0}\frac{\Theta(\!Ha(t''')\!-\!k')}{m^{D-1}}
\!\sum_{\vec{r}\neq 0}\frac{\Theta(\!Ha(t''')\!-\!k'')\,\Theta(\!Ha(t')\!-\!k'')}{r^{D-1}}
\cos\!{\left(\vec{k}''\!\cdot\!(\vec{x}\!-\!\vec{x}\,')\!\right)}\nonumber\\
&&\hspace{-0.5cm}+6\!\sum_{\vec{n}\neq 0}\frac{\Theta(\!Ha(t'')\!-\!k)\,\Theta(\!Ha(t''')\!-\!k)}{n^{D-1}}\!\sum_{\vec{m}\neq 0}\frac{\Theta(\!Ha(t'')\!-\!k')\,\Theta(\!Ha(t''')\!-\!k')}{m^{D-1}}\nonumber\\
&&\hspace{4cm}\times\!\sum_{\vec{r}\neq 0}\frac{\Theta(\!Ha(t''')\!-\!k'')\,\Theta(\!Ha(t')\!-\!k'')}{r^{D-1}}
\cos\!{\left(\vec{k}''\!\cdot\!(\vec{x}\!-\!\vec{x}\,')\!\right)}\!\Bigg\}\; .\label{stoch2loopscnd}
\ee
Now, we use Eqs.~(\ref{whitesum}) and (\ref{sumsingletheta}) to rewrite Eq.~(\ref{stoch2loopscnd}) as
\be
&&\hspace{-0.5cm}\Big\langle\!
\int_0^{t}\!\!dt''{\bar{\phi}}^2_0(t''\!,\vec{x})\!\!\!
\int_0^{t''}\!\!\!\!dt'''{\bar{\phi}}^3_0(t'''\!,\vec{x})
\bar{\phi}_0(t'\!,\vec{x}\,')\!
\Big\rangle\!=\!
\frac{H^{2D-4}}{2^{2D-2}\pi^{D}}\frac{\Gamma^2(\!D\!-\!1)}{\Gamma^2(\!\frac{D}{2})}\!
\int_0^{t}\!\!dt''\!\!\int_0^{t''}\!\!\!\!dt'''
\Bigg\{\nonumber\\
&&\hspace{0cm}6\Big\langle\!
\bar{\phi}_0(t''\!,\vec{x})\bar{\phi}_0(t'\!,\vec{x}\,')\!\Big\rangle\!\ln^2(a(t'''))\!+\!3\Big\langle\!
\bar{\phi}_0(t'''\!,\vec{x})\bar{\phi}_0(t'\!,\vec{x}\,')\!\Big\rangle\!\ln(a(t''))\!\ln(a(t'''))\nonumber\\
&&\hspace{5.3cm}+6\Big\langle\!
\bar{\phi}_0(t'''\!,\vec{x})\bar{\phi}_0(t'\!,\vec{x}\,')\!\Big\rangle\!\ln^2(a(t'''))
\!\Bigg\}\; .\label{stoch2lpscndfin}
\ee
Recall that the stochastic and quantum formalisms are equivalent at tree-order, see  Eq.~(\ref{stocquanttree}). Thus, Eq.~(\ref{stoch2lpscndfin}) is identical to Eq.~(\ref{twoloopsecondexpandequaltime}),
\be
&&\hspace{-1.2cm}\Big\langle\!
\int_0^{t}\!\!dt''{\bar{\phi}}^2_0(t''\!,\vec{x})\!\!\!
\int_0^{t''}\!\!\!\!dt'''{\bar{\phi}}^3_0(t'''\!,\vec{x})
\bar{\phi}_0(t'\!,\vec{x}\,')\!
\Big\rangle\!=\!\langle\Omega|\!\!
\int_0^{t}\!\!dt''{\bar{\varphi}}^2_0(t''\!,\vec{x})\!\!\!
\int_0^{t''}\!\!\!\!dt'''{\bar{\varphi}}^3_0(t'''\!,\vec{x})
\bar{\varphi}_0(t'\!,\vec{x}\,')
|\Omega\rangle\; .\label{eqstqnt2}
\ee

The final expectation value that contributes at $\mathcal{O}(\lambda^2)$ in Eq.~(\ref{expectstoachgenel}) is
\be
&&\hspace{-1cm}\Big\langle\!
\int_0^{t}\!\!dt''{\bar{\phi}}^3_0(t''\!,\vec{x})\!\!\!
\int_0^{t'}\!\!\!d\tilde{t}\,{\bar{\phi}}^3_0(\tilde{t},\vec{x}\,')
\!\Big\rangle\nonumber\\
&&\hspace{0.5cm}=\!\!\!\int_0^{t}\!\!dt''\!\!\int_0^{t'}\!\!\!d\tilde{t}
\,\,\mathcal{P}\prod_{{\vec{q}}\neq 0}\!\int_{-\infty}^{+\infty}\!\!\!\frac{d\alpha_{\vec{q}}}{\sqrt{2\pi}}
e^{-\frac{1}{2}\alpha_{\vec{q}}^2}\!\!\int_{-\infty}^{+\infty}\!\!\!
\frac{d{\beta_{\vec{q}}}}{\sqrt{2\pi}}e^{-\frac{1}{2}\beta_{\vec{q}}^2}\,
{\bar{\phi}}^3_0(t''\!,\vec{x})\,{\bar{\phi}}^3_0(\tilde{t},\vec{x}\,')\; .
\ee
We carry on evaluating the Gaussian integrals projecting out quartic and hexic combinations of the random variables and get,
\be
&&\hspace{-1cm}\Big\langle\!
\int_0^{t}\!dt''{\bar{\phi}}^3_0(t''\!,\vec{x})\!\!
\int_0^{t'}\!\!d\tilde{t}\,{\bar{\phi}}^3_0(\tilde{t},\vec{x}\,')
\!\Big\rangle\nonumber\\
&&\hspace{-1cm}=\!\frac{H^{3D-6}}{2^{6D-6}\pi^{3D-3}}
\frac{\Gamma^6\!\left(\!D\!-\!1\right)}{\Gamma^6\!\left(\!\frac{D}{2}\right)}
\!\!\int_0^{t}\!\!dt''\!\!\int_0^{t'}\!\!d\tilde{t}\Bigg\{6\!\sum_{\vec{n}\neq 0}\frac{\Theta(\!Ha(t'')\!-\!k)}{n^{D-1}}\nonumber\\
&&\hspace{-1cm}\times\!\!\sum_{\vec{m}\neq 0}\frac{\Theta(\!Ha(t'')\!-\!k')\,\Theta(\!Ha(\tilde{t})\!-\!k')}{m^{D-1}}
\cos\!{\left(\vec{k}'\!\cdot\!(\vec{x}\!-\!\vec{x}\,')\!\right)}\!\!\sum_{\vec{r}\neq 0}\frac{\Theta(\!Ha(\tilde{t})\!-\!k'')}{r^{D-1}}\nonumber\\
&&\hspace{-1cm}+3\!\sum_{\vec{n}\neq 0}\frac{\Theta(\!Ha(t'')\!-\!k)\,\Theta(\!Ha(\tilde{t})\!-\!k)}{n^{D-1}}
\cos\!{\left(\vec{k}\!\cdot\!(\vec{x}\!-\!\vec{x}\,')\!\right)}\!\!\sum_{\vec{m}\neq 0}\frac{\Theta(\!Ha(t'')\!-\!k')}{m^{D-1}}\!\sum_{\vec{r}\neq 0}\frac{\Theta(\!Ha(\tilde{t})\!-\!k'')}{r^{D-1}}\nonumber\\
&&\hspace{4cm}+6\Bigg[\!\sum_{\vec{n}\neq 0}\frac{\Theta(\!Ha(t'')\!-\!k)\,\Theta(\!Ha(\tilde{t})\!-\!k)}{n^{D-1}}
\cos\!{\left(\vec{k}\!\cdot\!(\vec{x}\!-\!\vec{x}\,')\!\right)}\Bigg]^3\Bigg\}\; .\label{stoch2looptrd}
\ee
At this stage we employ Eqs.~(\ref{whitesum}) and (\ref{sumsingletheta}) in Eq.~(\ref{stoch2looptrd}),
\be
&&\hspace{-1.5cm}\Big\langle\!
\int_0^{t}\!dt''{\bar{\phi}}^3_0(t''\!,\vec{x})\!\!
\int_0^{t'}\!\!d\tilde{t}\,{\bar{\phi}}^3_0(\tilde{t},\vec{x}\,')
\!\Big\rangle\nonumber\\
&&\hspace{-1.4cm}=\!\!\int_0^{t}\!\!dt''\!\!\!\int_0^{t'}\!\!\!d\tilde{t}\,\Bigg\{9\Big\langle\!
\bar{\phi}_0(t''\!,\vec{x})\bar{\phi}_0(\tilde{t},\vec{x}\,')\!\Big\rangle
\frac{H^{2D-4}}{2^{2D-2}\pi^D}\frac{\Gamma^2(\!D\!-\!1)}{\Gamma^2(\!\frac{D}{2})}\!
\ln(a(t''))\!\ln(a(\tilde{t}))\nonumber\\
&&\hspace{7cm}+6\!\left[\Big\langle\!
\bar{\phi}_0(t''\!,\vec{x})\bar{\phi}_0(\tilde{t},\vec{x}\,')\!\Big\rangle\right]^3
\!\Bigg\}\; .\label{stoch2loopthrdfin}
\ee
The final step is to make use of the identity~(\ref{stocquanttree}) which implies that Eq.~(\ref{stoch2loopthrdfin}) is identical to Eq.~(\ref{2loop3rdmidequaltime}),
\be
\Big\langle\!
\int_0^{t}\!dt''{\bar{\phi}}^3_0(t''\!,\vec{x})\!\!
\int_0^{t'}\!\!d\tilde{t}\,{\bar{\phi}}^3_0(\tilde{t},\vec{x}\,')
\!\Big\rangle\!=\!\langle\Omega|\!\!
\int_0^{t}\!dt''{\bar{\varphi}}^3_0(t''\!,\vec{x})\!\!
\int_0^{t'}\!\!d\tilde{t}\,{\bar{\varphi}}^3_0(\tilde{t},\vec{x}\,')
|\Omega\rangle \; .\label{eqstqnt3}
\ee
Equivalence of the individual stochastic expectation values to their corresponding quantum analogs, Eqs.~(\ref{eqstqnt1}), (\ref{eqstqnt2}) and (\ref{eqstqnt3}), yields
\be
\Big\langle\!
\bar{\phi}(t,\vec{x})\bar{\phi}(t'\!,\vec{x}\,')\!\Big\rangle_{\mathcal{O}(\lambda^2)}\!=\!\langle\Omega|
\bar{\varphi}(t,\vec{x})\bar{\varphi}(t'\!,\vec{x}\,')|\Omega\rangle_{\rm 2-loop}\; .\label{stocquant2lp}
\ee
Thus, Eqs.~(\ref{stocquanttree}), (\ref{stocquant1lp}) and (\ref{stocquant2lp}) show that the stochastic field theory gives the same two-point correlation function~(\ref{genelfullexpect}) of the IR truncated scalar obtained applying QFT
\beeq
\Big\langle\!
\bar{\phi}(t,\vec{x})\bar{\phi}(t'\!,\vec{x}\,')\!\Big\rangle\!=\!\langle\Omega|
\bar{\varphi}(t,\vec{x})\bar{\varphi}(t'\!,\vec{x}\,')|\Omega\rangle\; .\label{stocquantall}
\eneq
The variance of the stochastic field $\bar{\phi}(t,\vec{x})$ defined analogously to Eq.~(\ref{vargeneral}) also yields the same quantum field theoretical result whose equal time and equal space limits are given in Eqs.~(\ref{spatialvariance}) and (\ref{temporalvariance}), respectively. In the next section, we give a stochastic check of the two-point correlation function of the IR truncated scalar in the equal spacetime limit.

\subsection{A check of the $\lim_{x\rightarrow x'}\langle\Omega|
\bar{\varphi}(t,\vec{x})\bar{\varphi}(t'\!,\vec{x}\,')|\Omega\rangle$ via the Fokker-Planck equation}
\label{subsec:check}

Recall that we computed the two-point correlation function $\langle\Omega|
\bar{\varphi}(t,\vec{x})\bar{\varphi}(t'\!,\vec{x}\,')|\Omega\rangle$ in Sec.~\ref{sec:Quant} using QFT in $D$-dimensions. We have just shown in Sec.~\ref{sec:Stoch} that the same result is obtained when SFT is used. In this section, we want to give another stochastic check of our result for the correlation function.

The equal spacetime limits of Eqs.~(\ref{genelfullexpect}) and (\ref{stocquantall}) imply
\be
&&\hspace{-0.4cm}\langle\Omega|
\bar{\varphi}^2(t,\vec{x})|\Omega\rangle\!=\!\Big\langle\!
\bar{\phi}^2(t,\vec{x})\!\Big\rangle\!=\!\frac{H^{D-2}}{2^{D-1}\pi^{D/2}}
\frac{\Gamma(\!D\!-\!1)}{\Gamma(\!\frac{D}{2})}\!\ln(a(t))\!-\!\frac{\lambda}{3(\!D\!-\!1)}\frac{H^{2D-6}}{2^{2D-2}\pi^{D}}
\frac{\Gamma^2(\!D\!-\!1)}{\Gamma^2(\!\frac{D}{2})}\!\ln^3(a(t))\nonumber\\
&&\hspace{4.5cm}+\frac{\lambda^2}{5(\!D\!-\!1)^2}\frac{H^{3D-10}}{2^{3D-3}\pi^{3D/2}}
\frac{\Gamma^3(\!D\!-\!1)}{\Gamma^3(\!\frac{D}{2})}\!\ln^5(a(t))\!+\!\mathcal{O}(\lambda^3)\; .\label{limitofphiphi}
\ee
One can also calculate the expectation values of functionals of
a stochastic field $\bar\phi$ as
\begin{equation}
\Big\langle\!
F\!\left[\bar\phi (t, {\vec x}) \right]
\!\Big\rangle\!=\!\int_{-\infty}^{+\infty}
\!\!d\omega\varrho(t, \omega) F(\omega)
\; , \label{defdistr}
\end{equation}
where the probability density $\varrho(t, \bar\phi)$ is a solution of the Fokker-Planck equation
\be
\frac{\partial}{\partial t}
\varrho(t,\bar\phi)\!\!&=&\!\!\frac{\partial}{\partial \bar\phi}
\Bigl[\frac{V'(\bar\phi)}{(\!D\!-\!1)H} \varrho(t,\bar\phi) \Bigr] \!+\! \frac12 \frac{\partial^2}{\partial
{\bar\phi}^2} \Bigl[\Big\langle\!
f_{\!{\bar\phi}_0}(t,\vec{x}) f_{\!{\bar\phi}_0}(t,\vec{x})\!\Big\rangle \varrho(t,\bar\phi) \Bigr]\; .
\label{FP}\ee
The expectation value we want to check in Eq.~(\ref{limitofphiphi}) is of the form
\begin{equation}
\Big\langle\!
\bar\phi^{2n} (t, {\vec x})
\!\Big\rangle\!=\!\int_{-\infty}^{+\infty}
\!\!d\omega\varrho(t, \omega) \omega^{2n}
\; ,\label{expectviaFP}
\end{equation}
in $D$-dimensions with $n\!=\!1$. The computation of Eq.~(\ref{expectviaFP}) was done in Ref.~\cite{Wstocqgrav} for $D\!=\!4$. Here, we need to do the computation for an arbitrary $D$. Taking the derivative of both sides of Eq.~(\ref{expectviaFP}) with respect to $t$, using Eqs.~(\ref{stocwhite}) and (\ref{FP}), and then integrating the result by parts, we find
\be
&&\hspace{-2cm}\frac{\partial}{\partial t}\Big\langle\!
\bar\phi^{2n} (t, {\vec x})
\!\Big\rangle\!=\!\!\!\int_{-\infty}^{+\infty}
\!\!d\omega \omega^{2n}\frac{\partial}{\partial t}\varrho(t, \omega)\nonumber\\
&&\hspace{-1cm}=\!\!\int_{-\infty}^{+\infty}
\!\!d\omega \Bigg\{\!\!-\!2n\omega^{2n-1}\frac{V'(\omega)}{(\!D\!-\!1)H}\!+\!n(2n\!-\!1)
\omega^{2n-2}\frac{\Gamma(\!D\!-\!1)}{\Gamma(\!\frac{D}{2})}
\frac{H^{D-1}}{2^{D-1}\pi^{D/2}}\Bigg\}\varrho(t, \omega)\nonumber\\
&&\hspace{-1cm}=\!n(2n\!-\!1)\frac{\Gamma(\!D\!-\!1)}{\Gamma(\!\frac{D}{2})}
\frac{H^{D-1}}{2^{D-1}\pi^{D/2}}\Big\langle\!
{\bar\phi}^{2n-2}(t, {\vec x})\!\Big\rangle\!-\!\frac{2n}{(\!D\!-\!1)H}\Big\langle\!
{\bar\phi}^{2n-1}(t, {\vec x})V'(\bar\phi)
\!\Big\rangle\; .\label{stocquartic}
\ee
Defining new variables \beeq
\alpha\!\equiv\!\frac{1}{2^{D-1}\pi^{D/2}}\frac{\Gamma(\!D\!-\!1)}{\Gamma(\!\frac{D}{2})}\ln(a(t))\; , \qquad \bar\lambda\!\equiv\!2^{D-1}\pi^{D/2}\frac{\Gamma(\!\frac{D}{2})}{\Gamma(\!D\!-\!1)}\frac{H^{D-4}}{3(\!D\!-\!1)}\lambda\; ,\label{defininglambdabar}
\eneq
and choosing $V(\bar\phi)\!=\!\frac{\lambda}{4!}{\bar\phi}^4$  in Eq.~(\ref{stocquartic}) we obtain
\beeq
\frac{\partial}{\partial \alpha}\left\langle\!\left(
\!\frac{\bar\phi^2 (t, {\vec x})}{H^{D-2}}\!\right)^n
\right\rangle\!=\!n(2n\!-\!1)\left\langle\!\left(
\!\frac{\bar\phi^2 (t, {\vec x})}{H^{D-2}}\!\right)^{n-1}
\right\rangle\!-\!n\bar\lambda\left\langle\!\left(
\!\frac{\bar\phi^2 (t, {\vec x})}{H^{D-2}}\!\right)^{n+1}
\right\rangle\; .\label{fokker}
\eneq
This equation has the same form obtained in Ref.~\cite{Wstocqgrav}. So does its solution
\beeq
\left\langle\!\left(
\!\frac{\bar\phi^2 (t, {\vec x})}{H^{D-2}}\!\right)^n
\right\rangle\!=\!(2n\!-\!1)!!\alpha^n\!\!\left[1\!-\!\bar\lambda\frac{n}{2}(n\!+\!1)\alpha^2
\!+\!\bar\lambda^2\frac{n}{280}(35n^3\!\!+\!\!170n^2\!+\!225n\!+\!74)\alpha^4\!+\!\mathcal{O}(\bar\lambda^3)\right]\; ,\nonumber
\eneq
which yields,
\beeq
\left\langle\!\left(
\!\frac{\bar\phi^2 (t, {\vec x})}{H^{D-2}}\!\right)
\!\right\rangle\!=\!\alpha\!-\!\bar\lambda\alpha^3
\!+\!\bar\lambda^2\frac{9}{5}\alpha^5\!+\!\mathcal{O}(\lambda^3)\; ,\label{fokkern1}
\eneq
for $n\!=\!1$. Employing Eq.~(\ref{defininglambdabar}) in Eq.~(\ref{fokkern1}) reproduces Eq.~(\ref{limitofphiphi}). The equal spacetime limit of our result~(\ref{genelfullexpect}) is indeed correct.
\section{Conclusions}
\label{sec:conclusions}

We considered infrared truncated massless minimally coupled scalar field $\bar\varphi(t, \vec{x})$ with a quartic self-interaction on a $D$-dimensional locally de Sitter background of an inflating spacetime. In Sec.~\ref{sec:Quant}, we computed the two-point correlation function $\langle\Omega|\bar\varphi(t, \vec{x})
\bar\varphi(t'\!, \vec{x}\,')|\Omega\rangle$ of the scalar at one and two-loop order applying quantum field theory. The tree-order correlator at a fixed comoving separation, that is at increasing physical distance, approaches a constant nonzero value. It grows linearly with comoving time at fixed physical distance. This is because more and more particles are created out of vacuum during inflation which increases local field strength. The dominant (one-loop) quantum correction, however, grows negatively at a fixed comoving separation and at a fixed physical distance.

We used our correlation function to compute the variance
\be
&&\hspace{-1.cm}\sigma^2_{\Delta \bar{\varphi}}(t, t' ; \vec{x}, \vec{x}\,')\!\equiv\!\langle\Omega|\!
\left[\Delta\bar{\varphi} \!-\!\langle\Omega|\Delta\bar{\varphi}|\Omega\rangle\right]^2\!|\Omega\rangle\!=\!\langle\Omega|
\left(\Delta\bar{\varphi}\right)^2\!|\Omega\rangle\nonumber\\
&&\hspace{-1cm}=\langle\Omega|{\bar{\varphi}}^2(t, \vec{x})|\Omega\rangle \!-\!\langle\Omega|\bar{\varphi}(t, \vec{x})\bar{\varphi}(t'\!, \vec{x}\,')|\Omega\rangle\!-\!\langle\Omega|\bar{\varphi}(t'\!, \vec{x}\,')\bar{\varphi}(t, \vec{x})|\Omega\rangle \!+\!\langle\Omega|\bar{\varphi}^2(t'\!, \vec{x}\,')|\Omega\rangle\; ,\nonumber
\ee
and obtained corrections for the variance, in space and time, of one and two-loop correlators. Time dependence of the tree-order variance implies that the magnitude of spatial variation $\Delta \bar{\varphi}(t, \vec{x}, \vec{x}\,')\!=\!
\bar{\varphi}(t, \vec{x})-\bar{\varphi}(t, \vec{x}\,')$ increases with time and freezes in, for a fixed physical distance, to a not especially large value. This spatial variation is the main reason for some cosmologists to argue that using expectation values gives a misleading description of a physical processes because fluctuations about the mean value change the entire picture. A local observer detects either an increase or a decrease in the field strength as the fluctuations take place. The expectation value measures merely the average effect. The authors of Ref.~\cite{W0}, on the other hand, argue that one can roughly trust implications of expectation values under certain circumstances, {\it despite} the spatial and temporal {\it variation} in the real effects. The key issue in this paper is that whether the tree-order comparisons between expectation values and variances found in Ref.~\cite{W0} become better or worse when loop corrections are included. We found out that the one-loop variance grows negatively for fixed physical distance in our self-interacting model. This means that the spatial variance does not grow as large as the tree-order variance indicates. Thus, the contrast between the expectation value and variance decreases when we include the loop corrections.

In Sec.~\ref{sec:Stoch}, we examined the model by repeating the computations of Sec.~\ref{sec:Quant} applying a stochastic field theory where the annihilation and creation operators of the free quantum field are considered to be two complex conjugate stochastic random variables. The real and imaginary parts of the random variables are governed by a Gaussian probability distribution function. We computed the two-point correlation function of the IR truncated stochastic field in our model. The result we found is the same as the one in Sec.~\ref{sec:Quant}. Thus, the analogously defined variance in the context of stochastic field theory is in perfect agreement with the variance obtained using quantum field theory. Finally, we presented a check of our computation for the correlation function. Expectation values of functionals of a stochastic field can be calculated by integrating the functional weighted with a probability density function which satisfies a Fokker-Planck equation. Employing this technique, we computed the expectation value of the stochastic field squared $\langle
\bar\phi^{2} (t, {\vec x})\rangle$ in $D$-dimensions. The result is what our two-point correlation function yields in the equal spacetime limit.

\begin{acknowledgements}
I thank Richard P. Bessel-Woodard for stimulating discussions.
\end{acknowledgements}

\newpage
\begin{appendix}
\section{Computing the two-loop contribution to the two-point correlation function $\langle\Omega|
\bar{\varphi}(t,\vec{x})\bar{\varphi}(t'\!,\vec{x}\,')\!|\Omega\rangle$} \label{App:correlationfunc}

Two-loop contribution to the two-point correlation function of the IR truncated scalar full field consists of three terms each of which is proportional to a VEV; see Eq.~(\ref{expectequaltime}). The VEVs are computed in Sec.~\ref{subsec:2loop}. In this Appendix, we give the details of the computation for each of the VEVs in sections~\ref{App:VEV1}, \ref{App:VEV2} and \ref{App:VEV3}, respectively.
\subsection{Computing the VEV $\langle\Omega|
\bar{\varphi}_0(t,\vec{x})\!\int_0^{t'}\!\!d\tilde{t}\,{\bar{\varphi}}^2_0(\tilde{t},\vec{x}\,')
\!\int_0^{\tilde{t}}\!d\tilde{\tilde{t}}\, {\bar{\varphi}}^3_0(\tilde{\tilde{t}},\vec{x}\,')
|\Omega\rangle$}
\label{App:VEV1}
In Eq.~(\ref{2loop1stequaltime}) computation this VEV is reduced to the evaluations of three double integrals. The first double integral is
\be
&&\hspace{-1cm}\int_0^{t'}\!\!\!\!d\tilde{t}\!\!\int_0^{\tilde{t}}\!\!\!d\tilde{\tilde{t}}\,
2\!\cdot\!3\!\cdot\!1\!\!\left[\mathcal{C}(\!\Delta x)\!+\!\ln(a(\tilde{t}))\!+\!\!\sum_{n=1}^\infty\!
                 \frac{(-1)^{n}(a(\tilde{t})H\!\Delta x)^{2n}}{2n(2n\!+\!1)!}\right]\!\!\ln^2(a(\tilde{\tilde{t}}))\nonumber\\
&&\hspace{-1cm}=\frac{1}{2H^2}\Bigg\{\!\frac{4}{5}\!\ln^5(a(t'))\!+\!\mathcal{C}(\!\Delta x)\!\ln^4(a(t'))\!+\!\!\sum_{n=1}^\infty
\!\frac{(-1)^{n}(\!H\!\Delta x)^{2n}}{n^2(2n\!+\!1)!}\nonumber\\
&&\hspace{2cm}\times\left\{\!a^{2n}(t')\!\!\left[\ln^3(a(t'))
\!-\!\frac{3}{2n}\!\!\left(\!\ln^2(a(t'))\!-\!\frac{\ln(a(t'))}{n}\!\right)\!\!-\!\frac{3}{4n^3}\right]
\!\!+\!\frac{3}{4n^3}\!\right\}\!\!\Bigg\}\; .\label{2loop1st1stintequaltime}
\ee
The second double integral in Eq.~(\ref{2loop1stequaltime}) is
\be
&&\hspace{-1cm}\int_0^{t'}\!\!\!\!d\tilde{t}\!\!\int_0^{\tilde{t}}\!\!\!d\tilde{\tilde{t}}\,
3\!\cdot\!1\!\cdot\!1\!\!\left[\mathcal{C}(\!\Delta x)\!+\!\ln(a(\tilde{\tilde{t}}))\!+\!\!\sum_{n=1}^\infty\!
                 \frac{(-1)^{n}(a(\tilde{\tilde{t}})H\!\Delta x)^{2n}}{2n(2n\!+\!1)!}\right]\!\!\ln(a(\tilde{t}))\!\ln(a(\tilde{\tilde{t}}))\nonumber\\
&&\hspace{-1cm}=\frac{1}{2H^2}\Bigg\{\!\frac{2\!\ln^5(a(t'))}{5}\!+\!\frac{3\mathcal{C}(\!\Delta x)}{4}\!\ln^4(a(t'))\!+\!\frac{3}{4}\!\sum_{n=1}^\infty
\!\frac{(-1)^{n}(\!H\!\Delta x)^{2n}}{n^3(2n\!+\!1)!}\nonumber\\
&&\hspace{2cm}\times\!\left\{\!a^{2n}(t')\!\!\left[\ln^2(a(t'))
\!-\!\frac{3\!\ln(a(t'))}{2n}\!+\!\frac{3}{4n^2}\right]\!\!+\!\frac{\ln^2(a(t'))}{2}\!-\!\frac{3}{4n^2}
\!\right\}\!\!\Bigg\}\; .\label{2loop1st2ndintequaltime}
\ee
The third double integral in Eq.~(\ref{2loop1stequaltime}) is
\be
&&\hspace{-1cm}\int_0^{t'}\!\!\!\!d\tilde{t}\!\!\int_0^{\tilde{t}}\!\!\!d\tilde{\tilde{t}}\,
3\!\cdot\!2\!\cdot\!1\!\!\left[\mathcal{C}(\!\Delta x)\!+\!\ln(a(\tilde{\tilde{t}}))\!+\!\!\sum_{n=1}^\infty\!
                 \frac{(-1)^{n}(a(\tilde{\tilde{t}})H\!\Delta x)^{2n}}{2n(2n\!+\!1)!}\right]\!\!\ln^2(a(\tilde{\tilde{t}}))\nonumber\\
&&\hspace{-1cm}=\frac{1}{2H^2}\Bigg\{\!\frac{3\!\ln^5(a(t'))}{5}\!+\!\mathcal{C}(\!\Delta x)\!\ln^4(a(t'))\!+\!\frac{3}{2}\!\sum_{n=1}^\infty
\!\frac{(-1)^{n}(\!H\!\Delta x)^{2n}}{n^3(2n\!+\!1)!}\nonumber\\
&&\hspace{1cm}\times\!\left\{\!a^{2n}(t')\!\!\left[\ln^2(a(t'))
\!-\!\frac{2\!\ln(a(t'))}{n}\!+\!\frac{3}{2n^2}\right]\!-\!\frac{\ln(a(t'))}{n}
\!-\!\frac{3}{2n^2}\!\right\}\!\!\Bigg\}\; .\label{2loop1st3rdintequaltime}
\ee

\subsection{Computing the VEV $\langle\Omega|\!
\int_0^{t}\!dt''{\bar{\varphi}}^2_0(t''\!,\vec{x})\!
\int_0^{t''}\!\!dt'''{\bar{\varphi}}^3_0(t'''\!,\vec{x})
\bar{\varphi}_0(t'\!,\vec{x}\,')
|\Omega\rangle$}
\label{App:VEV2}
In Eq.~(\ref{twoloopsecondexpandequaltime}) computation this VEV is reduced to the evaluations of three double integrals. Let us evaluate the first one. Because $t'\!\leq\!t$, we can break up the double integral into two (to be able to use Eq.~(\ref{VEVgeneral})): \be\int_0^{t}\!\!dt''\!\!\int_0^{t''}\!\!\!dt'''\!=\!\int_0^{t'}\!\!\!dt''\!\!\int_0^{t''}\!\!\!dt'''
\!+\!\!\int_{t'}^{t}\!\!dt''\!\!\int_0^{t''}\!\!\!dt'''\; .\label{divideint}\ee
In the first double integral on the right we have $0\!\leq\!t''\!\leq\!t'$, whereas $t'\!\leq\!t''\!\leq\!t$ in the second double integral. Thus, using Eqs.~(\ref{VEVgeneral}) and (\ref{divideint}) we can write the first double integral in Eq.~(\ref{twoloopsecondexpandequaltime}) as\be
&&\hspace{-0.3cm}\frac{6H^{D-2}}{2^{D-1}\pi^{D/2}}\frac{\Gamma(\!D\!-\!1)}{\Gamma(\!\frac{D}{2})}
\Bigg\{\!\!\int_0^{t'}\!\!dt''\!\!\int_0^{t''}\!\!\!dt'''\!\!\left[\mathcal{C}(\!\Delta x)\!+\!\ln(a(t''))\!+\!\!\sum_{n=1}^\infty
\!\frac{(-1)^{n}(a(t'')H\!\Delta x)^{2n}}{2n\,(2n\!+\!1)!}\right]\!\!\ln^2(a(t'''))\nonumber\\
&&\hspace{2cm}+\!\!\int_{t'}^t\!dt''\!\!\int_0^{t''}\!\!\!dt'''\!\!\left[\mathcal{C}(\!\Delta x)\!+\!\ln(a(t'))\!+\!\!\sum_{n=1}^\infty
\!\frac{(-1)^{n}(a(t')H\!\Delta x)^{2n}}{2n\,(2n\!+\!1)!}\right]\!\!\ln^2(a(t'''))\Bigg\}\\
&&\hspace{-0.3cm}=\!\frac{H^{D-4}}{2^{D}\pi^{D/2}}\frac{\Gamma(\!D\!-\!1)}{\Gamma(\!\frac{D}{2})}
\Bigg\{\!\!-\!\frac{\ln^5(a(t'))}{5}\!+\!\Big[\!\ln(a(t'))\!+\!\mathcal{C}(\!\Delta x)\Big]\!\ln^4(a(t))
\!+\!\frac{3}{2}\!\sum_{n=1}^\infty
\!\frac{(-1)^{n}(\!H\!\Delta x)^{2n}}{n\,(2n\!+\!1)!}\nonumber\\
&&\hspace{-0.3cm}\times\!\left[a^{2n}(t')\!\left(\!\frac{\ln^4(a(t))\!-\!\ln^4(a(t'))}{3}
\!+\!\frac{2\ln^3(a(t'))}{3n}\!-\!\frac{\ln^2(a(t'))}{n^2}\!+\!\frac{\ln(a(t'))}{n^3}\!-\!\frac{1}{2n^4}\!\!\right)
\!\!+\!\!\frac{1}{2n^4}\right]\!\!\Bigg\}\label{firstint}\; .
\ee

Let us now evaluate the second double integral in Eq.~(\ref{twoloopsecondexpandequaltime}). To break up the double integral appropriately,  consider the three cases that we may have depending on $t'$:
(i) $0\!\leq\!t'\!\leq\!t'''\!\leq\!t''\!\leq\!t$, (ii) $0\!\leq\!t'''\!\leq\!t'\!\leq\!t''\!\leq\!t$ and (iii) $0\!\leq\!t'''\!\leq\!t''\!\leq\!t'\!\leq\!t$. In cases (i) and (ii) we have $t'\!\leq\!t''$, therefore, the second double integral in Eq.~(\ref{divideint}) can be written as \be\int_{t'}^{t}\!\!dt''\!\!\int_0^{t''}\!\!\!dt'''\!=\!\int_{t'}^{t}\!\!\!dt''\!\!\int_0^{t'}\!\!\!dt'''
\!+\!\!\int_{t'}^{t}\!\!dt''\!\!\int_{t'}^{t''}\!\!\!dt'''\; .\label{dividsecondterm}\ee
In case (iii), on the other hand, we have $t''\!\leq\!t'$. Thus, in this case, the second double integral in Eq.~(\ref{divideint}) can be written as\be\int_{t'}^{t}\!dt''\!\!\int_0^{t''}\!\!\!dt'''\!=\!\int_{t'}^{t}\!dt''\!\!\int_0^{t'}\!\!dt'''
\!-\!\!\int_{t'}^{t}\!dt''\!\!\int_{t''}^{t'}\!\!dt'''\!=\!\int_{t'}^{t}\!dt''\!\!\int_0^{t'}\!\!dt'''
\!+\!\!\int_{t'}^{t}\!dt''\!\!\int_{t'}^{t''}\!\!\!dt'''\; .\label{dividsectermcase3}\ee
Note that the last equality in Eq.~(\ref{dividsectermcase3}) is the same as the one in Eq.~(\ref{dividsecondterm}). Combining Eqs.~(\ref{divideint})-(\ref{dividsectermcase3}), we conclude that we can break up the double integral in Eq.~(\ref{divideint}), in all possible cases, as
\be\int_0^{t}\!dt''\!\!\int_0^{t''}\!\!\!dt'''\!=\!\int_0^{t'}\!\!dt''\!\!\int_0^{t''}\!\!\!dt'''
\!+\!\!\int_{t'}^{t}\!dt''\!\!\int_0^{t'}\!\!\!dt'''
\!+\!\!\int_{t'}^{t}\!dt''\!\!\int_{t'}^{t''}\!\!\!dt'''\; .\label{divisonfinal}\ee
Thus, using Eqs.~(\ref{VEVgeneral}) and (\ref{divisonfinal}) the second double integral in Eq.~(\ref{twoloopsecondexpandequaltime}) becomes
\be
&&\hspace{-0.4cm}\frac{3H^{D-2}}{2^{D-1}\pi^{D/2}}\frac{\Gamma(\!D\!-\!1)}{\Gamma(\!\frac{D}{2})}
\Bigg\{\!\!\int_0^{t'}\!\!\!\!\!dt''\!\ln(a(t''))\!\!\!\int_0^{t''}\!\!\!\!\!\!dt'''\!\!\left[\mathcal{C}(\!\Delta x)\!+\!\ln(a(t'''))\!+\!\!\sum_{n=1}^\infty
\!\frac{(-1)^{n}(a(t''')H\!\Delta x)^{2n}}{2n\,(2n\!+\!1)!}\right]\!\!\ln(a(t'''))\nonumber\\
&&\hspace{1.5cm}+\!\!\int_{t'}^t\!\!\!dt''\!\ln(a(t''))\!\!\!\int_0^{t'}\!\!\!\!dt'''\!\!\left[\mathcal{C}(\!\Delta x)\!+\!\ln(a(t'''))\!+\!\!\sum_{n=1}^\infty
\!\frac{(-1)^{n}(a(t''')H\!\Delta x)^{2n}}{2n\,(2n\!+\!1)!}\right]\!\!\ln(a(t'''))\nonumber\\
&&\hspace{1.5cm}+\!\!\int_{t'}^t\!\!\!dt''\!\ln(a(t''))\!\!\!\int_{t'}^{t''}\!\!\!\!\!dt'''\!\!\left[\mathcal{C}(\!\Delta x)\!+\!\ln(a(t'))\!+\!\!\sum_{n=1}^\infty
\!\frac{(-1)^{n}(a(t')H\!\Delta x)^{2n}}{2n\,(2n\!+\!1)!}\right]\!\!\ln(a(t'''))\Bigg\}\\
&&\hspace{-0.3cm}=\frac{H^{D-4}}{2^{D}\pi^{D/2}}\frac{\Gamma(\!D\!-\!1)}{\Gamma(\!\frac{D}{2})}
\Bigg\{\!\frac{3\ln^5(a(t'))}{20}\!-\!\frac{\ln^3(a(t'))}{2}\!\ln^2(a(t))
\!+\!\frac{3}{4}\Big[\!\ln(a(t'))\!+\!\mathcal{C}(\!\Delta x)\Big]\!\ln^4(a(t))\nonumber\\
&&\hspace{-0.3cm}+\frac{3}{8}\!\sum_{n=1}^\infty
\!\frac{(-1)^{n}(\!H\!\Delta x)^{2n}}{n\,(2n\!+\!1)!}\Bigg\{\!a^{2n}(t')\!\Bigg[\!\Big(\!\ln^2(a(t))\!-\!\ln^2(a(t'))\!\Big)^2
\!+\!\frac{2\ln(a(t'))}{n}\Big(\!\ln^2(a(t))\!-\!\ln^2(a(t'))\!\Big)\nonumber\\
&&\hspace{2.3cm}-\frac{\ln^2(a(t))}{n^2}\!+\!\frac{3\ln^2(a(t'))}{n^2}\!-\!\frac{3\ln(a(t'))}{n^3}
\!\Bigg]\!+\!\frac{\ln^2(a(t))}{n^2}\!+\!\frac{3}{2n^4}\Big(a^{2n}(t')\!-\!1\!\Big)\!\Bigg\}\!\Bigg\}\; .\label{secondint}
\ee

The third double integral in Eq.~(\ref{twoloopsecondexpandequaltime}) is obtained similarly. Using Eqs.~(\ref{VEVgeneral}) and (\ref{divisonfinal}) we obtain
\be
&&\hspace{-0.4cm}\frac{6H^{D-2}}{2^{D-1}\pi^{D/2}}\frac{\Gamma(\!D\!-\!1)}{\Gamma(\!\frac{D}{2})}
\Bigg\{\!\!\int_0^{t'}\!\!\!\!dt''\!\!\int_0^{t''}\!\!\!\!dt'''\!\!\left[\mathcal{C}(\!\Delta x)\!+\!\ln(a(t'''))\!+\!\!\sum_{n=1}^\infty
\!\frac{(-1)^{n}(a(t''')H\!\Delta x)^{2n}}{2n\,(2n\!+\!1)!}\right]\!\!\ln^2(a(t'''))\nonumber\\
&&\hspace{1.5cm}+\!\!\int_{t'}^t\!\!\!dt''\!\!\int_0^{t'}\!\!\!\!dt'''\!\!\left[\mathcal{C}(\!\Delta x)\!+\!\ln(a(t'''))\!+\!\!\sum_{n=1}^\infty
\!\frac{(-1)^{n}(a(t''')H\!\Delta x)^{2n}}{2n\,(2n\!+\!1)!}\right]\!\!\ln^2(a(t'''))\nonumber\\
&&\hspace{1.5cm}+\!\!\int_{t'}^t\!\!\!dt''\!\!\int_{t'}^{t''}\!\!\!\!\!dt'''\!\!\left[\mathcal{C}(\!\Delta x)\!+\!\ln(a(t'))\!+\!\!\sum_{n=1}^\infty
\!\frac{(-1)^{n}(a(t')H\!\Delta x)^{2n}}{2n\,(2n\!+\!1)!}\right]\!\!\ln^2(a(t'''))\Bigg\}\\
&&\hspace{-0.3cm}=\frac{H^{D-4}}{2^{D}\pi^{D/2}}\frac{\Gamma(\!D\!-\!1)}{\Gamma(\!\frac{D}{2})}
\Bigg\{\!\frac{3\!\ln^5(a(t'))}{5}\!-\!\ln^4(a(t'))\!\ln(a(t))\!+\!\Big[\!\ln(a(t'))\!+\!\mathcal{C}(\!\Delta x)\Big]\!\ln^4(a(t))\nonumber\\
&&\hspace{-0.3cm}+3\!\sum_{n=1}^\infty
\!\frac{(-1)^{n}(\!H\!\Delta x)^{2n}}{n\,(2n\!+\!1)!}\Bigg\{\!a^{2n}(t')\!\Bigg[
\frac{2\!\ln^4(a(t'))}{3}\Big(\!\!\ln(a(t'))\!-\!\ln(a(t))\!\Big)\!+\!\frac{\ln^4(a(t))\!-\!\ln^4(a(t'))}{6}\nonumber\\
&&\hspace{-0.3cm}+\frac{\ln^2(a(t'))}{n}\Big(\!\!\ln(a(t))\!-\!\ln(a(t'))\!\Big)\!+\!\frac{\ln(a(t'))}{n^2}\Big(\frac{3\!\ln(a(t'))}{2}
\!-\!\ln(a(t))\!\Big)\!+\!\frac{\ln(a(t))}{2n^3}
\!-\!\frac{3\!\ln(a(t'))}{2n^3}\!+\!\frac{3}{4n^4}\Bigg]\nonumber\\
&&\hspace{11cm}-\frac{\ln(a(t))}{2n^3}\!-\!\frac{3}{4n^4}\!\Bigg\}\!\Bigg\}\; .\label{thirdint}
\ee
\subsection{Computing the VEV $\langle\Omega|\!
\int_0^{t}\!dt''{\bar{\varphi}}^3_0(t''\!,\vec{x})\!
\int_0^{t'}\!\!d\tilde{t}\,{\bar{\varphi}}^3_0(\tilde{t},\vec{x}\,')
|\Omega\rangle$}
\label{App:VEV3}
In Eq.~(\ref{2loop3rdaftmidequaltime}) computation this VEV is reduced to the evaluations of four double integrals.

The first double integral in Eq.~(\ref{2loop3rdaftmidequaltime}) yields,
\be
&&\hspace{-1.3cm}\int_0^{\tilde{t}}\!\!\!dt''\!\!\int_0^{t'}\!\!\!d\tilde{t}\,9\!\!\left[\mathcal{C}(\!\Delta x)\!+\!\ln(a(t''))\!+\!\!\sum_{n=1}^\infty
\!\frac{(-1)^{n}(a(t'')H\!\Delta x)^{2n}}{2n\,(2n\!+\!1)!}\!\right]\!\!\ln(a(t''))\!\ln(a(\tilde{t}))\nonumber\\
&&\hspace{-1.3cm}=\!\frac{9}{8H^2}\Bigg\{\!\frac{8\!\ln^5(a(t'))}{15}\!+\!\mathcal{C}(\!\Delta x)\!\ln^4(a(t'))\nonumber\\
&&\hspace{-1cm}\!+\!\sum_{n=1}^\infty
\!\frac{(-1)^{n}(\!H\!\Delta x)^{2n}}{n^3(2n\!+\!1)!}\!\left\{\!a^{2n}(t')\!\!\left[\ln^2(a(t'))
\!-\!\frac{3\!\ln(a(t'))}{2n}\!+\!\frac{3}{4n^2}\right]\!\!+\!\frac{\!\ln^2(a(t'))}{2}
\!-\!\frac{3}{4n^2}\!\right\}\!\!\Bigg\}\; .\label{2loop3rdint1}
\ee
The second double integral in Eq.~(\ref{2loop3rdaftmidequaltime}) yields,
\be
&&\hspace{-0.4cm}\int_0^{\tilde{t}}\!\!\!dt''\!\!\int_0^{t'}\!\!\!d\tilde{t}\,6\!\!\left[\mathcal{C}(\!\Delta x)\!+\!\ln(a(t''))\!+\!\!\sum_{n=1}^\infty
\!\frac{(-1)^{n}(a(t'')H\!\Delta x)^{2n}}{2n\,(2n\!+\!1)!}\right]^3\nonumber\\
&&\hspace{-0.45cm}=\!\frac{3}{H^2}\Bigg\{\!\frac{\ln^5(a(t'))}{10}\!+\!\frac{\mathcal{C}(\!\Delta x)}{2}\!\ln^4(a(t'))\!+\!\mathcal{C}^2(\!\Delta x)\!\ln^3(a(t'))\!+\!\mathcal{C}^3(\!\Delta x)\!\ln^2(a(t'))\nonumber\\
&&\hspace{-0.45cm}+\frac{3}{2}\!\sum_{n=1}^\infty
\!\frac{(-1)^{n}(\!H\!\Delta x)^{2n}}{n^3(2n\!+\!1)!}\Bigg\{\!a^{2n}(t')\!\!\left[\frac{\ln^2(a(t'))}{2}
\!+\!\left(\!\mathcal{C}(\!\Delta x)\!-\!\frac{1}{n}\!\right)\!\ln(a(t'))\!+\!\frac{\mathcal{D}(\!\Delta x, n)}{2}\right]\nonumber\\
&&\hspace{7.5cm}-n\mathcal{E}(\!\Delta x, n)\ln(a(t'))\!-\!\frac{\mathcal{D}(\!\Delta x, n)}{2}\Bigg\}\nonumber\\
&&\hspace{-0.45cm}+\frac{3}{8}\!\sum_{p=2}^\infty\!\sum_{n=1}^{p-1}
\!\frac{(-1)^{p}(\!H\!\Delta x)^{2p}}{p^2n(p\!-\!n)(2n\!+\!\!1\!)![2(p\!-\!n)\!+\!\!1]!}\Bigg\{\!a^{2p}(t')\!\!\left[\ln(a(t'))\!+\!\mathcal{C}(\!\Delta x)\!-\!\frac{1}{p}\right]\!\!-\!\!\Big[\!2p\,\mathcal{C}(\!\Delta x)\!-\!1\!\Big]\!\!\ln(a(t'))\nonumber\\
&&\hspace{-0.45cm}-\mathcal{C}(\!\Delta x)\!+\!\frac{1}{p}\!\Bigg\}\!+\!\frac{1}{16}\!\sum_{q=3}^\infty\!\sum_{p=2}^{q-1}
\!\sum_{n=1}^{p-1}\!\frac{(-1)^{q}(\!H\!\Delta x)^{2q}\left[a^{2q}(t')\!-\!2q\!\ln(a(t'))\!-\!1\right]}
{q^2(q\!-\!p)[2(q\!-\!p)\!+\!\!1]!n(p\!-\!n)(2n\!+\!\!1)![2(p\!-\!n)\!+\!\!1]!}\Bigg\}\; ,\label{2loop3rdint2}
\ee
where \be
\mathcal{D}(\!\Delta x,\! n)\!\equiv\!\mathcal{C}^2(\!\Delta x)-\!\frac{2\mathcal{C}(\!\Delta x)}{n}+\!\frac{3}{2n^2}\qquad{\rm and}\qquad\mathcal{E}(\!\Delta x,\! n)\!\equiv\!\mathcal{C}^2(\!\Delta x)-\!\frac{\mathcal{C}(\!\Delta x)}{n}+\!\frac{1}{2n^2}\; .
\ee
The third double integral in Eq.~(\ref{2loop3rdaftmidequaltime}) yields
\be
&&\hspace{-1.2cm}\int_{\tilde{t}}^{t}\!\!\!dt''\!\!\int_0^{t'}\!\!\!d\tilde{t}\,9\!\!\left[\mathcal{C}(\!\Delta x)\!+\!\ln(a(\tilde{t}))\!+\!\!\sum_{n=1}^\infty
\!\frac{(-1)^{n}(a(\tilde{t})H\!\Delta x)^{2n}}{2n\,(2n\!+\!1)!}\!\right]\!\!\ln(a(t''))\!\ln(a(\tilde{t}))\nonumber\\
&&\hspace{-1cm}=\!\frac{9}{8H^2}\Bigg\{\!\!-\!\frac{4\ln^5(a(t'))}{5}
\!+\!\frac{4\ln^3(a(t'))\ln^2(a(t))}{3}\!+\!\mathcal{C}(\!\Delta x)\!\Big[2\!\ln^2(a(t'))\!\ln^2(a(t))\!-\!\ln^4(a(t'))\Big]\nonumber\\
&&\hspace{-1cm}+\!\sum_{n=1}^\infty\!\frac{(-1)^{n}(\!H\!\Delta x)^{2n}}{n^2(2n\!+\!1)!}\Bigg\{\!a^{2n}(t')\!\Bigg[\!\ln(a(t'))\Big(\!\ln^2(a(t))\!-\!\ln^2(a(t'))\!\Big)
\!+\!\frac{3\!\ln^2(a(t'))}{2n}\!-\!\frac{3\!\ln(a(t'))}{2n^2}\nonumber\\
&&\hspace{4.5cm}-\frac{\ln^2(a(t))}{2n}\!+\!\frac{3}{4n^3}\Bigg]\!\!+\!\frac{\!\ln^2(a(t))}{2n}
\!-\!\frac{3}{4n^3}\!\Bigg\}\!\Bigg\}\; .\label{2loop3rdint3}
\ee
The fourth double integral in Eq.~(\ref{2loop3rdaftmidequaltime}) yields
\be
&&\hspace{-0.34cm}\int_{\tilde{t}}^{t}\!\!\!dt''\!\!\int_0^{t'}\!\!\!d\tilde{t}\,6\!\!\left[\mathcal{C}(\!\Delta x)\!+\!\ln(a(\tilde{t}))\!+\!\!\sum_{n=1}^\infty
\!\frac{(-1)^{n}(a(\tilde{t})H\!\Delta x)^{2n}}{2n\,(2n\!+\!1)!}\!\right]^3\nonumber\\
&&\hspace{-0.34cm}=\!\frac{3}{H^2}\Bigg\{\!\!-\!\frac{2\!\ln^5(a(t'))}{5}\!+\!\frac{\ln^4(a(t'))}{2}\!\ln(a(t))
\!+\!\mathcal{C}(\!\Delta x)\!\!\left[2\!\ln^3(a(t'))\!\ln(a(t))\!-\!\frac{3\!\ln^4(a(t'))}{2}\right]\nonumber\\
&&\hspace{-0.34cm}+\mathcal{C}^2(\!\Delta x)\!\Big[3\!\ln^2(a(t'))\!\ln(a(t))\!-\!2\!\ln^3(a(t'))\Big]\!+\!\mathcal{C}^3(\!\Delta x)\!\Big[2\!\ln(a(t'))\!\ln(a(t))\!-\!\ln^2(a(t'))\Big]\nonumber\\
&&\hspace{-0.34cm}+\frac{3}{2}\!\sum_{n=1}^\infty
\!\frac{(-1)^{n}(\!H\!\Delta x)^{2n}}{n^2(2n\!+\!\!1)!}\Bigg\{\!a^{2n}(t')\!\Bigg[\!\Big(\!\!\ln(a(t))\!-\!\ln(a(t'))\!\Big)\!
\!\left(\!\ln^2(a(t'))
\!+\!\!\left[2\mathcal{C}(\!\Delta x)\!-\!\frac{1}{n}\right]\!\ln(a(t'))\!\right)\nonumber\\
&&\hspace{-0.34cm}+\frac{\ln^2(a(t'))}{2n}\!-\!\mathcal{D}(\!\Delta x,\!n)\!\ln(a(t'))
\!+\!\mathcal{E}(\!\Delta x,\!n)\!\ln(a(t))\!+\!\frac{\mathcal{D}(\!\Delta x,\! n)}{2n}\Bigg]\!\!-\!\mathcal{E}(\!\Delta x,\!n)\!\ln(a(t))\!-\!\frac{\mathcal{D}(\!\Delta x,\!n)}{2n}\!\Bigg\}\nonumber\\
&&\hspace{-0.34cm}+\frac{3}{4}\!\sum_{p=2}^\infty\!\sum_{n=1}^{p-1}
\!\frac{(-1)^{p}(\!H\!\Delta x)^{2p}}{pn(p\!-\!n)(2n\!+\!\!1\!)![2(p\!-\!n)\!+\!\!1]!}\Bigg\{\!a^{2p}(t')\!\Bigg[
\!\Big(\!\!\ln(a(t))\!-\!\ln(a(t'))\!\Big)\!
\!\left(\!\ln(a(t'))
\!+\!\mathcal{C}(\!\Delta x)\!-\!\frac{1}{2p}\!\right)\nonumber\\
&&\hspace{4cm}+\frac{\ln(a(t'))}{2p}
\!+\!\frac{\mathcal{C}(\!\Delta x)}{2p}\!-\!\frac{1}{2p^2}\Bigg]\!\!-\!\left[\mathcal{C}(\!\Delta x)\!-\!\frac{1}{2p}\right]\!\ln(a(t))
\!-\!\frac{\mathcal{C}(\!\Delta x)}{2p}\!+\!\frac{1}{2p^2}\!\Bigg\}\nonumber\\
&&\hspace{-0.34cm}+\frac{1}{8}\!\sum_{q=3}^\infty\!\sum_{p=2}^{q-1}
\!\sum_{n=1}^{p-1}\!\frac{(-1)^{q}(\!H\!\Delta x)^{2q}\!\left[a^{2q}(t')\!\left(\!\ln(a(t))\!-\!\ln(a(t'))\!+\!\frac{1}{2q}\!\right)\!-\!\ln(a(t))\!-\!\frac{1}{2q}\right]}
{q(q\!-\!p)[2(q\!-\!p)\!+\!\!1]!n(p\!-\!n)(2n\!+\!\!1)![2(p\!-\!n)\!+\!\!1]!}\Bigg\}\; .\label{2loop3rdint4}
\ee

\section{Computing the commutator $\left[\varphi_0(t, \vec{x}) , \dot{\varphi}_0(t, \vec{x}\,')\right]$} \label{App:commutator}
Here, we compute the commutator of the free scalar field given in Eq.~(\ref{expantruncwithu}) and its first time derivative
\be
\left[\varphi_0(t, \vec{x}) , \dot{\varphi}_0(t, \vec{x}\,')\right]\!\!&=&\!\!H^{D-1}\!\sum_{n\neq0}\Theta(Ha(t)\!-\!H2\pi n)
\left[u(t, k){\dot{u}}^*(t, k)\!-\!u^*(t, k)\dot{u}(t, k) \right]e^{i\vec{k}\cdot(\vec{x}-\vec{x}\,')}\nonumber\\
&=&\!\!\frac{iH^{D-1}}{a^{D-1}(t)}\!\sum_{n\neq0}\Theta(Ha(t)\!-\!H2\pi n)e^{i2\pi H\vec{n}\cdot(\vec{x}-\vec{x}\,')}\; .\label{comm}
\ee
Making the integral approximation to the discrete mode sum in Eq.~(\ref{comm}) we obtain
\be
\sum_{n\neq0}\Theta(Ha(t)\!-\!H2\pi n)e^{i2\pi H\vec{n}\cdot(\vec{x}-\vec{x}\,')}\!&=&\!\int\!\!d\Omega_{D-1}\!\!\int\!\! \frac{dk k^{D-2}}{(2\pi H)^{D-1}}\Theta(Ha(t)\!-\!k)e^{ik\Delta x \cos(\theta)}\nonumber\\
&=&\!\frac{\pi^{-\frac{D}{2}}}{H^{D-1}}\frac{\Gamma(\!\frac{D}{2})}{\Gamma(\!D\!-\!1)}\!\!\int_H^{Ha(t)}\!\! dk k^{D-2}\frac{\sin(k\Delta x)}{k\Delta x}\; .\label{commutdel}
\ee
Making a change of variable $k\Delta x\!\equiv\!y$ in the integral of Eq.~(\ref{commutdel}) we obtain
\be
\int_H^{Ha(t)}\!\! dk k^{D-2}\frac{\sin(k\Delta x)}{k\Delta x}=\frac{1}{(\Delta x)^{D-1}}\int_{H\!\Delta x}^{Ha(t)\Delta x}\!\!dy\, y^{D-3}\, \sin(y)\; ,\label{intmidlechange}
\ee
where
\be
\hspace{-0.5cm}\int\!dy\, y^{D-3}\sin(y)=(\!D\!-\!3)y^{D-4}\sin(y)\!-\!y^{D-3}\cos(y)\!-\!(\!D\!-\!3)(\!D\!-\!4)\!\!\int\!\!dy\, y^{D-5}\sin(y)\; .\label{intlong}
\ee
Finally, using Eqs.~(\ref{commutdel}), (\ref{intmidlechange}) and (\ref{intlong}) in Eq.~(\ref{comm}) we get the commutator as
\be
&&\hspace{-0.9cm}\left[\varphi_0(t, \vec{x}) , \dot{\varphi}_0(t, \vec{x}\,')\right]\!=\!\frac{i\pi^{-{D}/{2}}}{(a(t)\Delta x)^{D-1}}\frac{\Gamma\!\left(\!\frac{D}{2}\right)}{\Gamma\!\left(\!D\!-\!1\right)}
\Bigg\{\!(\!D\!-\!3)\left(a(t)H\!\Delta x\right)^{D-4}\sin\left(a(t)H\!\Delta x\right)\nonumber\\
&&\hspace{-0.9cm}+\!\left(\!H\!\Delta x\right)^{D-3}\cos\left(\!H\!\Delta x\right)
-\!\left(a(t)H\!\Delta x\right)^{D-3}\cos\left(a(t)H\!\Delta x\right)\!-\!(\!D\!-\!3)\left(\!H\!\Delta x\right)^{D-4}\sin\left(\!H\!\Delta x\right)\nonumber\\
&&\hspace{6.8cm}-(\!D\!-\!3)(\!D\!-\!4)\!\!\int_{H\!\Delta x}^{Ha(t)\Delta x}\!\!dy\, y^{D-5}\sin(y) \Bigg\}\; .
\ee

\end{appendix}


\begin{thebibliography}{99}

\bibitem{OW1} V. K. Onemli and R. P. Woodard, Classical Quantum Gravity {\bf 19}, 4607 (2002).

\bibitem{OW2} V. K. Onemli and R. P. Woodard, Phys. Rev. D {\bf70}, 107301 (2004).

\bibitem{KOW1} E. O. Kahya, V. K. Onemli and R. P. Woodard, Phys. Rev. D {\bf 81}, 023508 (2010).

\bibitem{BOW} T. Brunier, V. K. Onemli and R. P. Woodard, Classical Quantum Gravity {\bf 22}, 59 (2005).

\bibitem{KO} E. O. Kahya and V. K. Onemli, Phys. Rev. D {\bf 76}, 043512 (2007).

\bibitem{O}  V. K. Onemli, Phys. Rev. D {\bf 89}, 083537 (2014).

\bibitem{RPBW1} R. P. Woodard, in {\it Quantum Field Theory under the Influence of External
Conditions}, edited by K. A. Milton (Rinton Press, Princeton, 2004) p.
325, astro-ph/0310757.

\bibitem{rev1} R. P. Woodard, Rept. Prog. Phys. 72, 126002 (2009).

\bibitem{rev2} R. P. Woodard, Int. J. Mod. Phys. D {\bf 23}, 1430020 (2014).

\bibitem{KOW2} E. O. Kahya, V. K. Onemli and R. P. Woodard, Phys. Lett. B {\bf 694}, 101 (2010).

\bibitem{vascfl} References on fluctuations of scalar fields include:
S. Weinberg, Phys. Rev. D {\bf 72}, 043514 (2005); Phys. Rev. D {\bf74}, 023508 (2006);
K. Chaicherdsakul, Phys. Rev. D {\bf75}, 063522 (2007); P. Adshead, R. Easther and E. A. Lim, Phys. Rev. D {\bf79},
063504 (2009); D. Boyanovsky, H. J. de Vega and N. G. Sanchez, Nucl.
Phys. {\bf B747}, 25 (2006); Phys. Rev. D {\bf72}, 103006 (2005); M. Sloth, Nucl. Phys. {\bf B748}, 149 (2006); Nucl. Phys. {\bf B775}, 78 (2007); D. Seery, J. E. Lidsey and M. S. Sloth, J.~Cosmol. Astropart. Phys.~01 (2007) 027; M. van der Meulen and J. Smit, J.~Cosmol. Astropart. Phys.~11 (2007) 023; D. H. Lyth, J.~Cosmol. Astropart. Phys.~12 (2007) 016; D. Seery, J.~Cosmol. Astropart. Phys.~11 (2007) 025; J.~Cosmol. Astropart. Phys.~02 (2008) 006; J.~Cosmol. Astropart. Phys.~05 (2009) 021; Classical Quantum Gravity~{\bf 27},
124005 (2010); N. Bartolo, S. Matarrese, M. Pietroni, A. Riotto and D. Seery, J.~Cosmol. Astropart. Phys.~01 (2008) 015; Y. Urakawa and K. I. Maeda, Phys. Rev. D {\bf78}, 064004 (2008); A. Riotto and M. Sloth,
J.~Cosmol. Astropart. Phys.~04 (2008) 030; J.~Cosmol. Astropart. Phys.~10 (2011) 003; P. Adshead, R. Easther and E. A. Lim, Phys. Rev.
D {\bf 79}, 063504 (2009); Y. Urakawa and T. Tanaka, Prog. Theor. Phys. {\bf 122}, 779 (2009); Prog. Theor.
Phys. {\bf 122}, 1207 (2009); Phys. Rev. D {\bf82}, 121301 (2010);
Prog. Theor. Phys. {\bf 125}, 1067 (2011); J.~Cosmol. Astropart. Phys.~05 (2011)
014; Y. Urakawa, Prog. Theor. Phys. {\bf 126}, 961 (2011); S. B. Giddings and M. S. Sloth,
J.~Cosmol. Astropart. Phys.~07 (2010) 015; J.~Cosmol. Astropart. Phys.~01 (2011) 023; Phys. Rev.
D {\bf84}, 063528 (2011); Phys. Rev. D {\bf 86}, 083538 (2012); J.
Kumar, L. Leblond and A. Rajaraman, J.~Cosmol. Astropart. Phys.~04 (2010) 024; C.
P. Burgess, R. Holman, L. Leblond and S. Shandera, J.~Cosmol. Astropart. Phys.~03 (2010) 033;
D. Boyanovsky, Phys. Rev. D {\bf 85}, 123525 (2012); Phys. Rev. D {\bf 86}, 023509 (2012); K. Feng, Y.-F.
Cai and Y.-S. Piao, Phys. Rev. D {\bf 86}, 103515 (2012); A. Kaya, Int. J. Mod. Phys. D {\bf 17}, 2441 (2009); Phys. Rev. D {\bf 81}, 023521 (2010); Phys. Rev. D {\bf 90}, 043506 (2014); A. Kaya and E. S. Kutluk, arXiv:1409.2884;
E.T. Akhmedov, Int. J. Mod. Phys. D {\bf 23}, 1430001 (2014); K. Larjo and D. Lowe, Phys. Rev. D {\bf 87},
083506 (2013); J. Serreau and R. Parentani, Phys. Rev. D {\bf 87},
085012 (2013); J. Serreau, Phys. Lett. B {\bf 728}, 380 (2014); L. Lello, D. Boyanovsky and
R. Holman, Phys. Rev. D {\bf 89}, 063533 (2014); M. Herranen, T. Markkanen and A.
Tranberg, J. High Energy Phys. 05 (2014) 026.

\bibitem{recover} S. J. Rey, Nucl. Phys. {\bf B284}, 706 (1987);
M. Sasaki, Y. Nambu and K. I. Nakao, Nucl. Phys. {\bf B308}, 868 (1988);
S. Winitzki and A. Vilenkin, Phys. Rev. D {\bf 61}, 084008 (2000);
J. Martin and M. Musso, Phys. Rev. D {\bf 73}, 043517 (2006);
K. Enqvist, S. Nurmi, D. Podolsky and G. I. Rigopoulos, J.~Cosmol. Astropart. Phys.~04 (2008) 025.

\bibitem{initial} A. Vilenkin, Phys. Rev. D {\bf 27}, 2848 (1983);
Y. Nambu and M. Sasaki, Phys. Lett. B {\bf 219}, 240 (1989).

\bibitem{global} A. S. Goncharov, A. D. Linde and V. F. Mukhanov, Int. J.
Mod. Phys. A {\bf 2}, 561 (1987);
A. D. Linde and A. Mezhlumian, Phys. Lett. B {\bf 307}, 25 (1993).

\bibitem{nG} G. I. Rigopoulos, E. P. S. Shellard and B. J. W. van Tent,
Phys. Rev. D {\bf 72}, 083507 (2005); Phys. Rev. D {\bf 73}, 083521 (2006); Phys. Rev. D {\bf 73}, 083522 (2006).

\bibitem{Kuhnel} F.~Kuhnel and D.~J.~Schwarz, Phys. Rev. D {\bf 78}, 103501 (2008);
Phys. Rev. D {\bf 79}, 044009 (2009); Phys. Rev. Lett. {\bf 105}, 211302 (2010).

\bibitem{sari} C.-J. Feng, X.-Z. Li and E. N. Saridakis, Phys. Rev. D {\bf 82}, 023526 (2010); 
T.~Qiu and E.~N.~Saridakis, Phys. Rev. D {\bf 85}, 043504 (2012).

\bibitem{Staro} A.~A.~Starobinsky, in {\it Field Theory, Quantum
Gravity and Strings}, edited by H.~J.~de~Vega and N.~Sanchez
(Sringer-Verlag, Berlin, 1986), p. 107.

\bibitem{W1}
R.~P.~Woodard, Nucl. Phys.~B, Proc. Suppl. {\bf 148}, 108 (2005).

\bibitem{W4}
R.~P.~Woodard, J. Phys. Conf. Ser. {\bf 68}, 012032 (2007).

\bibitem{WRGflow}
R.~P.~Woodard, Phys. Rev. Lett. {\bf 101}, 081301 (2008).

\bibitem{W3}
S.~P.~Miao and R.~P.~Woodard, Phys. Rev.~D~{\bf 74},
044019 (2006).

\bibitem{Wstocsqed}
T.~Prokopec, N.~C.~Tsamis and R.~P.~Woodard, Ann. Phys. (Amsterdam)~{\bf 323}, 1324 (2008).

\bibitem{MiaoWood} S.~P.~Miao and R.~P.~Woodard, Classical Quantum Gravity {\bf 25}, 145009 (2008).

\bibitem{Kit1}
H.~Kitamoto and Y.~Kitazawa, Phys. Rev.~D~{\bf 83}, 104043 (2011).

\bibitem{Kit2}
H.~Kitamoto and Y.~Kitazawa, Phys. Rev.~D~{\bf 85}, 044062 (2012).

\bibitem{W0} N.~C.~Tsamis, A.~Tzetzias and R.~P.~Woodard, J.~Cosmol. Astropart. Phys.~09 (2010) 016.

\bibitem{Rig1}
B.~Garbrecht, F.~Gautier, G.~Rigopoulos and Y.~Zhu, Phys. Rev.~D~{\bf 89},
063506 (2014).

\bibitem{Rig2}
B.~Garbrecht, F.~Gautier, G.~Rigopoulos and Y.~Zhu, arXiv:1412.4893.

\bibitem{SY} A.~A.~Starobinsky and J.~Yokoyama, Phys. Rev.~D~{\bf 50},
6357 (1994).

\bibitem{Wstocqgrav}
N.~C.~Tsamis and R.~P.~Woodard, Nucl. Phys.~{\bf B724}, 295 (2005).

\bibitem{FMSVV1}
F.~Finelli, G.~Marozzi, A.~A.~Starobinsky, G.~P.~Vacca and G.~Venturi, Phys. Rev.~D~{\bf 79},
044007 (2009).

\bibitem{FMSVV2}
F.~Finelli, G.~Marozzi, A.~A.~Starobinsky, G.~P.~Vacca and G.~Venturi, Phys. Rev.~D~{\bf 82},
064020 (2010).

\bibitem{Tegmark}
M.~Tegmark, Phys. Rev.~D~{\bf 85}, 123517 (2012).


\end{thebibliography}
\end{document}